\documentclass{nature}

\usepackage[left=2.5cm, right=2.5cm,top=2.5cm,bottom=3cm]{geometry}
\newgeometry{top=1cm, bottom=2cm, left=2cm, right=2cm}

\usepackage{graphicx}
\usepackage[font=normal, skip=0pt, labelfont=bf]{caption}

\usepackage{bm}

\usepackage{xspace}
\usepackage{booktabs}

\usepackage{amsmath, amssymb}
\usepackage{times}
\usepackage{multirow}

\usepackage{fontawesome}

\usepackage[parfill]{parskip}

\usepackage[usenames]{xcolor}
\definecolor{mpl_red}{HTML}{D62728}

\usepackage[breaklinks, plainpages=false, colorlinks=true, anchorcolor=blue!50!black, citecolor=blue!50!black, linkcolor=blue!50!black, urlcolor=mpl_red, bookmarks=false]{hyperref}

\usepackage{lineno}

\makeatletter

\let\saved@includegraphics\includegraphics
\AtBeginDocument{\let\includegraphics\saved@includegraphics}
\makeatother

\renewcommand{\figurename}{\textbf{Fig.}}

\newcommand{\arcmin}{\mbox{$^\prime$}}
\newcommand{\arcsec}{\mbox{$^{\prime\prime}$}}

\usepackage[caption=false]{subfig}
\makeatletter
\renewcommand\setcaptiontype[1]{\edef\@captype{#1}\ignorespaces}
\makeatother

\usepackage{longtable}
\usepackage{supertabular}

\usepackage{multibib}

\makeatletter
\def\@mb@citenamelist{cite,citep,citet,citen,citealp,citealt,citepalias,citetalias}
\makeatother

\newcites{supplementary}{References (Supplementary Information)}

\title{Multiwavelength Constraints on the Origin of a Nearby \break Repeating Fast Radio Burst Source in a Globular Cluster}

\newcommand{\ABP}{\href{https://orcid.org/0000-0002-8912-0732}{\textcolor{blue!50!black}{Aaron~B.~Pearlman}}}
\newcommand{\PS}{\href{https://orcid.org/0000-0002-7374-7119}{\textcolor{blue!50!black}{Paul~Scholz}}}
\newcommand{\SB}{\href{https://orcid.org/0000-0003-1611-0063}{\textcolor{blue!50!black}{Suryarao~Bethapudi}}}
\newcommand{\JWTH}{\href{https://orcid.org/0000-0003-2317-1446}{\textcolor{blue!50!black}{Jason~W.~T.~Hessels}}}
\newcommand{\VMK}{\href{https://orcid.org/0000-0001-9345-0307}{\textcolor{blue!50!black}{Victoria~M.~Kaspi}}}
\newcommand{\FK}{\href{https://orcid.org/0000-0001-6664-8668}{\textcolor{blue!50!black}{Franz~Kirsten}}}
\newcommand{\KN}{\href{https://orcid.org/0000-0003-0510-0740}{\textcolor{blue!50!black}{Kenzie~Nimmo}}}
\newcommand{\LGS}{\href{https://orcid.org/0000-0002-3775-8291}{\textcolor{blue!50!black}{Laura~G.~Spitler}}}
\newcommand{\EF}{\href{https://orcid.org/0000-0001-8384-5049}{\textcolor{blue!50!black}{Emmanuel~Fonseca}}}
\newcommand{\BWM}{\href{https://orcid.org/0000-0001-8845-1225}{\textcolor{blue!50!black}{Bradley~W.~Meyers}}}
\newcommand{\IS}{\href{https://orcid.org/0000-0001-9784-8670}{\textcolor{blue!50!black}{Ingrid~H.~Stairs}}}
\newcommand{\CMT}{\href{https://orcid.org/0000-0001-7509-0117}{\textcolor{blue!50!black}{Chia~Min~Tan}}}
\newcommand{\MB}{\href{https://orcid.org/0000-0002-3615-3514}{\textcolor{blue!50!black}{Mohit~Bhardwaj}}}
\newcommand{\SC}{\href{https://orcid.org/0000-0002-2878-1502}{\textcolor{blue!50!black}{Shami~Chatterjee}}}
\newcommand{\AMC}{\href{https://orcid.org/0000-0001-6422-8125}{\textcolor{blue!50!black}{Amanda~M.~Cook}}}
\newcommand{\APC}{\href{https://orcid.org/0000-0002-8376-1563}{\textcolor{blue!50!black}{Alice~P.~Curtin}}}
\newcommand{\FAD}{\href{https://orcid.org/0000-0003-4098-5222}{\textcolor{blue!50!black}{Fengqiu~Adam~Dong}}}
\newcommand{\TE}{\href{https://orcid.org/0000-0003-0307-9984}{\textcolor{blue!50!black}{Tarraneh~Eftekhari}}}
\newcommand{\BMG}{\href{https://orcid.org/0000-0002-3382-9558}{\textcolor{blue!50!black}{B.~M.~Gaensler}}}
\newcommand{\TG}{\href{https://orcid.org/0000-0002-3531-9842}{\textcolor{blue!50!black}{Tolga~G\"uver}}}
\newcommand{\JK}{\href{https://orcid.org/0000-0003-4810-7803}{\textcolor{blue!50!black}{Jane~Kaczmarek}}}
\newcommand{\CL}{\href{https://orcid.org/0000-0002-4209-7408}{\textcolor{blue!50!black}{Calvin~Leung}}}
\newcommand{\KWM}{\href{https://orcid.org/0000-0002-4279-6946}{\textcolor{blue!50!black}{Kiyoshi~W.~Masui}}}
\newcommand{\DM}{\href{https://orcid.org/0000-0002-2551-7554}{\textcolor{blue!50!black}{Daniele~Michilli}}}
\newcommand{\TAP}{\href{https://orcid.org/0000-0002-8850-3627}{\textcolor{blue!50!black}{Thomas~A.~Prince}}}
\newcommand{\KRS}{\href{https://orcid.org/0000-0003-3154-3676}{\textcolor{blue!50!black}{Ketan~R.~Sand}}}
\newcommand{\KS}{\href{https://orcid.org/0000-0002-6823-2073}{\textcolor{blue!50!black}{Kaitlyn~Shin}}}
\newcommand{\KMS}{\href{https://orcid.org/0000-0002-2088-3125}{\textcolor{blue!50!black}{Kendrick~M.~Smith}}}
\newcommand{\SPT}{\href{https://orcid.org/0000-0003-2548-2926}{\textcolor{blue!50!black}{Shriharsh~P.~Tendulkar}}}

\newcommand{\ASTRON}{ASTRON, Netherlands Institute for Radio Astronomy, Dwingeloo, The Netherlands.}

\newcommand{\CALTECH}{Division of Physics, Mathematics, and Astronomy, California Institute of Technology, Pasadena, CA, USA.}
\newcommand{\CIFARAZGS}{CIFAR Azrieli Global Scholars Program, MaRS Centre, Toronto, Ontario, Canada.}
\newcommand{\CMU}{McWilliams Center for Cosmology, Department of Physics, Carnegie Mellon University, Pittsburgh, PA, USA.}
\newcommand{\CU}{Cornell Center for Astrophysics and Planetary Science, Cornell University, Ithaca, NY, USA.}
\newcommand{\CUT}{Department of Space, Earth and Environment, Chalmers University of Technology, Onsala Space Observatory, Onsala, Sweden.}
\newcommand{\CSIRO}{CSIRO Space and Astronomy, Parkes Observatory, Parkes, New South Wales, Australia.}
\newcommand{\DAA}{David A.~Dunlap Department of Astronomy and Astrophysics, University of Toronto, Toronto, Ontario, Canada.}
\newcommand{\DI}{Dunlap Institute for Astronomy and Astrophysics, University of Toronto, Toronto, Ontario, Canada.}

\newcommand{\ICRARC}{International Centre for Radio Astronomy Research, Curtin University, Bentley, Western Australia, Australia.}
\newcommand{\IU}{Department of Astronomy and Space Sciences, Science Faculty, Istanbul University, Istanbul, Turkey.}
\newcommand{\IUOBS}{Istanbul University Observatory Research and Application Center, Istanbul University, Istanbul, Turkey.}
\newcommand{\MCGILL}{Department of Physics, McGill University, Montr\'eal, Qu\'ebec, Canada.}
\newcommand{\MITK}{MIT Kavli Institute for Astrophysics and Space Research, Massachusetts Institute of Technology, Cambridge, MA, USA.}
\newcommand{\MITP}{Department of Physics, Massachusetts Institute of Technology, Cambridge, MA, USA.}
\newcommand{\MPIFR}{Max-Planck-Institut f{\"u}r Radioastronomie, Bonn, Germany.}

\newcommand{\NCRA}{National Centre for Radio Astrophysics, Pune, India.}

\newcommand{\NU}{Center for Interdisciplinary Exploration and Research in Astrophysics, Department of Physics and Astronomy, Northwestern University, Evanston, IL, USA.}
\newcommand{\PI}{Perimeter Institute for Theoretical Physics, Waterloo, Ontario, Canada.}
\newcommand{\TIFR}{Department of Astronomy and Astrophysics, Tata Institute of Fundamental Research, Mumbai, India.}
\newcommand{\TSI}{Trottier Space Institute, McGill University, Montr\'eal, Qu\'ebec, Canada.}
\newcommand{\UBC}{Department of Physics and Astronomy, University of British Columbia, Vancouver, British Columbia, Canada.}
\newcommand{\UBCO}{Department of Computer Science, Math, Physics, and Statistics, University of British Columbia, Okanagan Campus, Kelowna, British Columbia, Canada.}
\newcommand{\UCBASTRO}{Department of Astronomy, University of California Berkeley, Berkeley, CA, USA.}
\newcommand{\UCSC}{Department of Astronomy and Astrophysics, University of California Santa Cruz, Santa Cruz, CA, USA.}
\newcommand{\UVA}{Anton Pannekoek Institute for Astronomy, University of Amsterdam, Amsterdam, The Netherlands.}
\newcommand{\WVUPA}{Department of Physics and Astronomy, West Virginia University, Morgantown, WV, USA.}
\newcommand{\WVUGWAC}{Center for Gravitational Waves and Cosmology, Chestnut Ridge Research Building, West Virginia University, Morgantown, WV, USA.}
\newcommand{\YORK}{Department of Physics and Astronomy, York University, Toronto, Ontario, Canada.}

\author{\ABP$^{1, 2, 3}$, \PS$^{4, 5}$, \SB$^{6}$, \JWTH$^{1, 2, 7, 8}$, \break \VMK$^{1, 2}$, \FK$^{8, 9}$, \KN$^{10}$, \LGS$^{6}$, \EF$^{11, 12}$, \BWM$^{13, 14}$, \IS$^{14}$, \CMT$^{1,2,13}$, \MB$^{15}$, \SC$^{16}$, \break \AMC$^{4, 17}$, \APC$^{1, 2}$, \FAD$^{14}$, \TE$^{18}$, \break \BMG$^{4, 17, 19}$, \TG$^{20, 21}$, \JK$^{22, 23}$, \CL$^{24}$, \KWM$^{10, 25}$, \DM$^{10, 25}$, \TAP$^{3}$, \KRS$^{1, 2}$, \KS$^{10, 25}$, \KMS$^{26}$, and \SPT$^{27, 28, 29}$}

\begin{document}

\maketitle

\begin{affiliations}
	\item[] {{\small\faEnvelopeO}~}E-mail: \href{mailto:aaron.b.pearlman@physics.mcgill.ca}{aaron.b.pearlman@physics.mcgill.ca}
	\item \,\MCGILL
	\item \,\TSI
	\item \,\CALTECH
	\item \,\DI
	\item \,\YORK
	\item \,\MPIFR
	\item \,\UVA
	\item \,\ASTRON
	\item \,\CUT
	\item \,\MITK
	\item \,\WVUPA
	\item \,\WVUGWAC
	\item \,\ICRARC
	\item \,\UBC
	\item \,\CMU
	\item \,\CU
	\item \,\DAA
	\item \,\NU
	\item \,\UCSC
	\item \,\IU
	\item \,\IUOBS
	\item \,\CSIRO
	\item \,\UBCO
	\item \,\UCBASTRO
	\item \,\MITP
	\item \,\PI
	\item \,\TIFR
	\item \,\NCRA
	\item \,\CIFARAZGS
\end{affiliations}


\begin{abstract}

\textbf{The precise origins of fast radio bursts~(FRBs) remain unknown. Multiwavelength observations of nearby FRB~sources can provide important insights into the enigmatic FRB~phenomenon. Here, we present results from a sensitive, broadband \text{X-ray} and radio observational campaign of \break FRB~20200120E, the closest known extragalactic repeating FRB~source (located 3.63\,Mpc away in an $\sim$10-Gyr-old globular cluster). We place deep limits on the persistent and prompt \text{X-ray} emission from FRB~20200120E, which we use to constrain possible origins for the source. We compare our results with various classes of \text{X-ray} sources, transients, and FRB models. We find that FRB~20200120E is unlikely to be associated with ultraluminous \text{X-ray} bursts, magnetar-like giant flares, or an SGR~1935+2154-like intermediate flare. Although other types of bright magnetar-like intermediate flares and short \text{X-ray} bursts would have been detectable from FRB~20200120E during our observations, we cannot entirely rule them out as a class. We show that FRB~20200120E is unlikely to be powered by an ultraluminous \text{X-ray} source or a young extragalactic pulsar embedded in a Crab-like nebula. We also provide new constraints on the compatibility of FRB~20200120E with accretion-based FRB~models involving \text{X-ray} binaries. These results highlight the power of multiwavelength observations of nearby~FRBs for discriminating between FRB~models.}

\end{abstract}

\bigskip


FRBs are energetic, short-duration bursts of coherent radio emission\,\cite{Lorimer+2007} that typically emanate from objects located well outside the Milky Way\,\cite{Petroff+2019, Cordes+2019}. A wide variety of models have been proposed to explain the origins of FRBs\,\cite{Platts+2019}. However, models involving magnetars are presently favoured\,\cite{Kulkarni+2014, Pen+2015}.

FRBs and Galactic radio magnetars have been observed to produce radio emission with strikingly similar characteristics\,\cite{Pearlman+2018a, CHIME+2020a, Bochenek+2020c, Israel+2021}. The strongest link between FRBs and magnetars was established when an FRB-like radio burst was detected from the Galactic magnetar SGR~1935+2154 on 28~April~2020\,\cite{CHIME+2020a, Bochenek+2020c}, demonstrating that similar events could be detectable from extragalactic distances. A short \text{X-ray} burst was also detected near the time of this FRB-like radio burst from SGR~1935+2154\,\cite{Mereghetti+2020a, Li+2021, Ridnaia+2021, Tavani+2021}. The spectrum of the \text{X-ray} burst was harder than other short \text{X-ray} bursts detected from SGR~1935+2154\,\cite{Mereghetti+2020a, Ridnaia+2021}, but it was not unusually energetic relative to other short \text{X-ray} bursts from Galactic magnetars. These observations suggest that some fraction of the cosmological FRB population may be powered by extragalactic, \text{X-ray}-emitting magnetars.

FRB~20200120E was discovered using the Canadian Hydrogen Intensity Mapping Experiment~(CHIME) radio telescope and the CHIME/FRB system\,\cite{CHIME+2018, Bhardwaj+2021a}. Using radio telescopes from the European Very Long Baseline Interferometry~(VLBI) Network~(EVN), FRB~20200120E was subsequently localized to an $\sim$10-Gyr-old globular cluster within the M81~galactic system using VLBI\,\cite{Kirsten+2022}.

FRB~20200120E is the nearest known extragalactic repeating FRB source. Previous simultaneous \text{X-ray} and radio observations of more distant FRB sources (for example, see refs.~\citen{Pilia+2020, Scholz+2020a, Laha+2022}) have been unable to rule out most viable FRB source types because of the sensitivity limitations of current \text{X-ray} instruments. In addition, the radio bursts detected from FRB~20200120E sometimes display pulse components with widths of $\sim$\text{10--100}\,ns\,\cite{Majid+2021, Nimmo+2022}, which is the shortest-duration radio emission observed from an FRB source so far. These previous \text{X-ray} studies\,\cite{Pilia+2020, Scholz+2020a, Laha+2022} have not searched for prompt \text{X-ray} emission from FRBs on such ultrashort timescales.

In this Article, we present results from a simultaneous \text{X-ray} and radio observing campaign of \break FRB~20200120E, which yielded the deepest persistent and prompt \text{X-ray} luminosity limits from an FRB source so far in the soft \text{X-ray} band. We derive constraints on the possible origins of FRB~20200120E and compare our results with various types of \text{X-ray} sources, transients, and predictions for multiwavelength emission from different FRB models.


\bigskip

\section*{Observations}
\label{sec:observations}

High-time-resolution radio observations of FRB~20200120E were carried out using the Effelsberg \text{100-m} and CHIME radio telescopes. We recorded radio data at a central frequency of 1.4\,GHz with Effelsberg and in the \text{400--800}\,MHz frequency range using the CHIME/Pulsar system (Methods). We also performed \text{X-ray} observations of FRB~20200120E using the \textit{Neutron Star Interior Composition Explorer} (\textit{NICER}), \textit{XMM-Newton}, \textit{Chandra}, and \textit{NuSTAR} (Methods). These observations were coordinated to maximize our simultaneous \text{X-ray} and radio coverage of FRB~20200120E. A timeline of these observations is shown in Extended Data Fig.~\ref{fig:xray_radio_obs_timeline}, and the observations are listed in Extended Data Tables~\ref{tab:radio_obs} and~\ref{tab:xray_obs}.


\bigskip

\section*{Results}
\label{sec:results}

We detected nine radio bursts from FRB~20200120E during our Effelsberg observations. No radio bursts were detected during our CHIME/Pulsar observations. Throughout this Article, we label the radio bursts from FRB~20200120E as B$n$, and they are ordered chronologically by their time of arrival~(ToA). Bursts~\text{B1--B5} were detected during Pinpointing Repeating CHIME Sources with the EVN~(PRECISE) VLBI observations of FRB~20200120E\,\cite{Kirsten+2022, Nimmo+2022}, and bursts~\text{B6--B9} were detected during target of opportunity~(ToO) observations of FRB~20200120E with Effelsberg (Methods). The dynamic spectra, burst profiles, and frequency spectra of these radio bursts are shown in Fig.~\ref{fig:radio_bursts}. We provide measurements of the radio burst properties in Extended Data Table~\ref{tab:radio_burst_properties}.

Burst~B4 occurred during a simultaneous \textit{NICER} exposure (Extended Data Fig.~\ref{fig:nicer_b4_xray_radio_light_curves}a), and it was the most energetic radio burst detected. Together with measurements from \textit{NICER}~(Table~\ref{tab:nicer_xmm_b4_b9_xray_limits}, Extended Data Figs.~\ref{fig:nicer_b4_xray_radio_light_curves} and~\ref{fig:nicer_bkgd_xray_flux}, and Extended Data Table~\ref{tab:nicer_xray_bkg}), it provides the best constraints on prompt \text{X-ray} emission from FRB~20200120E. The radio burst profiles of~B4 are shown in Extended Data Fig.~\ref{fig:nicer_b4_xray_radio_light_curves}b--d at time resolutions of 8\,$\mu$s, 1\,$\mu$s, and 31.25\,ns, along with the barycentric time separations of the nearest photons from the peak of the radio burst. The nearest photon detected by \textit{NICER} in the \text{0.5--10}\,keV band occurred $\sim$300\,ms before the barycentric, infinite frequency peak time of burst~B4.

Bursts~\text{B6--B9} were detected during simultaneous ToO \text{X-ray} observations with \textit{XMM-Newton}. The \text{0.5--10}\,keV \textit{XMM-Newton} EPIC/pn light curves from these observations are shown in Extended Data Fig.~\ref{fig:xmm_b6_b7_b8_b9_xray_radio_light_curves}a, along with the ToAs of bursts~\text{B6--B9} (Extended Data Table~\ref{tab:radio_burst_properties}). The burst profiles of~\text{B6--B9}, times of the nearest photons, and their barycentric time separations from the peaks of the radio bursts are shown in Extended Data Fig.~\ref{fig:xmm_b6_b7_b8_b9_xray_radio_light_curves}\text{b--e}. Bursts~\text{B6--B8} occurred during times of high background flaring, so we provide prompt \text{X-ray} limits using \textit{XMM-Newton} EPIC/pn measurements at the time of burst~B9 only. The nearest photon occurred 58.4\,s before the barycentric, infinite frequency peak time of burst~B9.

We did not detect any statistically significant increase in \text{X-ray} emission near the times of bursts~\text{B1--B9} that could be attributed to FRB~20200120E. We also carried out a search for periodic \text{X-ray} pulsations but did not detect evidence of pulsed \text{X-ray} emission (Methods). In Table~\ref{tab:nicer_xmm_b4_b9_xray_limits}, we list 3$\sigma$ upper limits on the \text{0.5--10}\,keV \text{X-ray} fluence and \text{X-ray} energy at the times of bursts~B4 and~B9. For~B4, we calculated \text{X-ray} fluence limits using measurements from \textit{NICER} on 100\,ns to 10\,s timescales. \text{X-ray} fluence limits were also calculated for burst~B9 on 100\,ms to 100\,s timescales using measurements from \textit{XMM-Newton}'s EPIC/pn camera. The \text{X-ray} fluence limits were derived using standard 3$\sigma$ confidence intervals (Methods) and several fiducial spectral models that have been used to characterize prompt high-energy emission from magnetars (Table~\ref{tab:nicer_xmm_b4_b9_xray_limits}).

\begin{figure*}
	
	\vspace*{-0.7cm}
	
	\centering
	
	\includegraphics[trim=0cm 0cm 0cm 0cm, clip=true, scale=0.750, angle=0]{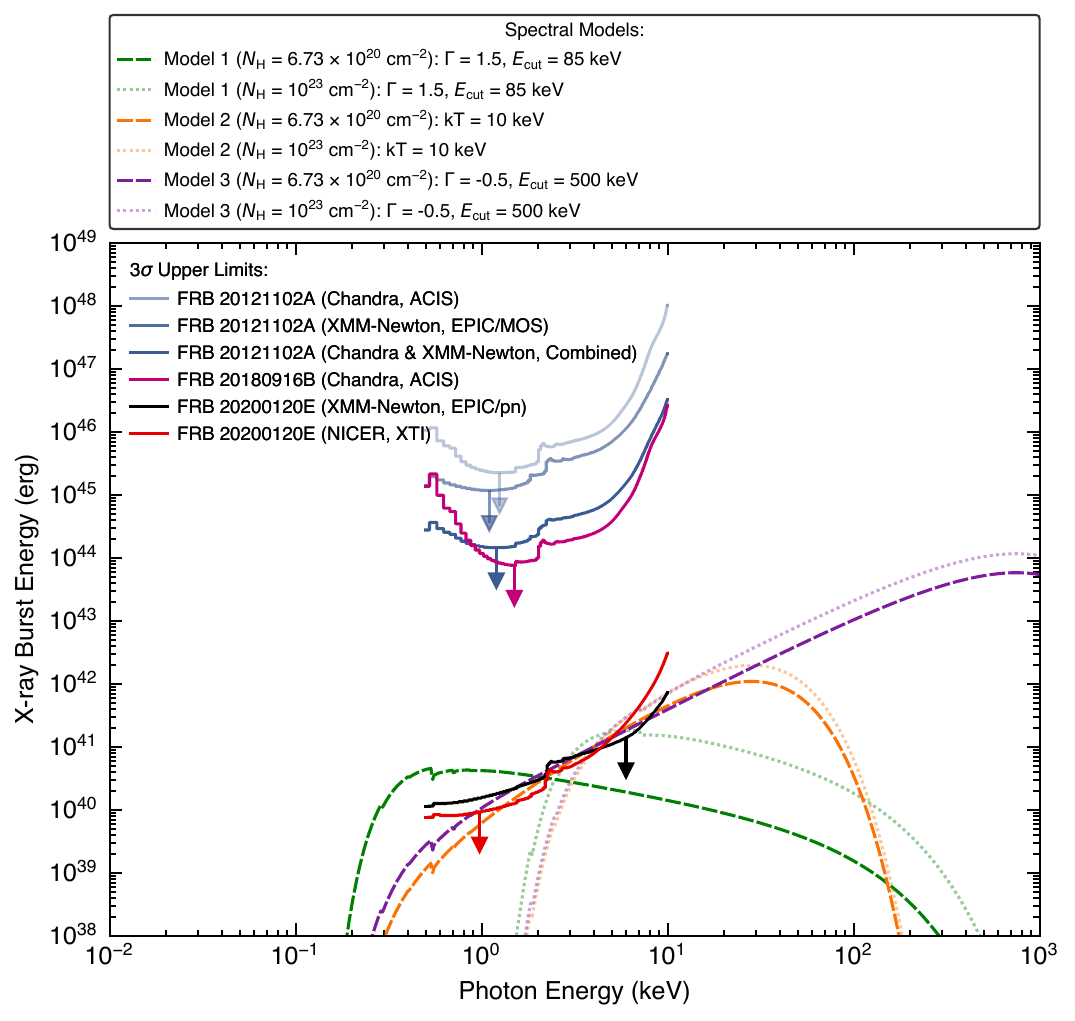}
	\caption{\textbf{Limits on the energy of \text{X-ray} bursts between 0.5~and~10\,keV at the times of radio bursts from FRB~20200120E, FRB~20121102A, and FRB~20180916B.} The red curve shows the 3$\sigma$ limits derived from \textit{NICER} observations of FRB~20200120E at the time of burst~B4, after incorporating the spectral response of the \textit{NICER}~XTI. The black curve shows the 3$\sigma$ limits derived from \textit{XMM-Newton} observations of FRB~20200120E at the time of burst~B9, after including the spectral response of the \textit{XMM-Newton} EPIC/pn camera. The light-blue curves correspond to previous 3$\sigma$ limits derived from \textit{Chandra} and \textit{XMM-Newton} observations at the times of multiple radio bursts from FRB~20121102A\,\cite{Scholz+2017a}. Stacked 3$\sigma$ limits for FRB~20121102A are shown in dark blue. The magenta curve shows previous 3$\sigma$ limits derived from a \textit{Chandra} observation at the time of a radio burst from FRB~20180916B\,\cite{Scholz+2020a}. Several fiducial spectral models are shown for comparison. The green curves correspond to exponentially cut-off power-law spectra with $\Gamma$\,$=$\,1.5 and $E_{\text{cut}}$\,$=$\,85\,keV, similar to the \text{X-ray} counterpart associated with the FRB-like radio burst detected from SGR~1935+2154 on 28~April~2020\,\cite{Li+2021, Mereghetti+2020a, Ridnaia+2021, Tavani+2021}. The orange curves represent blackbody spectra with $kT$\,$=$\,10\,keV, which have been observed from some magnetar hard \text{X-ray} bursts\,\cite{An+2014, CotiZelati+2021}. The purple curves show exponentially cut-off power-law spectra with \text{$\Gamma$\,$=$\,--0.5} and $E_{\text{cut}}$\,$=$\,500\,keV, similar to the spectrum observed during \text{SGR~1806--20's} giant flare\,\cite{Palmer+2005, Hurley+2005}. The dashed spectra were calculated using a hydrogen column density of $N_{\text{H}}$\,$=$\,6.73\,$\times$\,10$^{\text{20}}$\,cm$^{\text{--2}}$ towards FRB~20200120E\,\cite{HI4PI+2016}, and the dotted spectra were calculated using a hydrogen column density of $N_{\text{H}}$\,$=$\,10$^{\text{23}}$\,cm$^{\text{--2}}$ to illustrate the effects of high absorption local to the source. Larger hydrogen column densities yield harder \text{X-ray} spectra, with increased absorption at lower \text{X-ray} energies. Although high absorption towards FRB~20200120E is not expected, this scenario is shown for comparison. Each fiducial \text{X-ray} spectrum was normalized such that the integrated \text{X-ray} burst energy in the \text{0.5--10}\,keV energy range is equal to the \textit{NICER} \text{0.5--10}\,keV band-averaged energy limit from FRB~20200120E.}
	\label{fig:burst_energy_limits}
	
\end{figure*}

In Fig.~\ref{fig:burst_energy_limits}, we show our limits on the energy of \text{0.5--10}\,keV \text{X-ray} bursts at the times of bursts~B4 and~B9. For comparison, we also show previous limits from simultaneous \text{X-ray} and radio observations at the times of radio bursts detected from FRB~20121102A and FRB~20180916B\,\cite{Scholz+2017a, Scholz+2020a}, along with the fiducial spectral models used in Table~\ref{tab:nicer_xmm_b4_b9_xray_limits}. Figure~\ref{fig:burst_energy_limits} illustrates that our prompt \text{X-ray} burst energy limits are $\sim$10$^{\text{4}}$ times deeper than the \text{X-ray} limits obtained from the next nearest known repeating FRB source~(FRB~20180916B) for which such limits have been placed.

In Extended Data Table~\ref{tab:sgr1935_xray_fluences}, we list \text{X-ray} fluences for the \text{X-ray} burst associated with the \text{FRB-like} radio burst detected from SGR~1935+2154 on 28~April~2020, based on measurements from \textit{AGILE}\,\cite{Tavani+2021}, \textit{Insight-HXMT}\,\cite{Li+2021}, \textit{INTEGRAL}\,\cite{Mereghetti+2020a}, and \textit{Konus-Wind}\,\cite{Ridnaia+2021}. We also derive \text{X-ray} fluences in the \text{0.5--10}\,keV and \text{3--79}\,keV energy bands for an identical burst emitted from the location of FRB~20200120E.

\begin{figure*}
	
	\vspace*{-0.925cm}
	
	\centering
	\includegraphics[trim=0cm 0cm 0cm 0cm, clip=true, scale=0.5025, angle=0]{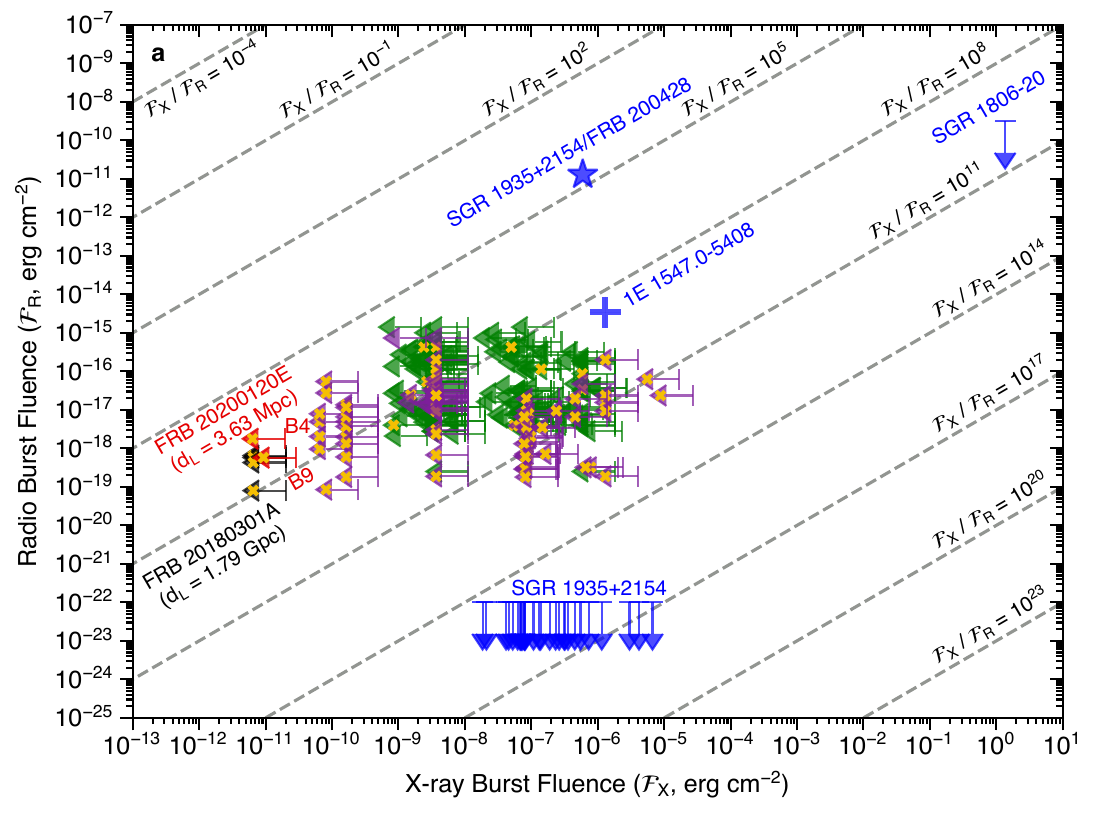}
	\includegraphics[trim=0cm 0cm 0cm 0cm, clip=true, scale=0.5025, angle=0]{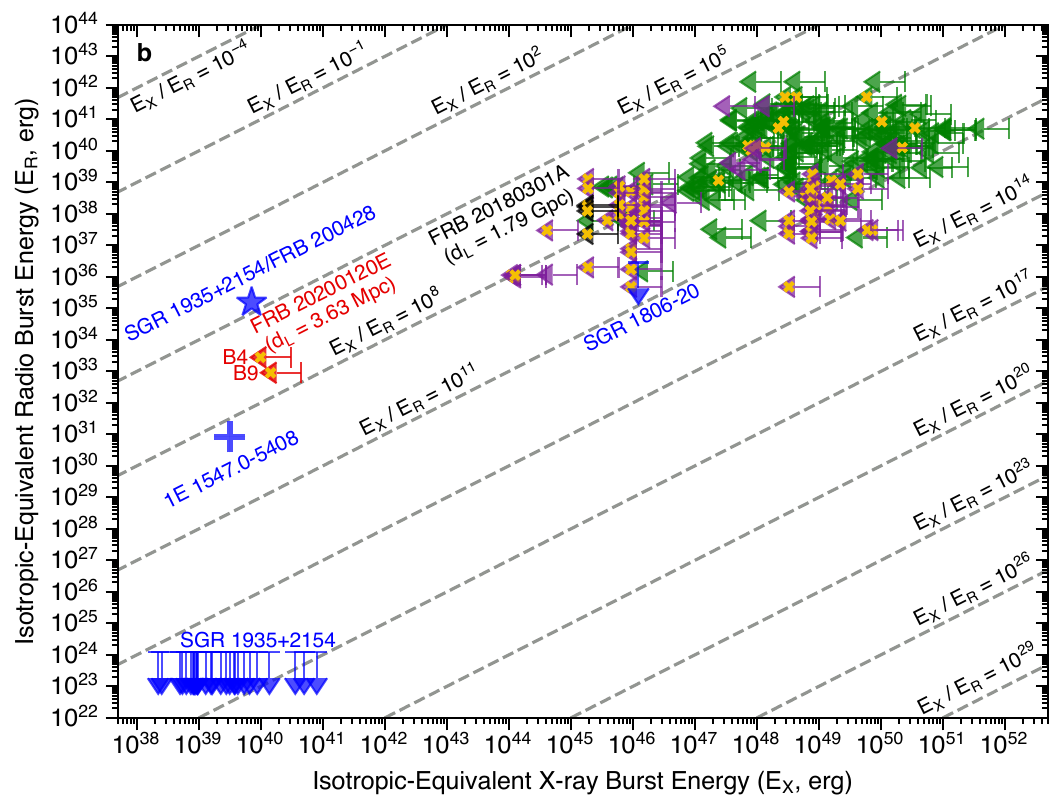}
	
	\caption{\textbf{\text{X-ray} and radio burst fluence and energy limits from observations of FRB~20200120E, along with previous measurements from simultaneous \text{X-ray} and radio observations of FRB sources and Galactic magnetars.} Panel~\textbf{a}: \text{X-ray} and radio fluence measurements of repeating FRB sources (purple), \text{non-repeating} FRB sources (green), and Galactic magnetars (blue). The red data points correspond to our \text{X-ray} and radio fluence measurements from \textit{NICER}, \textit{XMM-Newton}, and Effelsberg observations of FRB~20200120E at the times of bursts~B4 and~B9. The black data points highlight \text{X-ray} and radio fluence measurements from observations of FRB~20180301A (luminosity distance of $d_{\text{L}}$\,$=$\,1.79\,Gpc) with \textit{NICER} and the Five-hundred-meter Aperture Spherical radio Telescope~(FAST)\,\cite{Laha+2022}. The blue star indicates the \text{X-ray} and radio fluences of the \text{X-ray} and radio bursts detected from SGR~1935+2154 during an episode of FRB-like activity on 28~April~2020\,\cite{Li+2021, Bochenek+2020c, CHIME+2020a}. The blue upper limits from SGR~1935+2154 show constraints on prompt radio emission from FAST observations during a series of \text{X-ray} bursts detected with the \textit{Fermi} Gamma-ray Burst Monitor on 28~April~2020, close in time to the FRB-like radio burst\,\cite{Lin+2020}. The blue cross corresponds to \text{X-ray} and radio measurements of an FRB-like burst from the radio magnetar \text{1E~1547.0--5408}\,\cite{Israel+2021}. The blue upper limit for \text{SGR~1806--20} is derived from simultaneous radio and gamma-ray measurements at the time of the 27~December~2004 giant magnetar flare\,\cite{Tendulkar+2016}. The dashed grey lines indicate constant \text{X-ray}-to-radio fluence ratios ($\mathcal{F}_{\text{X}}/\mathcal{F}_{\text{R}}$). Panel~\textbf{b}: \text{X-ray} and radio energy measurements of repeating FRB sources, \text{non-repeating} FRB sources, and Galactic magnetars. \break The data points in panel \textbf{b} are derived from the same measurements shown in panel~\textbf{a}. Radio and \text{X-ray} energies were calculated using the distance and redshift of each object. \text{DM-inferred} distances were used for unlocalized FRB sources. The dashed grey lines indicate constant \text{X-ray}-to-radio energy ratios ($E_{\text{X}}/E_{\text{R}}$). In panels~\textbf{a} and~\textbf{b}, the yellow crosses highlight measurements from localized FRB sources with measured redshifts.}
	\label{fig:radio_xray_fluence_limits}
	
\end{figure*}

In Fig.~\ref{fig:radio_xray_fluence_limits}, we compare the \text{X-ray} and radio fluences of bursts from SGR~1935+2154 with other magnetars, previous limits from simultaneous \text{X-ray} and radio observations of repeating and non-repeating extragalactic FRBs, and our measurements from FRB~20200120E. The radio and \text{X-ray} fluences (or fluence limits) are shown in Fig.~\ref{fig:radio_xray_fluence_limits}a, and the corresponding isotropic-equivalent radio and \text{X-ray} energies (or energy limits) are plotted in Fig.~\ref{fig:radio_xray_fluence_limits}b. Figure~\ref{fig:radio_xray_fluence_limits} shows that our prompt \text{X-ray} fluence and \text{X-ray} energy limits from FRB~20200120E are deeper than all previous measurements from FRB sources.

We provide 3$\sigma$ persistent \text{X-ray} flux limits during each of our observations of FRB~20200120E with \textit{Chandra}, \textit{NICER}, \textit{NuSTAR}, and \textit{XMM-Newton} in Extended Data Table~\ref{tab:xray_obs}. The corresponding persistent \text{X-ray} luminosity limits are shown in Fig.~\ref{fig:xray_luminosity_upper_limits}a. We show the \text{X-ray} luminosity distributions of ultraluminous \text{X-ray}~(ULX) sources, low-mass \text{X-ray} binaries~(LMXBs), high-mass \text{X-ray} binaries~(HMXBs), and magnetars in Fig.~\ref{fig:xray_luminosity_upper_limits}b (Methods), along with our persistent \text{X-ray} luminosity limits derived from our observations of FRB~20200120E. Potential source types for FRB~20200120E with higher persistent \text{X-ray} luminosities than our limits are ruled out at the times of our \text{X-ray} observations.

In Fig.~\ref{fig:xray_transient_histogram}, we show the absorbed \text{0.5--10}\,keV \text{X-ray} fluences of different types of prompt emission from potential \text{X-ray} counterparts to FRBs, after translating to the distance of FRB~20200120E and accounting for the Galactic line-of-sight hydrogen column density ($N_{\text{H}}$\,$=$\,6.73\,$\times$\,10$^{\text{20}}$\,cm$^{\text{--2}}$)\,\cite{HI4PI+2016}. We refer to each translated \text{X-ray} fluence value as a `pseudo-fluence', since it corresponds to the \text{X-ray} fluence for an identical event emitted from the location of FRB~20200120E. To derive these pseudo-fluences, we translated each reported \text{X-ray} fluence to the \text{0.5--10}\,keV energy band using the \texttt{XSPEC} spectral fitting software package\,\cite{Arnaud1996} and the best-fit spectral model reported for each burst or flare (Methods). We show the pseudo-fluence distributions of magnetar giant flares, intermediate flares, and short bursts in Fig.~\ref{fig:xray_transient_histogram}, along with the distribution of type~I \text{X-ray} bursts from LMXBs. We also label the \text{X-ray} pseudo-fluences of \text{gamma-ray} bursts~(GRBs) from extragalactic magnetar giant flares, ultraluminous \text{X-ray} bursts~(ULXBs) from unknown objects in extragalactic globular clusters, bright type~II \text{X-ray} bursts from LMXBs, and predictions from the relativistic shock model described in refs.~\citen{Metzger+2019} and~\citen{Margalit+2020a} (Supplementary Information). Our best prompt 3$\sigma$ \text{X-ray} fluence limit for FRB~20200120E, derived from \textit{NICER} measurements on 100\,ns to 100\,ms timescales at the time of burst~B4, is also shown for comparison.


\bigskip

\section*{Discussion}
\label{sec:discussion}

Here, we compare our results to various types of \text{X-ray} sources and transients, as well as predictions for multiwavelength emission from popular FRB~models. In the Supplementary Information, we consider other possible scenarios, compare our results with multiwavelength observations of other repeating FRBs, and discuss the prospects of future searches for \text{X-ray} emission from FRBs using current and future \text{X-ray} telescopes~(Extended Data Fig.~\ref{fig:m81r_fluence_distance_constraints}).


\bigskip

\subsection{Magnetars}
\label{sec:magnetar_comparison}

High-energy magnetar bursts are typically classified into three types, based on their phenomenology and energetics\,\cite{Kaspi+2017, Enoto+2019}: (1)~giant flares (for example, see refs.~\citen{Hurley+1999a} and~\citen{Palmer+2005}), (2)~intermediate flares (for example, see refs.~\citen{Israel+2008} and~\citen{Kouveliotou+2001}), and (3)~short bursts (for example, see refs.~\citen{Younes+2020b} and~\citen{Younes+2021}). Giant flares from magnetars have been detected with peak \text{X-ray} luminosities of $L_{\text{X}}$\,$\approx$\,\text{10$^{\text{44}}$--10$^{\text{48}}$}\,erg\,s$^{\text{--1}}$ and have durations of approximately a few hundred seconds, whereas intermediate flares have typical peak luminosities of $L_{\text{X}}$\,$\approx$\,\text{10$^{\text{41}}$--10$^{\text{43}}$}\,erg\,s$^{\text{--1}}$ and durations of a few seconds. Short magnetar bursts, on the other hand, have been detected with lower peak \text{X-ray} luminosities ($L_{\text{X}}$\,$\approx$\,\text{10$^{\text{36}}$--10$^{\text{41}}$}\,erg\,s$^{\text{--1}}$) and durations ranging between a few milliseconds and a few seconds. We discuss the FRB--magnetar connection and the detectability of each of these transient radiative phenomena during our observations of FRB~20200120E.

An intermediate flare\,\cite{Kozlova+2016}, short \text{X-ray} bursts\,\cite{Mereghetti+2020a, Li+2021, Ridnaia+2021, Tavani+2021}, and FRB-like radio bursts\,\cite{CHIME+2020a, Bochenek+2020c} have been detected from SGR~1935+2154, suggesting that similar emission may be detectable from nearby FRB~sources. We rule out the possibility that an intermediate flare, similar to the event detected on 12~April~2015 from \text{SGR~1935+2154}\,\cite{Kozlova+2016}, was produced from FRB~20200120E at the times of bursts~B4 and~B9. The \text{0.5--10}\,keV \text{X-ray} pseudo-fluence of the 12~April~2015 intermediate flare from SGR~1935+2154 is above our prompt 3$\sigma$ \text{X-ray} fluence limits (Fig.~\ref{fig:xray_transient_histogram} and Table~\ref{tab:nicer_xmm_b4_b9_xray_limits}), indicating that such a flare would have been detectable if it were emitted from FRB~20200120E. The predicted \text{0.5--10}\,keV \text{X-ray} fluence for a short \text{X-ray} burst similar to the one accompanying the \text{FRB-like} radio burst from SGR~1935+2154 on 28~April~2020 (Fig.~\ref{fig:xray_transient_histogram} and Extended Data Table~\ref{tab:sgr1935_xray_fluences}) is approximately an order of magnitude below our deepest 3$\sigma$ \text{X-ray} fluence limit. However, \text{X-ray} bursts with larger ($\gtrsim$\,10$\times$) fluences than this \text{X-ray} burst from SGR~1935+2154 have been observed\,\cite{Mereghetti+2020a} and would have been detectable from FRB~20200120E during our \text{X-ray} observations.

Archival \text{X-ray} observations of M81 with \textit{INTEGRAL}, spanning nearly 18~years and covering the position of FRB~20200120E, have also been used to place shallow limits on powerful magnetar giant flares in the hard \text{X-ray} band\,\cite{Mereghetti+2021}, but not at times of simultaneous radio bursts. The absorbed \text{X-ray} pseudo-fluences of the magnetar giant flares from \text{SGR~1806--20}, \text{SGR~1900+14}, and \text{SGR~0526--66} are well above our prompt \text{X-ray} fluence limits derived at the times of bursts~B4 and~B9 (Fig.~\ref{fig:xray_transient_histogram} and Table~\ref{tab:nicer_xmm_b4_b9_xray_limits}). Thus, we rule out the possibility that a giant flare, similar to those observed from known magnetars, was produced from FRB~20200120E simultaneously with these radio bursts.

In Fig.~\ref{fig:xray_transient_histogram}, we show that most ($\sim$75\%) of the magnetar intermediate flares detected so far have absorbed \text{X-ray} pseudo-fluences that are above our prompt 3$\sigma$ \text{X-ray} fluence limits, derived at the times of bursts~B4 and~B9. Thus, similar intermediate flares would have been detectable during our observations of FRB~20200120E. A small fraction ($\sim$1\%) of the most energetic magnetar short bursts also have absorbed \text{X-ray} pseudo-fluences that are above our prompt 3$\sigma$ \text{X-ray} fluence limits, indicating that similar events also would have been detectable from FRB~20200120E. These include several bright short \text{X-ray} bursts from the radio magnetar 1E~1547.0--5408\,\cite{Mereghetti+2009}, which has been observed to produce \text{FRB-like} radio bursts\,\cite{Israel+2021}. While we cannot entirely rule out the possibility that an intermediate flare or short magnetar burst was emitted from FRB~20200120E, since events below our \text{X-ray} fluence detection threshold would not have been detected during our observations because of limited \text{X-ray} telescope sensitivity, we are able to rule out prompt \text{X-ray} emission similar to some of the most energetic intermediate flares and short bursts detected thus far from the known magnetar population.

Magnetars have been detected with persistent \text{X-ray} luminosities ranging between approximately 10$^{\text{29}}$ and 10$^{\text{36}}$\,erg\,s$^{\text{--1}}$ in the soft \text{X-ray} band\,\cite{Olausen+2014}. During bright outbursts, their persistent \text{X-ray} luminosities have been seen to reach up to $\sim$10$^{\text{36}}$\,erg\,s$^{\text{--1}}$\,\cite{CotiZelati+2018}. This maximum observed persistent \text{X-ray} luminosity is theoretically motivated by the dominance of neutrino cooling in the neutron star~(NS) crust at high temperatures\,\cite{Yakovlev+2004, Pons+2012}, so the persistent \text{X-ray} emission from magnetars is not expected to exceed this level. This maximum persistent \text{X-ray} luminosity is approximately five times fainter than the 3$\sigma$ \text{X-ray} luminosity limit obtained from our deepest observation of FRB~20200120E with \textit{Chandra} (Fig.~\ref{fig:xray_luminosity_upper_limits}). Magnetars hosted in FRB~20200120E's globular cluster would require a different formation channel from those of young magnetars in the Milky Way (for example, via accretion-induced collapse of a white dwarf~(WD) or a merger-induced collapse of a WD--WD, NS--WD, or NS--NS binary)\,\cite{Margalit+2019, Kremer+2021}. Since the maximum persistent \text{X-ray} luminosities of magnetars formed through core-collapse and delayed channels are predicted to be similar, we would have to probe at least an order of magnitude deeper to reach the level of persistent \text{X-ray} emission expected from an extragalactic, FRB-emitting magnetar.


\bigskip

\subsection{Young Extragalactic Pulsars}
\label{sec:young_pulsars}

Some models propose that FRBs may be supergiant radio pulses from young, rapidly-rotating extragalactic pulsars\,\cite{Cordes+2016a, Lyutikov+2016}, which may have similar properties to the Crab pulsar. Several of the radio bursts detected from FRB~20200120E have emission characteristics that are remarkably similar to radio pulses from the Crab pulsar. For example, some radio bursts from FRB~20200120E display multiple pulse components with widths shorter than 100\,ns\,\cite{Majid+2021, Nimmo+2022}, resembling the nanoshot radio emission observed from the Crab pulsar\,\cite{Hankins+2003, Hankins+2007}. However, we found no significant evidence of prompt \text{X-ray} emission from FRB~20200120E (down to timescales of 100\,ns using \textit{NICER} data) at the times of radio bursts or at other times during our multiwavelength campaign. Our \text{X-ray} observations did not have sufficient sensitivity to detect \text{X-ray} luminosities comparable to the pulsed \text{X-ray} emission from the Crab pulsar ($\sim$0.1$L_{\text{X}}^{\text{Crab}}$).

Our persistent \text{X-ray} luminosity limits (Fig.~\ref{fig:xray_luminosity_upper_limits}) show that FRB~20200120E is not associated with a supernova remnant~(SNR) or a pulsar wind nebula~(PWN) whose persistent \text{X-ray} luminosity is larger than or comparable to the Crab Nebula~($L_{\text{X}}^{\text{Crab}}$\,$\approx$\,10$^{\text{37}}$\,erg\,s$^{\text{--1}}$). However, a Crab-like SNR is not expected in FRB~20200120E's $\sim$10-Gyr-old globular cluster since all massive stars that could have produced this type of SNR would have died within the first $\sim$100\,Myr of the cluster's formation. Most PWNe have persistent \text{X-ray} luminosities of $\sim$\text{0.001--0.01}$L_{\text{X}}^{\text{Crab}}$\,\cite{Klingler+2018, Gotthelf+2021}, which is $\sim$\text{10--100} times below our persistent \text{X-ray} luminosity limits and would not have been detectable during our observations.


\bigskip

\subsection{Millisecond Pulsars}
\label{sec:msps}

FRB~20200120E may be powered by a giant pulse-emitting millisecond pulsar~(MSP), spun-up by accretion from a companion star and possibly formed via accretion-induced collapse of a massive WD in a close binary system\,\cite{Tauris+2013, Ye+2019}. Globular clusters have high stellar densities and host populations of MSPs that are produced through frequent dynamical interactions. Quasi-periodic microstructure ($P_{\mu}$\,$\approx$\,2--3\,$\mu$s) has been detected in a radio burst from FRB~20200120E\,\cite{Majid+2021}, which may indicate that the source is a rapidly spinning pulsar (with a rotational period of $P_{\text{spin}}$\,$\approx$\,2--3\,ms) based on the phenomenological \text{$P_{\mu}$--$P_{\text{spin}}$} relationship observed from Galactic pulsars.

Several Galactic MSPs, such as PSR~B1937+21\,\cite{Soglasnov+2004} and PSR~B1957+20\,\cite{Main+2017}, frequently produce giant radio pulses. Some of these giant pulses have nanosecond pulse widths ($w_{\text{t}}$\,$<$\,15\,ns) and high brightness temperatures ($T_{\text{b}}$\,$>$\,5\,$\times$10$^{\text{39}}$\,K)\,\cite{Soglasnov+2004}. In the soft \text{X-ray} band, a correlation has been observed between the persistent \text{X-ray} luminosity and spin-down luminosity ($L_{\text{spin}}$) of MSPs, that is, $L_{\text{X}}$\,$\propto$\,$L_{\text{spin}}^{1.3}$\,\cite{Lee+2018}. The persistent \text{X-ray} luminosities of Galactic MSPs are typically less than 10$^{\text{34}}$\,erg\,s$^{\text{--1}}$\,\cite{Lee+2018}, which is several orders of magnitude below our deepest persistent \text{X-ray} luminosity limits (Fig.~\ref{fig:xray_luminosity_upper_limits}). This level of persistent \text{X-ray} emission, from a putative population of similar MSPs in FRB~20200120E's globular cluster, was not detectable during our observations.


\bigskip

\subsection{ULXBs in Extragalactic Globular Clusters}
\label{sec:ulxbs_gcs}

ULXBs are a class of luminous, extragalactic \text{X-ray} flares that display fast rise times ($\sim$1\,min), high peak luminosities ($\sim$10$^{\text{39}}$--10$^{\text{43}}$\,erg\,s$^{\text{--1}}$), and slow decay timescales ($\sim$1\,h). Several ULXBs have been identified in nearby galaxies\,\cite{Sivakoff+2005, Jonker+2013, Irwin+2016}, but their origins remain unknown. Most sources of ULXBs have been found to be spatially coincident with globular clusters in their host galaxies\,\cite{Sivakoff+2005, Jonker+2013, Irwin+2016}. An association between repeating ULXBs and repeating FRBs has been proposed\,\cite{Chen+2022}, which is particularly relevant for FRB~20200120E because of its globular cluster location. If \text{X-ray} flares similar to ULXBs were produced by FRB~20200120E, their apparent soft \text{X-ray} fluences would be at least an order of magnitude larger than our \text{X-ray} fluence limits obtained using \textit{NICER} and \textit{XMM-Newton} at the times of bursts~B4 and~B9 (Fig.~\ref{fig:xray_transient_histogram} and Table~\ref{tab:nicer_xmm_b4_b9_xray_limits}). We conclude that there were no ULXBs emitted from FRB~20200120E during our \text{X-ray} observations, since we did not detect any astrophysical \text{X-ray} flares. These observations allow us to rule out a link between FRB~20200120E and ULXBs.


\bigskip

\subsection{ULX and Accretion-Powered Sources}
\label{sec:ulx_accretion}

\text{X-ray} binaries, involving accreting NSs or black holes (BHs), have also been proposed as a class of possible FRB sources\,\cite{Sridhar+2021, Deng+2021}, where mass transfer can occur near or exceeding the Eddington rate. ULX~sources have persistent \text{X-ray} luminosities exceeding 10$^{\text{39}}$\,erg\,s$^{\text{--1}}$, some of which have been found to be powered by pulsars in binary systems accreting at super-Eddington rates\,\cite{Bachetti+2014}. The accretion luminosity ($L_{\text{acc}}$) generated by a mass accretion rate, $\dot{M}_{\text{acc}}$, is given by \break $L_{\text{acc}}$\,$\approx$\,$\eta_{\text{acc}}\dot{M}_{\text{acc}}c^{2}$, where $\eta_{\text{acc}}$ is the accretion efficiency and $c$ is the speed of light. Assuming that any putative \text{X-ray} emission from FRB~20200120E is powered via accretion with an accretion efficiency of $\eta_{\text{acc}}$\,$=$\,0.1, our best persistent \text{X-ray} luminosity limit from \textit{Chandra} observations of FRB~20200120E (Fig.~\ref{fig:xray_luminosity_upper_limits} and Extended Data Table~\ref{tab:xray_obs}) corresponds to an upper limit on the mass transfer rate of \break $\dot{M}_{\text{acc}}$\,$<$\,0.06\,$\dot{M}_{\text{Edd}}$ for a canonical 1.4\,$M_{\odot}$~NS and $\dot{M}_{\text{acc}}$\,$<$\,0.003\,$\dot{M}_{\text{Edd}}$ for a 30\,$M_{\odot}$~BH, where $\dot{M}_{\text{Edd}}$ is the Eddington accretion rate of the compact object. Our \text{X-ray} observations rule out the possibility that FRB~20200120E is associated with an \text{X-ray} source accreting persistently at either \text{near-Eddington} or \text{super-Eddington} levels. Although an intermittently accreting \text{X-ray} source cannot be excluded, it is challenging to reconcile our observations with an \text{X-ray} system accreting transiently at hyper-Eddington rates ($\dot{M}_{\text{acc}}$\,$\gtrsim$\,10\,$\dot{M}_{\text{Edd}}$), even for a low-mass (1\,$M_{\odot}$) compact object, since \text{X-ray} luminosities ($L_{\text{X}}$\,$>$\,10$^{\text{39}}$\,erg\,s$^{\text{--1}}$) similar to ULX sources would be required. As shown in Fig.~\ref{fig:xray_luminosity_upper_limits}a, almost all of our \text{X-ray} observations yielded persistent \text{X-ray} luminosity limits below 10$^{\text{39}}$\,erg\,s$^{\text{--1}}$, which would have been sufficient to detect ULX sources at the location of FRB~20200120E.

The most energetic burst~(B4) shown in Fig.~\ref{fig:radio_bursts} has an isotropic-equivalent radio luminosity of approximately 10$^{\text{37}}$\,erg\,s$^{\text{--1}}$ and a burst duration of roughly 100\,$\mu$s (Extended Data Table~\ref{tab:radio_burst_properties}). The accretion-based \text{X-ray} binary FRB model in ref.~\citen{Sridhar+2021} predicts mass transfer rates of $\sim$0.1--10\,$\dot{M}_{\text{Edd}}$ for radio bursts with durations of 100\,$\mu$s. A 1.4\,$M_{\odot}$ NS accreting at the Eddington rate ($\dot{M}_{\text{acc}}$\,$=$\,$\dot{M}_{\text{Edd}}$) would produce an \text{X-ray} luminosity of 2\,$\times$\,10$^{\text{38}}$\,erg\,s$^{\text{--1}}$, which was not detected during most of our \text{X-ray} observations. Larger \text{X-ray} luminosities would be produced for more massive compact objects as a result of higher Eddington mass transfer rates. \text{X-ray} limits from FRBs more distant than FRB~20200120E have excluded ULX sources brighter than $L_{\text{X}}$\,$\gtrsim$\,10$^{\text{40}}$\,erg\,s$^{\text{--1}}$ from being associated with a subset of the FRB population\,\cite{Eftekhari+2023}. Our \text{X-ray} observations allow us to rule out an association between FRB~20200120E and ULX/accreting sources with persistent \text{X-ray} luminosities larger than $L_{\text{X}}$\,$>$\,9.8\,$\times$\,10$^{\text{36}}$\,erg\,s$^{\text{--1}}$. We also exclude transient, FRB-emitting ULX and accreting sources brighter than $L_{\text{X}}$\,$>$\,\text{(1.2--15.8)\,$\times$\,10$^{\text{37}}$}\,erg\,s$^{\text{--1}}$ that are \text{X-ray} active when radio emission is produced, based on the \text{X-ray} luminosity limits derived near the times of FRB~20200120E's bright radio bursts (Fig.~\ref{fig:xray_luminosity_upper_limits}a and Extended Data Table~\ref{tab:xray_obs}).


\bigskip

\section*{Conclusions}
\label{sec:conclusions}

Based on our prompt \text{X-ray} fluence limits, we rule out an association between FRB~20200120E and the following types of transient \text{X-ray} emission: (1)~magnetar giant flares, (2)~an SGR~1935+2154-like intermediate flare, and (3)~ULXBs. Although a significant fraction ($\sim$75\%) of known intermediate flares from magnetars would have been detectable from FRB~20200120E during our \text{X-ray} observations, along with some bright magnetar short bursts and type~I and type~II \text{X-ray} bursts from LMXBs~(Supplementary Information), we cannot entirely rule them out as a class.

In addition, our persistent \text{X-ray} limits allow us to rule out the following source types for \break FRB~20200120E: (1)~compact objects (for example, young extragalactic pulsars) embedded in a SNR or PWN, with persistent \text{X-ray} luminosities larger than or comparable to the Crab Nebula; (2)~ULX and accreting sources with persistent \text{X-ray} luminosities larger than $L_{\text{X}}$\,$>$\,9.8\,$\times$\,10$^{\text{36}}$\,erg\,s$^{\text{--1}}$; and (3)~transient ULX/accreting sources with $L_{\text{X}}$\,$>$\,\text{(1.2--15.8)\,$\times$\,10$^{\text{37}}$}\,erg\,s$^{\text{--1}}$ that produce bright, persistent \text{X-ray} emission near the times of radio bursts. Although LMXBs with persistent \text{X-ray} luminosities comparable to $\sim$30\% of the brightest LMXBs in the Milky Way were also detectable from FRB~20200120E's location during our \text{X-ray} observations, an LMXB~origin for FRB~20200120E cannot be entirely excluded.

A giant radio pulse-emitting pulsar or a magnetar formed through a delayed channel are both potential source types that could explain the physical origin of FRB~20200120E, given its localization to a globular cluster~\cite{Kirsten+2022}. Deeper \text{X-ray} observations in the future, particularly with sensitive, \text{next-generation} \text{X-ray} instruments, could help further constrain the nature of the astrophysical source powering FRB~20200120E.


\newpage

\section*{Methods}
\label{sec:methods}


\bigskip

\section*{Radio Observations}
\label{sec:radio_obs}


\bigskip

\subsection{EVN}
\label{sec:evn_obs}

FRB~20200120E was observed with the \text{100-m} Effelsberg radio telescope during interferometric observations with radio telescopes from the~EVN on 20~February~2021; 7, 10, 16, and 22~March~2021; 10 and 28~April~2021; and 2~May~2021. These observations were carried out as part of the PRECISE VLBI programme\,\cite{Kirsten+2022, Nimmo+2022}. During each observation, 2-bit sampled dual circular polarization baseband voltage data were recorded between 1254 and 1510\,MHz in 16~contiguous 16-MHz subbands using the Effelsberg radio telescope, which provided a total bandwidth of 256\,MHz and a native time resolution of 31.25\,ns. These voltage data were saved in VLBI Data Interchange Format~(VDIF)\,\cite{Whitney+2010}. Total intensity filterbank data were also simultaneously recorded using the PSRIX pulsar backend\,\cite{Lazarus+2016} between 1255 and 1505\,MHz with a time and frequency resolution of 102.4\,$\mu$s and 0.49\,MHz, respectively. These observations are listed in Extended Data Table~\ref{tab:radio_obs}, and additional details are provided in refs.~\citen{Kirsten+2022}, \citen{Nimmo+2022}, and~\citen{Nimmo+2023}.

A series of simultaneous \text{X-ray} observations with \textit{NICER} and \textit{XMM-Newton} (Extended Data Fig.~\ref{fig:xray_radio_obs_timeline} and Extended Data Table~\ref{tab:xray_obs}) were performed during all of these PRECISE observing sessions, except for those occurring on 20~February~2021 and 28~April~2021. In total, five radio bursts were detected using the Effelsberg radio telescope during this campaign. Bursts~B1 and~B2 were detected on 20~February~2021, B3 and~B4 were detected on 7~March~2021, and B5 was detected on 28~April~2021. These radio bursts~(\text{B1--B5}) were discovered by first converting the baseband voltage data from~VDIF to filterbank format using \texttt{digifil}\,\cite{vanStraten+2011}, with a time and frequency resolution of 64\,$\mu$s and 125\,kHz, respectively. The total intensity filterbank data were then dedispersed using dispersion measure~(DM) trials within $\pm$50\,pc\,cm$^{\text{--3}}$ of the expected DM (87.818\,pc\,cm$^{\text{--3}}$), based on radio bursts detected from FRB~20200120E using CHIME/FRB\,\cite{Bhardwaj+2021a}. Candidate radio bursts were identified using a \text{\texttt{Heimdall}-based}\,\cite{Barsdell+2012} FRB search pipeline, which were then classified using a deep neural network, available in the Fast Extragalactic Transient Candidate Hunter (\texttt{FETCH}) software package\,\cite{Agarwal+2020}. This burst search pipeline is described in more detail in refs.~\citen{Kirsten+2022}, \citen{Nimmo+2022}, and~\citen{Kirsten+2021}.

The radio data containing \text{B1--B5} were processed using the Super~FX Correlator~(\texttt{SFXC})\,\cite{Keimpema+2015}, coherently dedispersed within each \text{16-MHz} subband using a~DM of 87.75\,pc\,cm$^{\text{--3}}$, and incoherently shifted between the subbands\,\cite{Kirsten+2022, Nimmo+2022}. The temporal width and frequency extent of each burst was measured by performing a nonlinear \text{least-squares} fit of a \text{two-dimensional} Gaussian function to the \text{two-dimensional} autocorrelation function of the dynamic spectrum\,\cite{Nimmo+2022}. The peak signal-to-noise ratio~(S/N), peak flux density, fluence, isotropic-equivalent luminosity, and isotropic-equivalent energy of bursts~\text{B1--B5} were determined using dedispersed total intensity data, sampled at 8\,$\mu$s, and the $\pm$2$\sigma$ burst width region (Fig.~\ref{fig:radio_bursts} and Extended Data Table~\ref{tab:radio_burst_properties}). Additional details are provided in refs.~\citen{Kirsten+2022} and~\citen{Nimmo+2022}. The measured properties of bursts~\text{B1--B5} are provided in Extended Data Table~\ref{tab:radio_burst_properties}.


\bigskip

\subsection{Effelsberg}
\label{sec:effelsberg_obs}

We carried out additional radio observations of FRB~20200120E using the Effelsberg radio telescope and the central beam of the \text{seven-beam} \text{21-cm} receiver\,\cite{Barr+2013}. Two radio observations were performed on 16 and 17~December~2020, using the Pulsar Fast Fourier Transform Spectrometer~(PFFTS) backend\,\cite{Barr+2013} and the Effelsberg Direct Digitization~(EDD) backend, during simultaneous \text{X-ray} observations of FRB~20200120E with \textit{NuSTAR} and \textit{Chandra} (Extended Data Fig.~\ref{fig:xray_radio_obs_timeline} and Extended Data Table~\ref{tab:xray_obs}). Total intensity data were recorded using 512~frequency channels in a digital polyphase filterbank between 1210 and 1510\,MHz, with a time resolution of 54\,$\mu$s and a frequency resolution of 0.59\,MHz. The Effelsberg observations on 16 and 17~December~2020 were conducted using a preliminary position ($\alpha_{\text{J2000}}$\,$=$\,09$^{\text{h}}$57$^{\text{m}}$56.40$^{\text{s}}$,  $\delta_{\text{J2000}}$\,$=$\,68$^{\circ}$49$\arcmin$04.8$\arcsec$) for FRB~20200120E, which was derived from baseband voltage data of radio bursts detected by CHIME/FRB. These radio observations are listed in Extended Data Table~\ref{tab:radio_obs}.

Two additional radio observations of FRB~20200120E were performed using Effelsberg in the \break \text{1200--1600}\,MHz frequency range with the EDD~backend on 30~April~2021 and 14~May~2021. These observations were conducted during simultaneous \text{X-ray} observations of FRB~20200120E with \textit{XMM-Newton} (Extended Data Fig.~\ref{fig:xray_radio_obs_timeline} and Extended Data Table~\ref{tab:xray_obs}). We used a preliminary position \break ($\alpha_{\text{J2000}}$\,$=$\,09$^{\text{h}}$57$^{\text{m}}$56.00$^{\text{s}}$,  $\delta_{\text{J2000}}$\,$=$\,68$^{\circ}$49$\arcmin$32.0$\arcsec$) from the EVN for the Effelsberg observations of \break FRB~20200120E conducted on 30~April~2021 and 14~May~2021. During both of these observations, we recorded channelized, full Stokes data in 512~frequency channels, with a time resolution of 64\,$\mu$s and a frequency resolution of 0.78125\,MHz. Data between 1210 and 1510\,MHz, coinciding with the bandwidth of the \text{21-cm} receiver, were extracted and used for all of our analyses. A list of these radio observations is provided in Extended Data Table~\ref{tab:radio_obs}.

We searched each of these Effelsberg observations for radio bursts using a custom FRB search pipeline, which utilized the routines provided in the \texttt{PRESTO} pulsar search software package\,\cite{Ransom2001}. We did not perform any mitigation of radio frequency interference~(RFI), since the band was relatively unaffected by~RFI. The data from each of these observations were coherently and incoherently dedispersed using a DM of 87.82\,pc\,cm$^{\text{--3}}$\,\cite{Bhardwaj+2021a}. A single pulse search of the dedispersed data was performed to identify radio bursts at this single DM~trial. We searched for radio bursts with temporal widths between 64\,$\mu$s and 20\,ms and saved all candidates with a detection~S/N above~5. These candidates were each individually inspected for evidence of radio bursts from FRB~20200120E. Four radio bursts (B6, B7, B8, and~B9) were identified during the Effelsberg radio observation conducted on 30~April~2021. No other radio bursts were detected during the other Effelsberg radio observations described above.

\begin{figure*}
	
	\centering
	
	\hspace*{-1.75cm}
	\includegraphics[trim=0cm 0cm 0cm 0cm, clip=true, scale=0.4, angle=0]{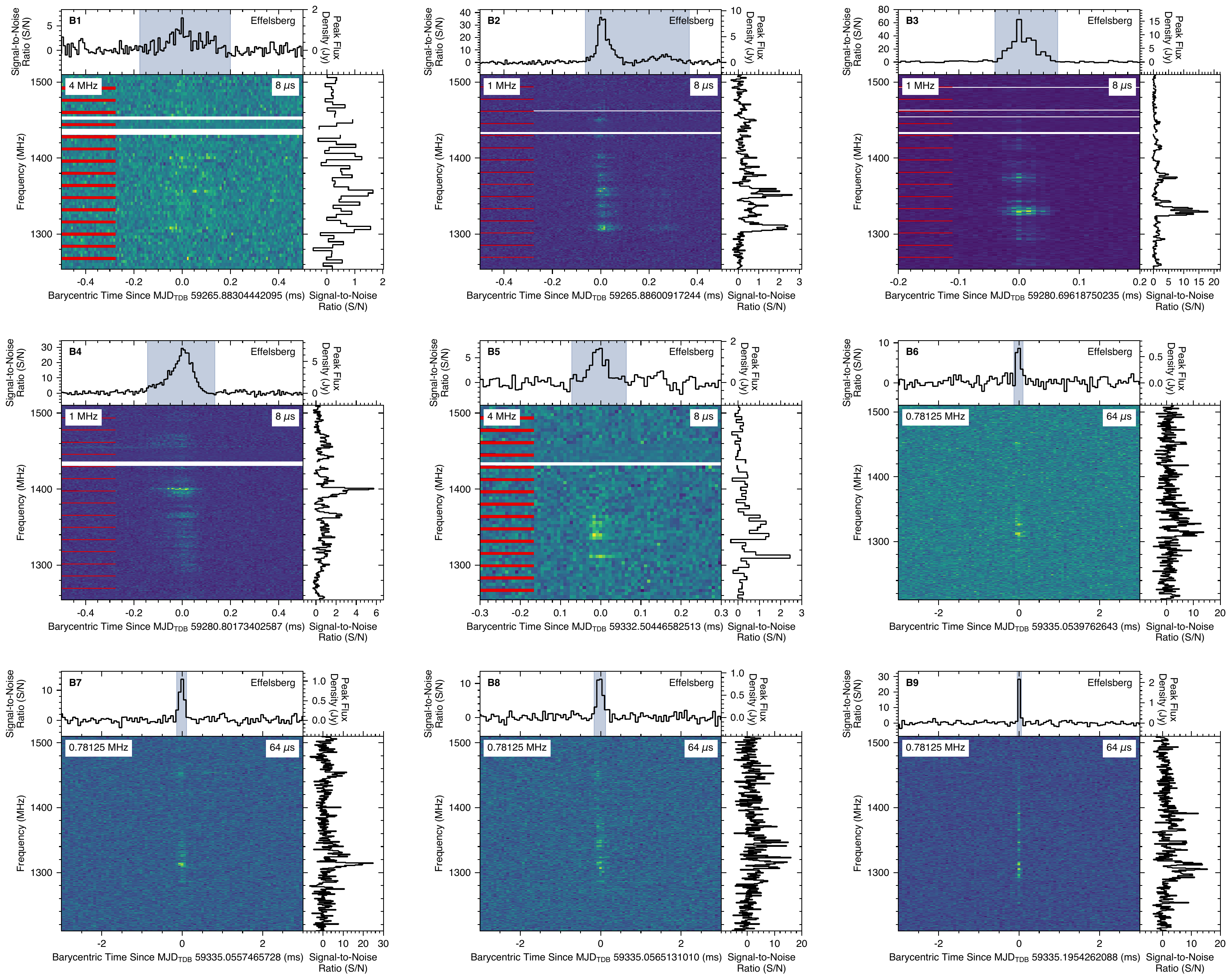}
	\caption{\textbf{Radio bursts detected from FRB~20200120E using the Effelsberg radio telescope.} Top panels: total intensity (Stokes~I) frequency-summed burst profiles. The shaded blue region in the top panels corresponds to the $\pm$2$\sigma$ burst width, which was used to calculate the S/N and peak flux density. Middle panels: dynamic spectra of the radio bursts. For bursts~\text{B1--B5}, the red lines mark the edges of the individual \text{16-MHz} subbands, and the white lines indicate frequency channels that were masked to mitigate~RFI. Right panels: time-averaged frequency spectra of the bursts. The time resolution and frequency resolution of the data are labelled in the top right and top left corners of the middle panels, respectively.}
	\label{fig:radio_bursts}
	
\end{figure*}

In Extended Data Table~\ref{tab:radio_burst_properties}, we provide measurements of the burst properties of~B6, B7, B8, and~B9. The properties of these bursts were measured after incoherently dedispersing the data with the S/N-maximizing DM obtained from~B9. The DM obtained from~B9 was used for dedispersion since it was the brightest and narrowest radio burst detected during the observation. The temporal widths of bursts~\text{B6--B9} were determined by fitting a Gaussian function to the dedispersed, frequency-summed burst profiles using a nonlinear \text{least-squares} fitting procedure. The burst width is defined as the full-width at half-maximum of the Gaussian fit divided by $\sqrt{2}$. We used the $\pm$2$\sigma$ burst width region (Fig.~\ref{fig:radio_bursts} and Extended Data Table~\ref{tab:radio_burst_properties}) to determine the peak S/N, peak flux density, fluence, isotropic-equivalent luminosity, and isotropic-equivalent energy of bursts~\text{B6--B9}. Our measurements of the frequency extent of these bursts were obtained from the number of frequency channels displaying a statistically significant `on-pulse' signal above the noise. The peak flux densities and fluences of bursts~\text{B6--B9} were calculated using the radiometer equation and assuming a system equivalent flux density~(SEFD) of 15\,Jy\,\cite{Barr+2013}, since a flux calibrator was not observed.

The fluence detection threshold for a given burst width, $w_{\text{t}}$, can be calculated using
\begin{equation}
	\mathcal{F}_{\text{R,\,min}} = \text{S/N}\times\frac{\text{SEFD}}{\sqrt{n_{\text{p}}\Delta\nu}}\times\sqrt{w_{\text{t}}},
	\label{eqn:fluence_eqn}
\end{equation}
where SEFD\,$=$\,$T_{\text{sys}}/G$, $T_{\text{sys}}$ is the system temperature, $G$ is the telescope gain, $n_{\text{p}}$ is the number of summed polarizations, and $\Delta\nu$ is the observing bandwidth. Using equation~(\ref{eqn:fluence_eqn}) and assuming a fiducial burst width of 100\,$\mu$s, we find that the 7$\sigma$ fluence detection threshold was $\mathcal{F}_{\text{R,\,min}}^{\text{7}\sigma}$\,$=$\,0.04\,Jy\,ms\,$\sqrt{\frac{w_{\text{t}}}{\text{100}\,\mu\text{s}}}$ during these Effelsberg observations. We estimate a conservative 15\% error on this measurement, based on the uncertainty on the SEFD at the time of these observations.

As part of a radio monitoring programme of FRB~20200120E, additional radio observations were carried out with the Effelsberg telescope on 11, 16, and 18~April~2021 in the 1233--1483\,MHz frequency range using the PSRIX pulsar backend\,\cite{Lazarus+2016} in baseband mode. During each observation, total intensity filterbank data were recorded with a time and frequency resolution of 65.5\,$\mu$s and 0.24\,MHz, respectively. Dual circular polarization raw voltages were also simultaneously recorded. No radio bursts from FRB~20200120E were detected during these observations. Additional details and results from these radio observations are provided in ref.~\citen{Nimmo+2023}. Coordinated \text{X-ray} observations with \textit{NICER} and \textit{XMM-Newton} were also performed simultaneously during these Effelsberg observations (Extended Data Fig.~\ref{fig:xray_radio_obs_timeline} and Extended Data Tables~\ref{tab:radio_obs} and~\ref{tab:xray_obs}).


\bigskip

\subsection{CHIME}
\label{sec:chime_psr_obs}

We observed FRB~20200120E in the \text{400--800}\,MHz frequency range with the~CHIME radio telescope and the CHIME/Pulsar system\,\cite{CHIME+2021a} during multiple simultaneous \text{X-ray} exposures with \textit{NICER} and \textit{XMM-Newton} (Extended Data Table~\ref{tab:xray_obs}). A list of the radio observations performed using CHIME/Pulsar is provided in Extended Data Table~\ref{tab:radio_obs}. The start times, end times, and durations of these observations are displayed in Extended Data Fig.~\ref{fig:xray_radio_obs_timeline}. Total intensity data, sampled at 8~bits per sample, were recorded from FRB~20200120E using the CHIME/Pulsar backend. The data from each observation were channelized, coherently dedispersed in real time, and saved in a digital polyphase filterbank with 1024~frequency channels for FRB searches. The CHIME/Pulsar radio observations performed on 15, 24, 26, and 28~February~2021, 11~April~2021, and 1~May~2021 were coherently dedispersed using a DM of 87.78\,pc\,cm$^{\text{--3}}$. A refined DM of 87.757\,pc\,cm$^{\text{--3}}$ was used to coherently dedisperse the radio observations carried out on 23 and 29~May~2021.

Before searching the channelized filterbank data for FRBs, the data were reduced using the algorithms described in refs.~\citen{Pearlman+2018a}, \citen{Pearlman+2019a}, \citen{Pearlman+2020a}, and~\citen{Pearlman+2020b}. We first masked frequency channels that were known to be regularly contaminated by RFI. The data from each observation were then corrected for the bandpass slope across the frequency band. Next, we removed low-frequency temporal variations from the channelized data by subtracting the moving average from each data value in each frequency channel, which was calculated using a sliding window spanning 0.5\,s around each time sample. The data in each frequency channel were then rescaled to have a mean of zero and a standard deviation of one. To further mitigate narrowband and wideband~RFI, we used the \texttt{rfifind} routine, available in the \texttt{PRESTO} pulsar search software package\,\cite{Ransom2001}, to generate a mask that was applied to excise additional data corrupted by~RFI.

To search for radio bursts from FRB~20200120E, the filterbank data from each observation were incoherently dedispersed with DMs between 87.70 and 87.90\,pc\,cm$^{\text{--3}}$ using 201 DM trials, with \break $\Delta$DM$_{\text{trial}}$\,$=$\,0.001\,pc\,cm$^{\text{--3}}$. FRB candidates were identified by convolving each dedispersed time series with boxcar functions, with logarithmically-spaced widths between the time resolution of each observation and 100\,ms, using a Fourier-domain matched-filtering algorithm. The detection S/N of each candidate was calculated using
\begin{equation}
	\text{S/N} = \frac{\sum_{i}(f_{i} - \bar{\mu})}{\bar{\sigma}\sqrt{w_{\text{b}}}},
	\label{eqn:frb_snr}
\end{equation}
where $f_{i}$ is the value of the time series at the $i^{\text{th}}$ bin of the boxcar function, $\bar{\mu}$ and $\bar{\sigma}$ are the local mean and root mean square~(RMS) noise after normalization, and $w_{\text{b}}$ is the boxcar width in number of bins. Before calculating the detection S/N of each candidate, the dedispersed, \text{frequency-summed} data were detrended and normalized so that $\bar{\mu}$\,$\approx$\,0 and $\bar{\sigma}$\,$\approx$\,1. Candidates with S/N\,$\geq$\,7 from each DM trial were recorded and visually inspected to determine if the event was astrophysical. No radio bursts were detected from FRB~20200120E during our observations with CHIME/Pulsar.

The SEFD was approximately 23\,Jy during these CHIME/Pulsar observations\,\cite{CHIME+2021a}. Using equation~(\ref{eqn:fluence_eqn}), we place 7$\sigma$ upper limits of $\mathcal{F}_{\text{R,\,min}}^{\text{7}\sigma}$\,$<$\,0.06\,Jy\,ms\,$\sqrt{\frac{w_{\text{t}}}{\text{100}\,\mu\text{s}}}$ on the fluence of radio bursts from FRB~20200120E in the 400--800\,MHz band during the times of our radio observations with the CHIME/Pulsar system, assuming a fiducial burst width of $w_{\text{t}}$\,$=$\,100\,$\mu$s. However, we note that the data from our CHIME/Pulsar observations on 20, 24, 26, and 28~February~2021 were recorded using a time resolution of 327.68\,$\mu$s (Extended Data Table~\ref{tab:radio_obs}), making them suboptimal for detecting radio bursts with widths smaller than the sampling time. We estimate a conservative 50\% error on these radio fluence limits, based on the uncertainty on the SEFD during these radio observations with the CHIME/Pulsar system.


\bigskip

\section*{\text{X-ray} Observations}
\label{sec:xray_obs}


\bigskip

\subsection{\textit{NICER}}
\label{sec:nicer}

\textit{NICER} is an \text{X-ray} telescope operating on the International Space Station\,\cite{Arzoumanian+2014, Gendreau+2016}. The \text{X-ray} Timing Instrument~(XTI) is the primary instrument onboard \textit{NICER}, which is composed of 56 \text{co-aligned} cameras (52~operational on orbit) that are sensitive to detecting photons in the \text{0.2--15}\,keV energy band\,\cite{Prigozhin+2016}. Each camera contains an \text{X-ray} concentrator~(XRC) and a non-imaging silicon drift detector, which is referred to as a Focal Plane Module~(FPM)\,\cite{Okajima+2016}. The FPMs are arranged in a 7\,$\times$\,8 array and provide a field of view of roughly 30\,arcmin$^{\text{2}}$. \text{X-ray} photons detected by the XTI are time-tagged with an absolute accuracy, relative to Global Positioning System~(GPS) time, of better than 100\,ns~(RMS), and the photon energies are measured to a precision of $\sim$2\%\,\cite{Gendreau+2017}. The nominal effective area of the XTI, combining all operational detectors, is 1840\,cm$^{\text{2}}$ at 1.5\,keV.

We performed observations of FRB~20200120E with \textit{NICER} in the \text{0.2--15}\,keV energy band between 15~February~2021 and 30~May~2021. A list of these \textit{NICER} observations is provided in Extended Data Table~\ref{tab:xray_obs}, and a timeline of these observations is shown in Extended Data Fig.~\ref{fig:xray_radio_obs_timeline}. These \textit{NICER} observations were each processed using the \texttt{FTOOLS}\,\cite{Blackburn+1999} and \texttt{HEASoft}\,\cite{Heasoft+2014} (version~6.31.1) software packages, along with the \textit{NICER} Data Analysis Software~(\texttt{nicerdas}; version~10.0, 2022-12-16\_V010a) and \textit{NICER} calibration
database~(\texttt{CALDB}; version~xti20221001). Cleaned event files were produced from the raw data using the \texttt{nicerl2} processing pipeline, which applies `Level2' calibration and performs data screening. The \texttt{nicerl2} pipeline executes five primary subtasks: (1)~\texttt{nicercal} (applies the standard \textit{NICER} calibration to each observation), (2) \texttt{niprefilter2} (creates augmented `Level2' filter files for filtering data and generating background estimates), (3)~\texttt{nimaketime} (generates a Good Time Interval~(GTI) file for event screening), (4)~\texttt{nicermergeclean} (combines \textit{NICER} data from independent Measurement Power Unit slices~(MPUs) and applies screening criteria to the data), and (5)~\texttt{niautoscreen} (performs additional screening of FPMs and MPUs to exclude bad data from detectors and time ranges that may have been missed by the screening applied by the \texttt{nimaketime} routine).

Standard screening criteria were used to generate the cleaned event files. We selected data during time intervals when the angular distance between FRB~20200120E and \textit{NICER}'s boresight was $<$\,54\arcsec, the elevation angle was $>$\,30$^{\circ}$ with respect to the bright Earth limb, and \textit{NICER} was outside of the South Atlantic Anomaly. The \texttt{underonly\_range} parameter, which is used to screen for high optical loading, was set to \text{`0--500'}. The \texttt{overonly\_range} parameter, which is used to mitigate high particle background, was set to \text{`0--30'}. We selected photons in the \text{0.5--10}\,keV energy range for most of our analyses. Cleaned event files were also generated in the \text{0.4--4}\,keV and \text{0.2--15}\,keV energy bands for \text{X-ray} burst and pulsation searches.

We excluded data from two~FPMs~(\texttt{DET\_IDs}~14 and~34), since they have been observed to display episodes of increased detector noise. We also excluded \text{X-ray} photons from `hot' detectors, which were determined for each Observation~ID by identifying detectors whose photon counts were more than 3$\sigma$ above the mean number of photons, calculated using all of the `good' detectors. The set of `good' detectors was determined for each Observation~ID as FPMs whose behaviour was not significantly different from the detector group.

The photon arrival times in each cleaned event file were corrected to the Solar System barycentre using the \texttt{barycorr} tool, the Jet Propulsion Laboratory~(JPL)~DE405 ephemeris, and the position of FRB~20200120E determined from VLBI with the EVN\,\cite{Kirsten+2022}. The photon times were determined using the Barycentric Dynamical Time~(TDB) system. After barycentring, the relative accuracy of the \textit{NICER} photon timestamps is $\sim$80\,ns.


\bigskip

\subsection{\textit{XMM-Newton}}
\label{sec:xmm_obs}

Observations of FRB~20200120E were performed using \textit{XMM-Newton} on \text{10--11}~April~2021 (Observation~ID 0872392401), 30~April to 1~May~2021 (Observation~ID 0872392501), and \text{14--15}~May~2021 (Observation~ID 0872392601). These observations (Extended Data Fig.~\ref{fig:xray_radio_obs_timeline} and Extended Data Table~\ref{tab:xray_obs}) were targeted at FRB~20200120E using a preliminary position ($\alpha_{\text{J2000}}$\,$=$\,09$^{\text{h}}$57$^{\text{m}}$56.00$^{\text{s}}$,  $\delta_{\text{J2000}}$\,$=$\,68$^{\circ}$49$\arcmin$32.0$\arcsec$) determined using the~EVN. The EPIC/MOS and EPIC/pn detectors were operated in Large Window mode, providing time resolutions of 0.9\,s and 47.7\,ms, respectively. We opted to use Large Window mode during these \textit{XMM-Newton} observations, instead of Timing mode, because it provided a lower background level with spatial information and high live time.

The \textit{XMM-Newton} data were reduced and analysed using standard tools from the \textit{XMM-Newton} Science Analysis System~(\texttt{XMM-SAS}; version~19.1.0)\,\cite{Gabriel+2004} and \texttt{HEASoft}\,\cite{Heasoft+2014} (version~6.28) software packages. Standard filters were applied to retain good events in the \text{0.5--10}\,keV energy band from the EPIC/MOS and EPIC/pn detectors. The \textit{XMM-Newton} EPIC/pn \text{X-ray} photons were extracted using a point-spread function~(PSF) centred on FRB~20200120E's position from VLBI\,\cite{Kirsten+2022}. Source events were extracted from a circular region of radius 18\arcsec, centred on the source position, and background events were extracted from an annular region with an inner diameter of 18\arcsec\ and outer diameter of 72\arcsec. The photon event arrival times were corrected to the Solar System barycentre using the \text{VLBI-determined} position of FRB~20200120E\,\cite{Kirsten+2022}.


\bigskip

\subsection{\textit{Chandra}}
\label{sec:chandra_obs}

We observed FRB~20200120E using \textit{Chandra} on 16 and 17~December~2020 (Observation~IDs 23544 and 24894), using the High Resolution Camera~(HRC) in Timing mode (Extended Data Fig.~\ref{fig:xray_radio_obs_timeline} and Extended Data Table~\ref{tab:xray_obs}). This mode provides a time resolution of 16\,$\mu$s\,\cite{Wilkes+2022}. We selected this mode because FRB~20200120E was localized to only arcminute precision\,\cite{Bhardwaj+2021a} at the time, and thus, a large field of view and high time resolution were required at the expense of effective area. These \textit{Chandra} observations of FRB~20200120E were targeted at a preliminary position ($\alpha_{\text{J2000}}$\,$=$\,09$^{\text{h}}$57$^{\text{m}}$56.40$^{\text{s}}$,  $\delta_{\text{J2000}}$\,$=$\,68$^{\circ}$49$\arcmin$04.8$\arcsec$), derived using baseband voltage data from CHIME/FRB. We also analysed an archival \textit{Chandra} ACIS observation from 24~August~2008 (Observation~ID 9540), previously presented in ref.~\citen{Kirsten+2022}, where FRB~20200120E was located 14\arcmin\ from boresight.

We used the \textit{Chandra} Interactive Analysis of Observations~ (\texttt{CIAO}; version~4.12) software package\,\cite{Fruscione+2006} and standard procedures to reduce and analyse the \textit{Chandra} observations. For the HRC observations, source photons were extracted from a circular region of radius 1\arcsec, centred on the VLBI position of FRB~20200120E\,\cite{Kirsten+2022}. Since FRB~20200120E was located 14\arcmin\ \text{off-axis} in Observation~ID 9540, events were extracted from a \text{10\arcsec-radius} region at the position of FRB~20200120E, and background events were extracted from a \text{60\arcsec-radius} region at a similar \text{off-axis} angle. The photon event arrival times were then corrected to the Solar System barycentre.


\bigskip

\subsection{\textit{NuSTAR}}
\label{sec:nustar}

\textit{NuSTAR} observations of FRB~20200120E (Extended Data Fig.~\ref{fig:xray_radio_obs_timeline} and Extended Data Table~\ref{tab:xray_obs}) were performed nearly simultaneously with the \textit{Chandra} observations on 16 and 17~December~2021 (Observation~IDs 80701506002 and 80701506004). The data were reduced using the standard \texttt{nupipeline} and \texttt{nuproducts} routines available in the \texttt{HEASoft} (version~6.28) software package. We corrected the photon arrival times to the Solar System barycentre using the \texttt{barycorr} tool. Source events were extracted from a circular region of radius 2\arcmin, centred on the position of FRB~20200120E determined from VLBI\,\cite{Kirsten+2022}. Background events were extracted from a region of the same size positioned away from the source.


\bigskip

\section*{\text{X-ray} Burst Searches and Prompt \text{X-ray} Fluence Limits}
\label{sec:xray_burst_search_limits}


\bigskip

\subsection{\textit{NICER}}
\label{sec:nicer_bursts}

We carried out a blind search for \text{X-ray} bursts during each of the \textit{NICER} observations of \break FRB~20200120E listed in Extended Data Table~\ref{tab:xray_obs}. We independently searched for \text{X-ray} bursts using the unbinned photon event times during GTIs in three different energy ranges: \text{0.4--4}\,keV, \text{0.5--10}\,keV, and \text{0.2--15}\,keV.  These energy bands were chosen to maximize the sensitivity to detecting \text{X-ray} bursts, both within the most sensitive part of \textit{NICER}'s photon energy bandpass and over a broader range of photon energies.

We used a Bayesian Blocks technique\,\cite{Scargle+2013} to search for \text{X-ray} bursts with durations up to 100\,s during each \textit{NICER} Observation ID. Bayesian Blocks is a dynamic programming algorithm that determines the optimal segmentation of the data and an optimal histogram characterizing the data with adaptive bin widths. This algorithm can detect and characterize statistically significant variations using event data, binned data, or measurements at arbitrary times with known error distributions, and it is not limited by signal amplitude, time resolution, or sampling of the data. In the Bayesian Blocks algorithm, the optimal block partition is obtained by maximizing the likelihood function ($\mathcal{L}$)\,\cite{Scargle+2013} using
\begin{equation}
	\ln\mathcal{L}_{k} = N_{k}\ln\lambda_{k} - \lambda_{k}T_{k},
	\label{eqn:bayesian_blocks_likelihood}
\end{equation}
where $N_{k}$ is the number of events in block $k$, $T_{k}$ is the length of block $k$, and $\lambda_{k}$\,$=$\,$N_{k}$/$T_{k}$ is the expected count rate in block $k$. The following prior distribution (\texttt{ncp\_prior})\,\cite{Scargle+2013} for the number of blocks was used when the Bayesian Blocks algorithm was applied to the unbinned \textit{NICER} event data from FRB~20200120E:
\begin{equation}
	\texttt{ncp\_prior} = 4 - \ln(73.53 p_{0}N^{-0.478}),
	\label{eqn:bayesian_blocks_ncp_prior}
\end{equation}
where $N$ is the total number of events, and we have chosen $p_{0}$\,$=$\,0.05 for the false positive probability.

The detection significance of each candidate burst was determined by calculating the Poisson probability of detecting the observed number of photons in the block, given a \text{non-burst} count rate estimated from nearby 1\,s time intervals. In all cases, no \text{X-ray} bursts were detected above a 3$\sigma$ probability detection threshold, corresponding to a false alarm probability ($P_{\text{fa}}$) of $P_{\text{fa}}$\,$<$\,0.0027.


\bigskip

\subsection{\textit{XMM-Newton}, \textit{Chandra}, and \textit{NuSTAR}}
\label{sec:xmm_chandra_nustar_bursts}

Several radio bursts (\text{B6--B9}) were detected from \break FRB~20200120E using the Effelsberg radio telescope during the \textit{XMM-Newton} observations performed between 31~April and 1~May~2021 (Observation~ID 087238501) (Extended Data Fig.~\ref{fig:xray_radio_obs_timeline}). Unfortunately, these observations were contaminated by soft proton flares, so we only consider the last radio burst~(B9) detected using Effelsberg, which occurred at a time when the contamination was low. We searched for \text{X-ray} bursts near the time of burst~B9 by considering the nearest photons, detected using \textit{XMM-Newton}'s EPIC/pn camera, to the barycentric ToA of the radio burst at infinite frequency. The nearest photon occurred 58.4\,s  before burst~B9 (Extended Data Fig.~\ref{fig:xmm_b6_b7_b8_b9_xray_radio_light_curves}e). Based on the count rate measured from the background region, the false alarm probability of a photon arriving in such a time window is $P_{\text{fa}}$\,$=$\,0.33. We also considered the ten nearest photons from the ToA of B9 to probe for \text{X-ray} bursts on longer timescales (up to $\sim$600\,s). For time windows spanning the second to tenth nearest photons to B9, there was no statistically significant \text{X-ray} emission above a 3$\sigma$ probability detection threshold ($P_{\text{fa}}$\,$<$\,0.0027). We repeated this search using data from the EPIC/MOS1 and EPIC/MOS2 cameras, and we also found no statistically significant \text{X-ray} excess above a 3$\sigma$ probability detection threshold ($P_{\text{fa}}$\,$<$\,0.0027).

In addition, we carried out searches for \text{X-ray} bursts at all times during the \textit{XMM-Newton} observations by searching for excesses from the Poisson background rate on 0.1\,s, 1\,s, 10\,s, and 100\,s timescales, where the minimum search timescale was set by the 47.7\,ms time resolution of the \textit{XMM-Newton} EPIC/pn data. We searched for evidence of enhanced \text{X-ray} emission by comparing the binned light curves of the source and background regions. After taking into account the number of search trials, no bins in the source light curves exceeded a 3$\sigma$ probability detection threshold ($P_{\text{fa}}$\,$<$\,0.0027).

We performed a blind search for \text{X-ray} bursts at all times during the \textit{Chandra} and \textit{NuSTAR} observations of FRB~20200120E. Following the same search procedure as described above for the \textit{XMM-Newton} EPIC/pn data, we searched for statistically significant deviations from the background rate on 0.1\,s, 1\,s, 10\,s, and 100\,s timescales. However, since the background count rate did not show any significant variability, we compared with a single background rate derived from the entire observation, instead of using a binned background light curve. No statistically significant excess \text{X-ray} emission was found above a 3$\sigma$ probability detection threshold ($P_{\text{fa}}$\,$<$\,0.0027) at any of these timescales.


\bigskip

\subsection{Prompt \text{X-ray} Fluence Limits}
\label{sec:xray_burst_fluence_limits}

We used the Bayesian method described in ref.~\citen{Kraft+1991} to derive \text{X-ray} fluence limits at the times of bursts~B4 and~B9. Using the number of detected photons and the expected number of background photons, we obtained standard 3$\sigma$ confidence intervals describing the number of observed photons on a variety of timescales. Confidence intervals were calculated for burst~B4 on timescales of 100\,ns to 10\,s using measurements from \textit{NICER} and on timescales of 100\,ms to 100\,s using measurements from \textit{XMM-Newton}'s EPIC/pn camera for burst~B9. These confidence intervals were then translated into an \text{X-ray} fluence limit for each spectral model listed in Table~\ref{tab:nicer_xmm_b4_b9_xray_limits}. For burst~B4, we report the most conservative limit after considering each of the \textit{NICER} background estimates listed in Extended Data Table~\ref{tab:nicer_xray_bkg}. Background estimates for burst~B9 were derived from the background count rate measured using \textit{XMM-Newton}'s EPIC/pn camera.


\bigskip

\subsection{Sensitivity to Detecting Short \text{X-ray} Bursts with \textit{NICER}}
\label{sec:nicer_sensitivity}

To characterize \textit{NICER}'s nominal sensitivity to detecting short \text{X-ray} bursts with temporal widths between 100\,ns and 10\,s, we performed an analysis of observations of six blank sky regions that are free of detectable \text{X-ray} point sources. The observations used in this analysis are labelled `BKGD\_RXTE\_$B_{i}$', where $B_{i}$\,$=$\,\{1, 2, 3, 4, 5, 6, 8\}. These fields are routinely observed by \textit{NICER}, and they were previously used by the \textit{Rossi \text{X-ray} Timing Explorer~(RXTE)} for background modelling\,\cite{Remillard+2022}. We note that one of these background fields, `BKGD\_RXTE\_6' ($\alpha_{\text{J2000}}$\,$=$\,10$^{\text{h}}$40$^{\text{m}}$00$^{\text{s}}$,  $\delta_{\text{J2000}}$\,$=$\,72$^{\circ}$34$\arcmin$12$\arcsec$), is only $\sim$5$^{\circ}$ offset from the VLBI position of FRB~20200120E\,\cite{Kirsten+2022}. Observations of `BKGD\_RXTE\_7' were not used in our analysis because a bright, soft \text{X-ray-emitting} star is located in the field. These data can be accessed through NASA's High Energy Astrophysics Science Archive Research Center~(HEASARC) archive. We used these observations to empirically measure \textit{NICER}'s detector and sky background on short timescales.

We processed all of the available \textit{NICER} blank sky observations (`BKGD\_RXTE\_$B_{i}$') performed between 25~June~2017 and 6~February~2022 using the \texttt{nicerl2} pipeline with standard calibration and data screening parameters. \text{X-ray} photons in the \text{0.5--10}\,keV energy range were selected for our background analysis. We excluded photons from two active FPMs (\texttt{DET\_IDs} 14 and 34) that are known to exhibit episodes of increased detector noise. We also excluded \text{X-ray} photons from `hot' detectors from each exposure, using the method described above (see \text{X-ray} Observations). We included all exposures with GTIs spanning at least 200\,s in our analysis. Using standard parameters, the \texttt{nicerl2} pipeline was unable to completely remove all of the contamination from some of the blank sky exposures near the boundaries of the GTIs. To mitigate this problem, we removed photons from each exposure with arrival times occurring within 1~min of the GTI start and end times. In addition, a small fraction ($\sim$3\%) of the background observations needed to be excluded due to missing data or poor data quality. After filtering, our analysis included blank sky data from 4265 independent exposures (total exposure time of 2\,Ms), obtained from 1658~\textit{NICER} Observation IDs. 

The background count rate distribution was modelled separately for each blank sky region using a Bayesian Markov chain Monte Carlo~(MCMC) parameter estimation procedure. We generated light curves using data from each background exposure, binned at each of the following time resolutions: 100\,ns, 1\,$\mu$s, 10\,$\mu$s, 100\,$\mu$s, 1\,ms, 10\,ms, 100\,ms, 1\,s, and 10\,s. For each background field, we measured the distribution of count rates at each of the above-mentioned time resolutions and then carried out an unbinned likelihood analysis to measure the mean background count rate. Posterior probability distributions for the mean background count rate, $\lambda_{b}$, at each timescale were calculated from MCMC analyses, where we used an uninformed, flat prior for $\lambda_{b}$ and a Poisson likelihood function, $\mathcal{L}(\left.\lambda_{b}\right|b_{i})$, given by
\begin{equation}
	\ln\mathcal{L}(\left.\lambda_{b}\right|b_{i})=-N_{b}\lambda_{b}+\ln(\lambda_{b})\sum_{i=1}^{N_{b}}b_{i}-\sum_{i=1}^{N_{b}}\ln(b_{i}!).
	\label{eqn:poisson_likelihood}
\end{equation}
Here, $b_{i}$ represents the measured background count rate in each time bin, and the number of background count rate measurements is denoted by $N_{b}$.

The posterior probability density functions were sampled using an affine-invariant MCMC ensemble sampler\,\cite{Goodman+2010}, implemented in the \texttt{emcee} software package\,\cite{ForemanMackey+2013}. The parameter spaces were explored using 200 walkers and a chain length of 5000 steps per walker. The walkers were initialized to start from a small Gaussian ball centred around the maximum likelihood estimator. The first 100 steps in each chain were treated as the initial burn-in phase and were removed from the analysis. The posterior probability distributions were calculated using the remaining 4900 steps in each chain. Best-fit values for the mean background count rate, $\lambda_{b}$, were obtained at each timescale for each blank sky region from the median of the posterior distributions, along with 1$\sigma$ uncertainties derived from Bayesian credible intervals.

We used these measurements to calculate a single value for $\lambda_{b}$ at each timescale from the weighted average of the mean background count rates from all of the blank sky regions. The weighted average values of $\lambda_{b}$ were treated as the nominal background rate at each timescale when calculating confidence intervals describing the number of \text{X-ray} counts near the times of radio bursts from FRB~20200120E, using the Bayesian method described in ref.~\citen{Kraft+1991}. In Extended Data Fig.~\ref{fig:nicer_bkgd_xray_flux}a, we show the average, absorbed \text{X-ray} background flux of \textit{NICER} in the \text{0.5--10}\,keV energy band for integration times between 100\,ns and 10\,s using these background count rates, assuming $N_{\text{H}}$\,$=$\,6.73\,$\times$\,10$^{\text{20}}$\,cm$^{\text{--2}}$ (the Galactic hydrogen column density contribution along the line of sight to FRB~20200120E\,\cite{HI4PI+2016}) and an \text{X-ray} spectrum with a photon index of $\Gamma$\,$=$\,1.4 (similar to the diffuse \text{X-ray} background\,\cite{Mushotzky+2000}). The corresponding \text{X-ray} fluences are shown in Extended Data Fig.~\ref{fig:nicer_bkgd_xray_flux}b on timescales of 100\,ns to 10\,s.

We compare these results with two independent methods used to estimate the \textit{NICER} background near the time of burst~B4. First, we consider the `Space Weather' background model\,\cite{Remillard+2022}, which uses a combination of the cut-off rigidity~(\texttt{COR\_SAX})\,\cite{Campana+2014} and the planetary $K$-index~($K_{\text{p}}$)\,\cite{Bartels1949} to obtain an estimate of the space weather environment. This model also uses the \texttt{SUN\_ANGLE} parameter to account for the \text{low-energy} background produced by optical loading. On timescales of 100\,ns to 10\,s, we find that the predicted \textit{NICER} background in the \text{0.5--10}\,keV energy range from our analysis of blank sky regions is larger and more conservative than the background estimates obtained from the \textit{NICER} `Space Weather' model near the time of burst~B4\,\cite{Remillard+2022} (Extended Data Table~\ref{tab:nicer_xray_bkg}). These measurements are also more conservative than our background estimates obtained from \textit{NICER} using photons within $\pm$100\,s of burst~B4 (Extended Data Table~\ref{tab:nicer_xray_bkg}).


\bigskip

\section*{Spectral Analysis and Persistent \text{X-ray} Emission}
\label{sec:persistent_xray_emission}


\bigskip

\subsection{\textit{Chandra}, \textit{NuSTAR}, and \textit{XMM-Newton}}
\label{sec:chandra_nustar_xmm_persistent_xray_emission}

For all of the \textit{Chandra}, \textit{NuSTAR}, and \textit{XMM-Newton} observations listed in Extended Data Table~\ref{tab:xray_obs}, the number of counts in the source extraction regions was consistent with the background count rate. To place limits on the count rate of a putative \text{X-ray} source coincident with FRB~20200120E, we used the Bayesian method described in ref.~\citen{Kraft+1991}. We then translated the count rate limit to a flux limit by assuming a photoelectrically absorbed power-law source spectrum, with a photon index of $\Gamma$\,$=$\,2 and a hydrogen column density of $N_{\text{H}}$\,$=$\,6.73\,$\times$\,10$^{\text{20}}$\,cm$^{\text{--2}}$ towards FRB~20200120E\,\cite{HI4PI+2016}, while also accounting for the spectral response of the source. The spectral response was derived using \texttt{CIAO}'s \texttt{specextract} tool for the \textit{Chandra} observations, the \texttt{nuproducts} routine in the \texttt{HEASoft} software package for the \textit{NuSTAR} observations, and \texttt{XMM-SAS}'s \texttt{arfgen} tool for the \textit{XMM-Newton} observations. In Extended Data Table~\ref{tab:xray_obs}, we list 3$\sigma$ upper limits on the persistent \text{X-ray} flux, derived using this method, in the \text{0.5--10}\,keV energy band for each \textit{Chandra} and \textit{XMM-Newton} observation, and in the \text{3--79}\,keV energy band for each \textit{NuSTAR} observation.


\bigskip

\subsection{\textit{NICER}}
\label{sec:nicer_persistent_xray_emission}

After the data from each \textit{NICER} Observation ID (Extended Data Table~\ref{tab:xray_obs}) were processed with the \texttt{FTOOLS}\,\cite{Blackburn+1999} and \texttt{HEASoft} (version~6.31.1) software packages\,\cite{Heasoft+2014}, using the \texttt{nicerl2} standard calibration and filtering pipeline, we used the \texttt{nicerl3-spect} spectral product pipeline to produce source and background spectra, along with ancillary response files~(ARFs) and response matrix files~(RMFs), for each observation. Background spectra were generated using three different models: (1)~an empirical `3C50' background model\,\cite{Remillard+2022}, which can be produced using the \texttt{nibackgen3C50} \texttt{FTOOLS} routine\,\cite{Remillard+2022}; (2)~the `SCORPEON' background model, which is available through the \texttt{niscorpeon} \texttt{FTOOLS} routine and allows the background to be fit as a parameterized model simultaneously with the data; and (3)~the `Space Weather' background model, which is implemented in the \texttt{niswbkgspect} \texttt{FTOOLS} routine. The \texttt{niswbkgspect} routine estimates the space weather conditions during a given observation and produces a \textit{NICER} background spectrum based on blank sky data matching these conditions. Each of the spectra was optimally rebinned, as described in ref.~\citen{Kaastra+2016}, with a minimum of ten counts per bin using the \texttt{ftgrouppha} \texttt{FTOOLS} routine.

We obtained source and background spectra from each \textit{NICER} observation in the \text{0.5--10}\,keV energy range using each of the background models described above. All of the \textit{NICER} observations yielded background-dominated spectra, based on comparisons with background spectra generated using these background models. The source spectra from Observation~IDs 3658040103, 3658040104, and 3658040105 were marginally inconsistent with the background spectra. However, for these Observation~IDs, we attribute this to incompleteness in modelling the time-variable background using the currently available background models. In Extended Data Table~\ref{tab:xray_obs}, we provide 3$\sigma$ limits on the \text{0.5--10}\,keV persistent \text{X-ray} flux of FRB~20200120E during each \textit{NICER} observation, since we did not find evidence of significant persistent \text{X-ray} emission.

The 3$\sigma$ persistent \text{X-ray} flux limits from each \textit{NICER} observation were derived based on a fiducial photoelectrically absorbed \text{power-law} spectral model, with a hydrogen column density of \break $N_{\text{H}}$\,$=$\,6.73\,$\times$\,10$^{\text{20}}$\,cm$^{\text{--2}}$\,\cite{HI4PI+2016} and a photon index of $\Gamma$\,$=$\,2. For each Observation~ID, we used the \texttt{fakeit} routine in the \texttt{XSPEC} (version~12.13.0c) spectral fitting software package\,\cite{Arnaud1996} to generate simulated spectra using the \texttt{tbabs(po)} spectral model. Simulated spectra for each \textit{NICER} observation were constructed using the source spectrum, a background spectrum produced using the `3C50' background model, and the ARF and RMF from each observation. The hydrogen column density and photon index were fixed to the above values, and a wide range of trial normalizations were used to produce multiple simulated spectra for each observation. The simulated spectra from each observation were then fit using the source spectrum as an empirical background model, while using the ARF and RMF from each observation. During the fitting process, the hydrogen column density and photon index were fixed, and the normalization was left as a free parameter. The \texttt{flux} routine in \texttt{XSPEC} was used to determine the largest normalization parameter value that yielded a 3$\sigma$ confidence interval for the \text{0.5--10}\,keV \text{X-ray} flux that was consistent with non-detection. The 3$\sigma$ persistent \text{X-ray} flux limit on FRB~20200120E during each \textit{NICER} observation was obtained using the \texttt{tbabs(po)} spectral model, with this normalization parameter value, $N_{\text{H}}$\,$=$\,6.73\,$\times$\,10$^{\text{20}}$\,cm$^{\text{--2}}$\,\cite{HI4PI+2016}, and $\Gamma$\,$=$\,2. The 3$\sigma$ persistent \text{X-ray} luminosity limits shown in Fig.~\ref{fig:xray_luminosity_upper_limits} for FRB~20200120E were calculated using the 3$\sigma$ persistent \text{X-ray} flux limits listed in Extended Data Table~\ref{tab:xray_obs} and the distance to FRB~20200120E.


\bigskip

\section*{Periodicity Searches}
\label{sec:periodicity}


\bigskip

\subsection{\text{X-ray} Pulsations}
\label{sec:nicer_periodicity}

We carried out several independent searches for \text{X-ray} pulsations using the \textit{NICER} data. Each of our analyses was repeated after selecting photons in three different energy bands: \text{0.4--4}\,keV, \text{0.5--10}\,keV, and \text{0.2--15}\,keV. These energy ranges were chosen to maximize the sensitivity to detecting \text{X-ray} pulsations, both using photons in the most sensitive part of \textit{NICER}'s photon energy bandpass and over a broader range of photon energies.

First, we used all of the \textit{NICER} observations in Extended Data Table~\ref{tab:xray_obs} to generate evenly sampled light curves, sampled at 65.536\,kHz (time resolution of $\sim$15.3\,$\mu$s), using \text{X-ray} photon arrival times that were corrected to the Solar System barycentre using the position of FRB~20200120E determined from VLBI\,\cite{Kirsten+2022}. These light curves spanned a total duration of 8.94\,Ms (103.5\,days) and contained $\sim$5.9\,$\times$\,10$^{\text{11}}$ time bins. We calculated power spectra from fast Fourier transforms~(FFTs) of each light curve, which were searched for \text{X-ray} pulsations. No astrophysical candidates were found above a 3$\sigma$ detection threshold, after accounting for the number of independent frequency trials.

Next, we repeated the search described above using separate light curves derived from \textit{NICER} observations between: (1)~22~February~2021 and 29~February~2021 (Observation IDs: 3658040102, 3658040103, 3658040104, 3658040105, 3658040106, and 3658040107) and (2)~22~May~2021 and 30~May~2021 (Observation IDs: 3658040109, 3658040110, 3658040111, and 3658040112). These time periods were chosen because they correspond to time intervals when \textit{NICER} was used to perform \text{X-ray} observations of FRB~20200120E with both the highest cadence and largest cumulative exposure over an approximately 1~week period. Searches for \text{X-ray} pulsations in these shorter combined datasets provide greater sensitivity to detecting periodicities that may be smeared in the power spectrum of the full light curve spanning 103.5\,days, for example, if FRB~20200120E harbours a rotating NS that exhibits changes in its rotational frequency over time. We did not detect any periodic signals above a significance threshold of 3$\sigma$ in the power spectra derived from these light curves spanning $\sim$1~week.

We also performed `semi-coherent' searches for \text{X-ray} pulsations using power spectra computed from the light curves of each \textit{NICER} Observation~ID. The light curves from each Observation~ID were evenly sampled at a time resolution of $\sim$15.3\,$\mu$s and \text{zero-padded} to have the same length as the light curve from Observation~ID 3658040105, which spanned the longest duration. This yielded a power spectrum from each \textit{NICER} Observation~ID with the same frequency resolution. The powers in each spectral bin of the power spectra were summed to create a stacked power spectrum, which was then searched for statistically significant peaks. We refer to this technique as a `semi-coherent' method since powers are added, instead of complex amplitudes, and the overall phase information of the signal between observations is not incorporated. The power spectra used in this analysis were computed from an FFT and also using the Rayleigh statistic\,\cite{Buccheri+1983}, $Z_{n}^{\text{2}}$, via
\begin{equation}
	\small
	Z_{n}^{2}=\frac{2}{N}\sum_{k=1}^{n}\left[\left(\sum_{j=1}^{N}\cos k\phi_{j}\right)^{2}+\left(\sum_{j=1}^{N}\sin k\phi_{j}\right)^{2}\right],
	\label{eqn:z2}
\end{equation}
where $n$ denotes the number of harmonics summed, $N$ is the number of photon~ToAs, and $\phi_{j}$ is the phase of each ToA calculated at a trial frequency, $\nu$. We used $n$\,$=$\,1 to derive stacked power spectra with the $Z_{\text{1}}^{\text{2}}$ statistic. No statistically significant signals were detected above a 3$\sigma$ detection threshold in the stacked power spectra obtained using the FFT and $Z_{\text{1}}^{\text{2}}$ statistics.

Assuming a \text{pulsar-like} FRB~source, we place 3$\sigma$ limits on FRB~20200120E's pulsed \text{X-ray} flux ($F^{\text{pulsed}}_{\text{X}}$). For a duty cycle of 10\%, the 3$\sigma$ upper limits on the pulsed \text{X-ray} flux are given by \break $F^{\text{pulsed}}_{\text{X}}$\,$=$\,$\sqrt{\delta_{\text{PSR}} / (1 - \delta_{\text{PSR}})}F_{\text{X}}$\,$<$\,0.33$F_{\text{X}}$, where $\delta_{\text{PSR}}$ is the assumed duty cycle of the putative \text{pulsar-like} FRB~source and $F_{\text{X}}$ is the persistent \text{X-ray} flux limit derived at the time of each \textit{NICER} observation (Extended Data Table~\ref{tab:xray_obs}). The 3$\sigma$ limits on FRB~20200120E's pulsed \text{X-ray} flux range between 2.5\,$\times$10$^{\text{--15}}$ and 3.5\,$\times$10$^{\text{--13}}$\,erg\,cm$^{\text{--2}}$\,s$^{\text{--1}}$ at the times of our \textit{NICER} observations.


\bigskip

\section*{Comparisons with \text{X-ray} Sources and Transients}
\label{sec:xray_comparison}


\bigskip

\subsection{Persistent \text{X-ray} Luminosity}
\label{sec:comparison_persistent_xray_limits}

In Fig.~\ref{fig:xray_luminosity_upper_limits}, we compare our persistent \text{X-ray} luminosity limits for \break FRB~20200120E, from \textit{Chandra}, \textit{NICER}, \textit{NuSTAR}, and \textit{XMM-Newton} observations, to the persistent \text{X-ray} luminosities of ULX sources, Galactic LMXBs, Galactic HMXBs, and magnetars. In each case, we calculated the \text{X-ray} luminosity ($L_{\text{X}}$) from the \text{X-ray} flux ($F_{\text{X}}$) using $L_{\text{X}}$\,$=$\,$4\pi d_{\text{L}}^2 F_{\text{X}}/(1 + z)$, where $d_{\text{L}}$ is the luminosity distance of each object and $z$ is the corresponding redshift. Redshifts were determined for each source using the \texttt{Planck18} routine in the \texttt{astropy.cosmology} Python software package\,\cite{Astropy+2013}, which uses the cosmological parameters measured in ref.~\citen{Planck+2020}.

The histogram of persistent \text{X-ray} luminosities for ULX~sources was calculated using values from the ULX~source catalogue available in ref.~\citen{Bernadich+2022}. We adopt a typical luminosity definition of $L_{\text{X}}$\,$\ge$\,10$^{\text{39}}$\,erg\,s$^{\text{--1}}$ for ULX~sources\,\cite{Kaaret+2017, Bernadich+2022}. In the pink histogram shown in Fig.~\ref{fig:xray_luminosity_upper_limits}b, we only included objects from the ULX~catalogue with an \text{X-ray} luminosity above this threshold. We also only included ULX sources whose luminosity measurement was significant, with respect to its uncertainty. We further selected only ULX~sources that were associated with a single potential host galaxy to mitigate contamination from unreliable distance measurements. The persistent \text{X-ray} luminosity values were calculated in the \text{0.2--12}\,keV energy band, using the \text{0.2--12}\,keV \text{X-ray} flux of each object and the host galaxy luminosity distances listed in the ULX~catalogue.

The grey histogram in Fig.~\ref{fig:xray_luminosity_upper_limits}b shows the persistent \text{X-ray} luminosities of Galactic~LMXBs, obtained from measurements provided in the Galactic~LMXB catalogue available in ref.~\citen{Avakyan+2023}. We calculated \text{X-ray} luminosities in the \text{0.2--12}\,keV energy range using the maximum \text{0.2--12}\,keV \text{X-ray} flux of each source and the average distance listed for each object in the LMXB catalogue. Similarly, the green histogram in Fig.~\ref{fig:xray_luminosity_upper_limits}b shows the persistent \text{X-ray} luminosities of Galactic~HMXBs, derived using measurements from the Galactic~HMXB catalogue provided in ref.~\citen{Neumann+2023}. The \text{X-ray} luminosities were also calculated in the \text{0.2--12}\,keV energy band using the maximum \text{0.2--12}\,keV \text{X-ray} flux of each source and the average distance listed for each object in the HMXB~catalogue. We excluded LMXB and HMXB~sources that were missing either a maximum \text{X-ray} flux value or a distance value in their respective catalogue. Since maximum \text{X-ray} flux values were used to generate these histograms, the grey and green histograms in Fig.~\ref{fig:xray_luminosity_upper_limits}b show the distributions of Galactic~LMXBs and HMXBs in their high \text{X-ray} state.

The brown histogram in Fig.~\ref{fig:xray_luminosity_upper_limits}b shows the distribution of persistent \text{X-ray} luminosities for magnetars. We used the unabsorbed \text{X-ray} flux values in \text{2--10}\,keV energy band and the distance measurements listed in the McGill Magnetar Catalog\,\cite{Olausen+2014} to calculate an \text{X-ray} luminosity for each magnetar. We also included \text{X-ray} luminosities from several recently discovered magnetars\,\cite{Younes+2020b}\textsuperscript{,}\cite{Rea+2009, Hu+2020, Enoto+2021b, Younes+2022a}, determined by translating the reported persistent \text{X-ray} flux to the \text{2--10}\,keV energy band using the \texttt{XSPEC} spectral fitting software package\,\cite{Arnaud1996} and the \text{best-fit} spectral parameters describing the persistent \text{X-ray} emission for each source.


\bigskip

\subsection{Prompt \text{X-ray} Pseudo-Fluence}
\label{sec:comparison_prompt_xray_limits}

\begin{figure*}
	
	\vspace*{-1.0cm}
	
	\centering
	
	\includegraphics[trim=0cm 0cm 0cm 0cm, clip=true, scale=0.795, angle=0]{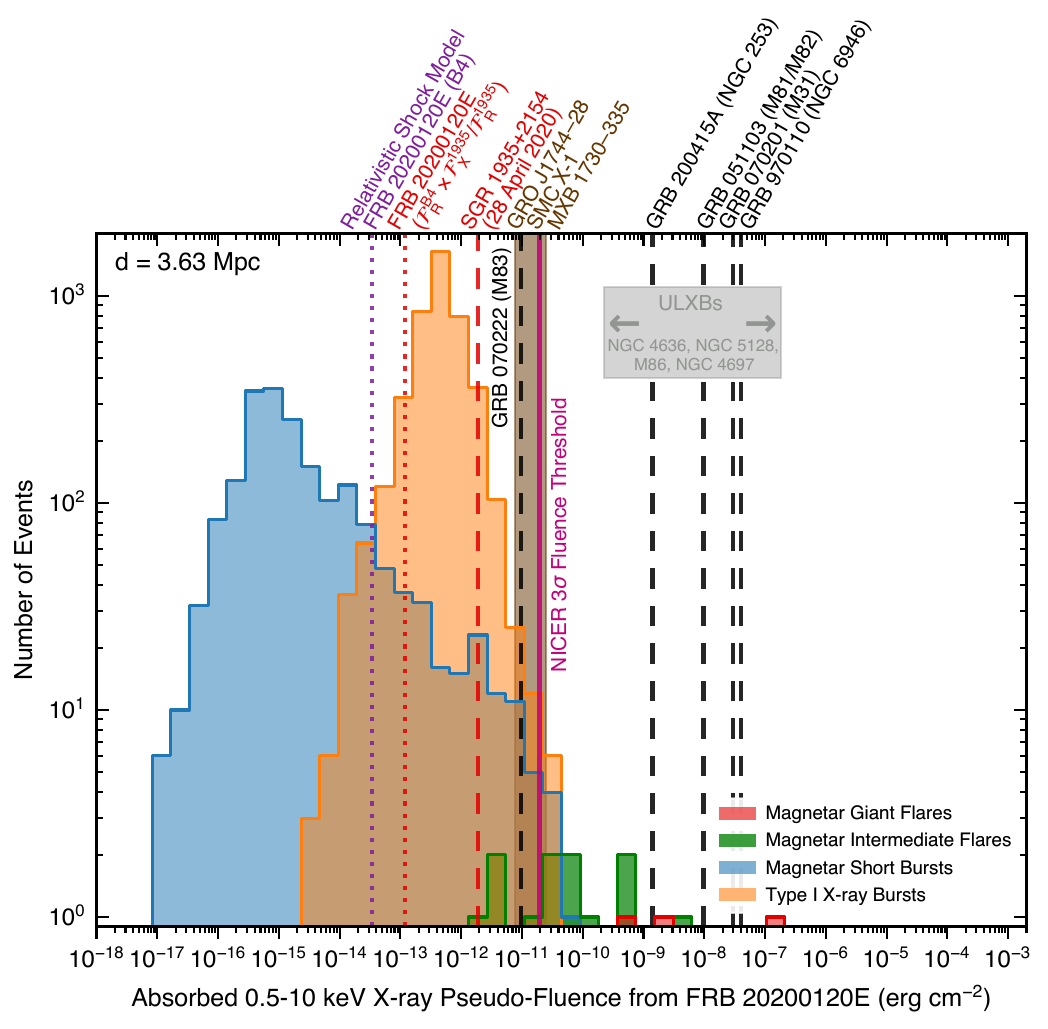}
	
	\caption{\textbf{Absorbed \text{X-ray} pseudo-fluence distributions of \text{X-ray} transients from the distance of FRB~20200120E in the \text{0.5--10}\,keV energy range, compared with our best prompt \text{X-ray} fluence limit from FRB~20200120E.} The red, green, blue, and orange histograms show the \text{X-ray} pseudo-fluence distributions of magnetar giant flares, magnetar intermediate flares, magnetar short bursts, and type~I \text{X-ray} bursts from LMXBs\,\cite{Galloway+2020}, respectively. The vertical magenta line corresponds to our best prompt \text{X-ray} fluence limit from FRB~20200120E, derived from \textit{NICER} observations at the time of burst~B4. The \text{X-ray} fluence limit derived from \textit{XMM-Newton} observations at the time of burst~B9 is similar to the \textit{NICER} \text{X-ray} fluence limit derived at the time of~B4 (Table~\ref{tab:nicer_xmm_b4_b9_xray_limits}). The \textit{NICER} \text{X-ray} fluence limit is only shown for clarity. The grey rectangle shows the range of \text{X-ray} pseudo-fluences derived from~ULXBs detected from nearby galaxies\,\cite{Irwin+2016, Jonker+2013, Sivakoff+2005}. The vertical dashed black lines indicate the \text{X-ray} pseudo-fluences of GRBs believed to be associated with giant flares from extragalactic magnetars\,\cite{Crider2006, Ofek+2006, Frederiks+2007, Mazets+2008, Burns+2021, Svinkin+2021}. The vertical dashed red line shows the predicted \text{X-ray} fluence of the \text{X-ray} burst associated with the FRB-like radio burst detected from SGR~1935+2154 on 28~April~2020\,\cite{Li+2021}, if it were emitted from the location of FRB~20200120E. The vertical dotted red line indicates the predicted \text{X-ray} fluence of burst~B4, assuming the same \text{X-ray}-to-radio fluence ratio as the FRB-like radio burst detected on 28~April~2020 from SGR~1935+2154\,\cite{Li+2021, Bochenek+2020c, CHIME+2020a}. The shaded brown region corresponds to the range of \text{X-ray} pseudo-fluences derived from bright type~II \text{X-ray} bursts from GRO~J1744--28\,\cite{Younes+2015} and MXB~1730--335\,\cite{Bagnoli+2015}, and \text{X-ray} flares resembling type~II \text{X-ray} bursts from SMC~X-1\,\cite{Moon+2003}. The vertical dotted purple line shows the \text{X-ray} pseudo-fluence of the flare predicted for burst~B4 from FRB~20200120E, based on predictions from the relativistic shock model\,\cite{Metzger+2019, Margalit+2020a} (Methods and Supplementary Information). The \text{X-ray} pseudo-fluences were calculated using a distance of $d$\,$=$\,3.63\,$\pm$\,0.34\,\cite{Freedman+1994} to FRB~20200120E.}
	\label{fig:xray_transient_histogram}
	
\end{figure*}

The \text{X-ray} \text{pseudo-fluence} distributions of magnetar giant flares, magnetar intermediate flares, magnetar short bursts, and type~I \text{X-ray} bursts from Galactic~LMXBs are shown in Fig.~\ref{fig:xray_transient_histogram}. We also show \text{X-ray} \text{pseudo-fluence} values for GRBs from extragalactic magnetar  giant flares, ULXBs from an unknown class of astrophysical sources in extragalactic globular clusters, bright type~II \text{X-ray} bursts from Galactic~LMXBs, and predictions from the relativistic shock model in refs.~\citen{Metzger+2019} and~\citen{Margalit+2020a} (Supplementary Information). The \text{X-ray} \text{pseudo-fluence} value for each flare or burst was determined by correcting the observed or predicted \text{X-ray} fluence for the distance to FRB~20200120E, the expected hydrogen column density towards FRB~20200120E ($N_{\text{H}}$\,$=$\,6.73\,$\times$\,10$^{\text{20}}$\,cm$^{\text{--2}}$\,\cite{HI4PI+2016}), and the \text{best-fit} spectral model reported for each event. We calculated \text{X-ray} \text{pseudo-fluences} in the \text{0.5--10}\,keV energy band using the \texttt{XSPEC} spectral fitting software package\,\cite{Arnaud1996}.

The absorbed \text{0.5--10}\,keV \text{X-ray} \text{pseudo-fluences} of magnetar flares and magnetar bursts were determined using values reported in the literature for each event (for example, see refs.~\citen{Palmer+2005, Kozlova+2016, Mereghetti+2009}). The corresponding values for GRBs from extragalactic magnetar giant flares\,\cite{Crider2006, Ofek+2006, Frederiks+2007, Mazets+2008, Burns+2021, Svinkin+2021}, ULXBs\,\cite{Irwin+2016, Jonker+2013, Sivakoff+2005}, and bright type~II \text{X-ray} bursts\,\cite{Younes+2015, Bagnoli+2015, Moon+2003} were obtained similarly. The absorbed \text{X-ray} \text{pseudo-fluence} distribution of type~I \text{X-ray} bursts was derived from the type~I \text{X-ray} burst Multi-INstrument Burst ARchive~(MINBAR) catalogue\,\cite{Galloway+2020}, using the \text{best-fit} blackbody temperature ($kT$) at burst peak.


\bigskip

\section*{Predictions from Relativistic Shock Model}
\label{sec:shock_model_predictions}

Following the formalism described in ref.~\citen{Margalit+2020a}, we calculate the predicted Lorentz factor of the shocked gas~($\Gamma_{\text{shock}}$), shock radius~($R_{\text{shock}}$), upstream external density at the shock radius~($n_{\text{ext}}(R_{\text{shock}})$), and energy of the relativistic flare~($E_{\text{flare}}$) using the measured duration ($w_{\text{t}}$) and energy ($E_{\text{R}}$) of burst~B4 from FRB~20200120E (Extended Data Table~\ref{tab:radio_burst_properties}) as follows:
\begin{equation}
	\small
	\Gamma_{\text{shock}}\approx51\left(\frac{m_{*}}{m_{e}}\right)^{1/30}\left(\frac{f_{e}}{0.5}\right)^{1/15}\left(\frac{f_{\xi}}{10^{-3}}\right)^{-1/15}\left(\frac{\nu_{\text{centre}}}{1382\,\text{MHz}}\right)^{-7/30}\left(\frac{w_{\text{t}}}{117\,\text{\ensuremath{\mu}s}}\right)^{-2/5}\left(\frac{E_{\text{R}}}{2.8\times10^{33}\,\text{erg}}\right)^{1/6}
	\label{eqn:lorentz_factor_shock_model}
\end{equation}
\begin{equation}
	\resizebox{0.948\textwidth}{!}{
		$R_{\text{shock}}\approx2\Gamma^{2}cw_{\text{t}}\approx1.8\times10^{10}\,\text{cm}\left(\frac{m_{*}}{m_{e}}\right)^{1/15}\left(\frac{f_{e}}{0.5}\right)^{2/15}\left(\frac{f_{\xi}}{10^{-3}}\right)^{-2/15}\left(\frac{\nu_{\text{centre}}}{1382\,\text{MHz}}\right)^{-7/15}\left(\frac{w_{\text{t}}}{117\,\text{\ensuremath{\mu}s}}\right)^{1/5}\left(\frac{E_{\text{R}}}{2.8\times10^{33}\,\text{erg}}\right)^{1/3}$
	}
	\label{eqn:shock_radius_shock_model}
\end{equation}
\begin{equation}
	\resizebox{0.948\textwidth}{!}{
		$n_{\text{ext}}(R_{\text{shock}})\approx9.0\times10^{4}\,\text{cm}^{-3}\left(\frac{m_{*}}{m_{e}}\right)^{2/15}\left(\frac{f_{e}}{0.5}\right)^{-11/15}\left(\frac{f_{\xi}}{10^{-3}}\right)^{-4/15}\left(\frac{\nu_{\text{centre}}}{1382\,\text{MHz}}\right)^{31/15}\left(\frac{w_{\text{t}}}{117\,\text{\ensuremath{\mu}s}}\right)^{2/5}\left(\frac{E_{\text{R}}}{2.8\times10^{33}\,\text{erg}}\right)^{-1/3}$
	}
	\label{eqn:external_density_shock_model}
\end{equation}
\begin{equation}
	\resizebox{0.936\textwidth}{!}{
		$E_{\text{flare}}\approx5.3\times10^{37}\,\text{erg}\left(\frac{m_{*}}{m_{e}}\right)^{2/5}\left(\frac{f_{e}}{0.5}\right)^{-1/5}\left(\frac{f_{\xi}}{10^{-3}}\right)^{-4/5}\left(\frac{\nu_{\text{centre}}}{1382\,\text{MHz}}\right)^{1/5}\left(\frac{w_{\text{t}}}{117\,\text{\ensuremath{\mu}s}}\right)^{1/5}\left(\frac{E_{\text{R}}}{2.8\times10^{33}\,\text{erg}}\right)$
	}
	\label{eqn:flare_energy_shock_model}
\end{equation}
From equation~(\ref{eqn:flare_energy_shock_model}), the ratio of the relativistic flare energy to radio burst energy is given by
\begin{equation}
	\resizebox{0.936\textwidth}{!}{
		$\eta_{\text{shock}}=\frac{E_{\text{flare}}}{E_{\text{R}}}=\frac{\mathcal{F}_{\text{flare}}}{\mathcal{F}_{\text{R}}}\approx1.9\times10^{4}\left(\frac{m_{*}}{m_{e}}\right)^{2/5}\left(\frac{f_{e}}{0.5}\right)^{-1/5}\left(\frac{f_{\xi}}{10^{-3}}\right)^{-4/5}\left(\frac{\nu_{\text{centre}}}{1382\,\text{MHz}}\right)^{1/5}\left(\frac{w_{\text{t}}}{117\,\text{\ensuremath{\mu}s}}\right)^{1/5}$,
	}
	\label{eqn:fluence_ratio_shock_model}
\end{equation}
which is also valid for the \text{X-ray} flare fluence to radio fluence ratio. Here, $m_{*}$ is the particle mass in the upstream medium, $m_{e}$ is the electron mass (we assume $m_{*}$\,$=$\,$m_{e}$, which is valid for a pair plasma medium), $f_{e}$\,$\equiv$\,$n_{e}/n_{\text{ext}}$ is the ratio of the electron density to ion density in the upstream medium (we consider $f_{e}$\,$=$\,0.5 for an electron--ion medium), $f_{\xi}$ is the efficiency of the maser (normalized to a characteristic value of $f_{\xi}$\,$=$\,10$^{\text{--3}}$), and $\nu_{\text{centre}}$ is the central frequency of the radio burst. We selected burst~B4 for these calculations, instead of the other radio bursts in Extended Data Table~\ref{tab:radio_burst_properties}, since it was the most energetic burst detected from FRB~20200120E, had a well-measured duration, and was covered simultaneously in the soft \text{X-ray} band with \textit{NICER}.


\clearpage


\newpage

\begin{addendum}


\item[Data Availability]

The \textit{NICER}, \textit{XMM-Newton}, \textit{Chandra}, and \textit{NuSTAR} data can be accessed from the publicly available HEASARC~archive. The data containing radio bursts~\text{B1--B9} from FRB~20200120E are available via Zenodo at \href{https://doi.org/10.5281/zenodo.13359005}{https://doi.org/10.5281/zenodo.13359005}\,\cite{Pearlman+2024b}.


\item[Code Availability]

The following software packages were used to analyse the data presented in this paper: \break \texttt{astropy}\,\cite{Astropy+2013}, \texttt{CIAO}\,\cite{Fruscione+2006}, \texttt{dspsr}\,\cite{vanStraten+2011}, \texttt{emcee}\,\cite{ForemanMackey+2013}, \texttt{FETCH}\,\cite{Agarwal+2020}, \texttt{FTOOLS}\,\cite{Blackburn+1999}, \texttt{HEASOFT}\,\cite{Heasoft+2014}, \texttt{Heimdall}\,\cite{Barsdell+2012}, \break \texttt{PRESTO}\,\cite{Ransom2001}, \texttt{SFXC}\,\cite{Keimpema+2015}, \texttt{XMM-SAS}\,\cite{Gabriel+2004}, and \texttt{XSPEC}\,\cite{Arnaud1996}. The codes used for data processing and producing the figures may be made available from the corresponding author upon reasonable request.


\item[Acknowledgments]

We are very grateful to B.~Margalit for valuable and insightful discussions. We thank Z.~Arzoumanian, K.~Gendreau, and E.~Ferrara for prompt scheduling of these \textit{NICER} observations. We also thank M.~Snelders and D.~Hewitt for their help with scheduling PRECISE/EVN observations.

A.B.P. is a Banting Fellow, a McGill Space Institute~(MSI) Fellow, and a Fonds de Recherche du Qu\'ebec -- Nature et Technologies~(FRQNT) postdoctoral fellow. V.M.K. holds the Lorne Trottier Chair in Astrophysics and Cosmology, a Distinguished James McGill Professorship, and receives support from an Natural Sciences and Engineering Research Council of Canada~(NSERC) Discovery grant (RGPIN 228738-13), from an R.~Howard Webster Foundation Fellowship from the Canadian Institute for Advanced Research~(CIFAR), and from the FRQNT CRAQ. K.N. is an MIT Kavli Fellow. L.G.S. is a Lise-Meitner Max Planck independent group leader and acknowledges funding from the Max Planck Society. M.B. is a McWilliams fellow and International Astronomical Union Gruber fellow. M.B. receives support from the McWilliams seed grant. S.C. acknowledges support provided by NASA through grant HST-GO-16664 from the Space Telescope Science Institute, which is operated by the Association of Universities for Research in Astronomy, Inc., under NASA contract NAS5-26555. A.M.C. is funded by an NSERC Doctoral Postgraduate Scholarship. A.P.C. is a Vanier Canada Graduate Scholar. F.A.D. is supported by the UBC Four Year Fellowship. T.E. is supported by NASA through the NASA Hubble Fellowship grant HST-HF2-51504.001-A awarded by the Space Telescope Science Institute, which is operated by the Association of Universities for Research in Astronomy, Inc., for NASA, under contract NAS5-26555. B.M.G. acknowledges the support of NSERC, through grant RGPIN-2022-03163, and support from the Canada Research Chairs programme. T.G. is supported by the Turkish Republic, Presidency of Strategy and Budget project, 2016K121370. C.L. is supported by NASA through the NASA Hubble Fellowship grant \text{HST-HF2-51536.001-A} awarded by the Space Telescope Science Institute, which is operated by the Association of Universities for Research in Astronomy, Inc., for NASA, under contract NAS5-26555. K.W.M. holds the Adam J. Burgasser Chair in Astrophysics and is supported by NSF grants (2008031 and 2018490). K.R.S. acknowledges support from a FRQNT doctoral fellowship. K.S. is supported by the NSF Graduate Research Fellowship Program. S.P.T. is a CIFAR Azrieli Global Scholar in the Gravity and Extreme Universe Program.

A.B.P. acknowledges partial support for this work through NASA Grants 80NSSC21K0215 and 80NSSC21K2028. The AstroFlash research group at McGill University, University of Amsterdam, ASTRON, and JIVE is supported by: a Canada Excellence Research Chair in Transient Astrophysics~(CERC-2022-00009); the European Research Council~(ERC) under the European Union's Horizon~2020 research and innovation programme~(`EuroFlash': grant agreement no.~101098079); and an NWO-Vici grant~(`AstroFlash': VI.C.192.045). Pulsar and FRB research at UBC is funded by an NSERC Discovery Grant and by CIFAR. The Dunlap Institute is funded through an endowment established by the David Dunlap family and the University of Toronto. We acknowledge that CHIME is located on the traditional, ancestral, and unceded territory of the Syilx/Okanagan people.


\item[Author Contributions]

A.B.P. is the lead author of the paper and wrote the majority of the manuscript. A.B.P. prepared the figures, with significant contributions from P.S. A.B.P. and P.S. led the astrophysical interpretation of the results from this study. A.B.P. coordinated the data analyses. A.B.P. led the analysis of the \textit{NICER} data of FRB~20200120E, with significant contributions from P.S. P.S. led the analysis of the \textit{XMM-Newton}, \textit{NuSTAR}, and \textit{Chandra} data of FRB~20200120E, with significant contributions from A.B.P., and wrote the corresponding sections of the paper. S.B. and L.G.S. searched the ToO Effelsberg data for radio bursts from FRB~20200120E. A.B.P., P.S., S.B., and L.G.S. analysed the properties of the radio bursts detected from FRB~20200120E during the ToO Effelsberg observations. P.S., L.G.S., and A.B.P. coordinated the simultaneous \text{X-ray} and radio observations with \textit{XMM-Newton} and Effelsberg. P.S. and L.G.S. coordinated the simultaneous \text{X-ray} and radio observations with \textit{NuSTAR}, \textit{Chandra}, and Effelsberg. A.B.P. and K.N. measured the properties and arrival times of the radio bursts detected from FRB~20200120E during the PRECISE observations. A.B.P. analysed the simultaneous \textit{NICER} and Effelsberg data obtained during the PRECISE observational campaign. A.B.P., F.K., and J.W.T.H. coordinated the \textit{NICER} and Effelsberg observations during the PRECISE campaign. A.B.P. and K.N. coordinated the \textit{NICER} observations during an independent radio monitoring campaign of FRB~20200120E with Effelsberg. A.B.P. performed the searches for radio bursts from FRB~20200120E in the CHIME/Pulsar radio data. A.B.P., C.M.T., and E.F. coordinated the radio observations of FRB~20200120E with CHIME/Pulsar. A.B.P. and B.W.M. performed the flux calibration of the CHIME/Pulsar data. V.M.K. played a significant leadership role that enabled these results. All other co-authors contributed comments that helped improve the manuscript.

A.B.P. is the Principal Investigator~(PI) of the Guest Observer~(GO) Cycle~2 and~3 \textit{NICER} programmes dedicated to performing simultaneous \text{X-ray} and radio observations of nearby repeating FRBs. P.S. is the PI of the \textit{XMM-Newton}, \textit{NuSTAR}, and \textit{Chandra} \text{X-ray} follow-up programmes of FRB~20200120E. P.S. is the PI of the ToO Effelsberg radio observing programme of nearby repeating FRB sources.


\item[Competing Interests]

The authors declare no competing interests.


\item[Correspondence and Requests for Materials]

Correspondence and requests for materials should be addressed to: Aaron B. Pearlman (\href{mailto:aaron.b.pearlman@physics.mcgill.ca}{aaron.b.pearlman@physics.mcgill.ca}).


\end{addendum}


\clearpage

\begin{table*}
	\caption{3$\sigma$ \text{X-ray} fluence and \text{X-ray} energy upper limits from FRB~20200120E at the times of bursts~B4 and~B9, derived from \textit{NICER} and \textit{XMM-Newton} observations.}
	
	\hspace{-1.5cm}
	\resizebox{1.15\textwidth}{!}{
		\begin{tabular}{cccccc}
			\toprule
			Timescale & $N_{\text{obs}}$$^{\mathrm{a}}$ & Absorbed \text{0.5--10}\,keV & Absorbed \text{0.5--10}\,keV & Absorbed \text{0.5--10}\,keV & Unabsorbed \text{0.5--10}\,keV \\
			($\Delta t$) & (Counts) & Fluence Limit & Fluence Limit & Fluence Limit & Energy Limit$^{\mathrm{e}}$ \\
			& & (Model~1$^{\mathrm{b}}$: & (Model~2$^{\mathrm{c}}$: & (Model~3$^{\mathrm{d}}$: & ($E_{\text{X}}$, 10$^{\text{41}}$\,erg) \\
			& & $\Gamma$\,$=$\,1.5, $E_{\text{cut}}$\,$=$\,85\,keV) & $kT$\,$=$\,10\,keV) & $\Gamma$\,$=$\,--0.5, $E_{\text{cut}}$\,$=$\,500\,keV) & \\
			& & ($\mathcal{F}_{\text{X}}$, 10$^{\text{--10}}$\,erg\,cm$^{\text{--2}}$) & ($\mathcal{F}_{\text{X}}$, 10$^{\text{--10}}$\,erg\,cm$^{\text{--2}}$) & ($\mathcal{F}_{\text{X}}$, 10$^{\text{--10}}$\,erg\,cm$^{\text{--2}}$) & \\
			\hline
			\multicolumn{6}{c}{\textit{NICER} Upper Limits at the Time of Burst~B4} \\
			\hline
			10\,s & 7 & $<$\,0.4 & $<$\,3 & $<$\,3 & $<$\,0.5--5 \\
			1\,s & 1 & $<$\,0.3 & $<$\,2 & $<$\,2 & $<$\,0.3--3 \\
			\text{100\,ns--100\,ms}$^{\mathrm{f}}$ & 0 & $<$\,0.2 & $<$\,1 & $<$\,1 & $<$\,0.3--2 \\
			\hline
			\multicolumn{6}{c}{\textit{XMM-Newton} Upper Limits at the Time of Burst~B9} \\
			\hline
			100\,s & 1 & $<$\,0.4 & $<$\,1 & $<$\,1 & $<$\,0.6--2 \\
			\text{100\,ms--10\,s}$^{\mathrm{g}}$ & 0 & $<$\,0.3 & $<$\,1 & $<$\,1 & $<$\,0.5--2 \\
			\bottomrule
			\multicolumn{6}{l}{\text{X-ray} fluence and \text{X-ray} energy upper limits are derived from Poisson 3$\sigma$ (99.73\%) confidence intervals, calculated using the} \\
			\multicolumn{6}{l}{Bayesian method described in ref.~\citen{Kraft+1991}. The background measurements in Extended Data Table~\ref{tab:nicer_xray_bkg} were used to obtain confidence} \\
			\multicolumn{6}{l}{intervals for \textit{NICER}.} \\
			\multicolumn{6}{l}{$^{\mathrm{a}}$ The number of photons detected with \textit{NICER} or \textit{XMM-Newton} in the \text{0.5--10}\,keV energy band within $\pm$$\Delta t / 2$ of the barycentric,} \\
			\multicolumn{6}{l}{infinite frequency peak time of burst~B4 ($t_{\text{peak}}^{\text{B4}}$) or burst~B9 ($t_{\text{peak}}^{\text{B9}}$). The number of photons reported for \textit{XMM-Newton} are within the} \\
			\multicolumn{6}{l}{PSF of the EPIC/pn camera, which was centred on FRB~20200120E's position from VLBI\,\cite{Kirsten+2022}.} \\
			\multicolumn{6}{l}{$^{\mathrm{b}}$ Model~1 corresponds to an exponentially cut-off power-law spectrum, similar to the \text{X-ray} counterpart of the FRB-like radio burst} \\
			\multicolumn{6}{l}{from SGR~1935+2154\,\cite{Li+2021, Mereghetti+2020a, Ridnaia+2021, Tavani+2021}.} \\
			\multicolumn{6}{l}{$^{\mathrm{c}}$ Model~2 corresponds to a typical blackbody spectrum observed from magnetar hard \text{X-ray} bursts\,\cite{An+2014, CotiZelati+2021}.} \\
			\multicolumn{6}{l}{$^{\mathrm{d}}$ Model~3 corresponds to an exponentially cut-off power-law spectrum, similiar to the spectrum measured from the SGR~1806--20} \\
			\multicolumn{6}{l}{giant flare\,\cite{Palmer+2005, Hurley+2005}.} \\
			\multicolumn{6}{l}{$^{\mathrm{e}}$ The unabsorbed energy limits were calculated assuming a distance of 3.63\,$\pm$\,0.34\,Mpc\,\cite{Freedman+1994} to FRB~20200120E. The range} \\
			\multicolumn{6}{l}{indicates the minimum and maximum values obtained using the unabsorbed fluence limits from models~1, 2, and 3.} \\
			\multicolumn{6}{l}{$^{\mathrm{f}}$ The absorbed fluence and unabsorbed energy limits were calculated on timescales of 100\,ns, 1\,$\mu$s, 10\,$\mu$s, 100\,$\mu$s, 1\,ms, 10\,ms, and} \\
			\multicolumn{6}{l}{100\,ms. Identical values were obtained for these timescales.} \\
			\multicolumn{6}{l}{$^{\mathrm{g}}$ The absorbed fluence and unabsorbed energy limits were calculated on timescales of 100\,ms, 1\,s, and 10\,s. Identical values were} \\
			\multicolumn{6}{l}{obtained for these timescales.} \\
	\end{tabular}}
	
	\label{tab:nicer_xmm_b4_b9_xray_limits}
\end{table*}


\clearpage

\setlength{\footskip}{40pt}

\clearpage

\setlength{\footskip}{53.5pt}

\clearpage

\setlength{\footskip}{30pt}

\begin{figure*}
	
	\hspace*{-1.7cm}
	\includegraphics[trim=0cm 0cm 0cm 0cm, clip=false, scale=0.75, angle=0]{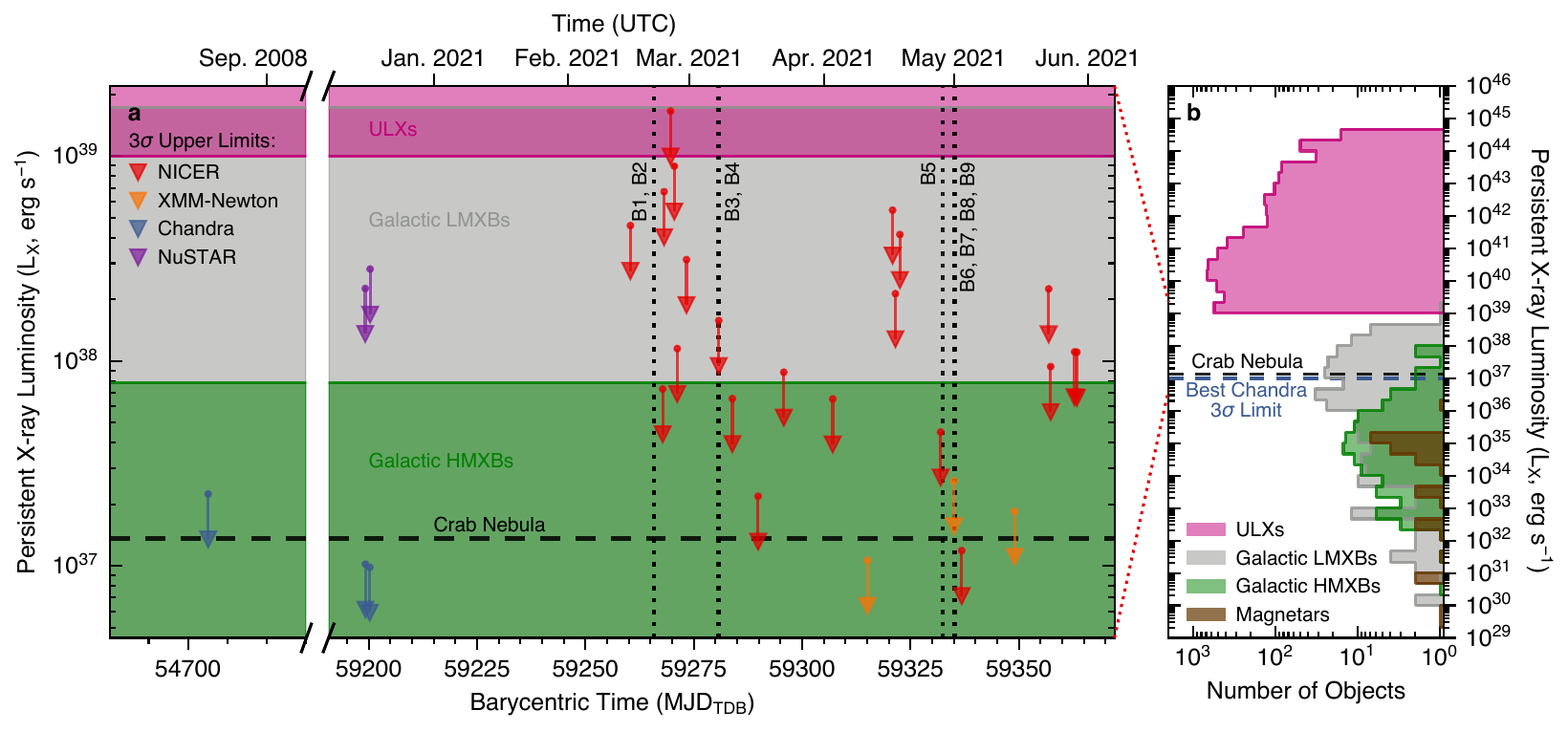}
	
	\caption{\textbf{Persistent \text{X-ray} luminosity limits from observations of FRB~20200120E, compared with the persistent \text{X-ray} luminosities of ULX~sources, Galactic~LMXBs, Galactic~HMXBs, and magnetars.} Panel~\textbf{a}: 3$\sigma$ persistent isotropic-equivalent \text{X-ray} luminosity limits from \textit{NICER}~(red), \textit{XMM-Newton}~(orange), \textit{Chandra}~(blue), and \textit{NuSTAR}~(purple) observations of FRB~20200120E. These \text{X-ray} luminosity limits are calculated using the 3$\sigma$ persistent \text{X-ray} flux upper limits listed in Extended Data Table~\ref{tab:xray_obs} and the distance to FRB~20200120E\,\cite{Freedman+1994}. The luminosity limits from \textit{NICER}, \textit{XMM-Newton}, and \textit{Chandra} observations are derived in the \text{0.5--10}\,keV energy band, and the luminosity limits from \textit{NuSTAR} observations are derived in the 3--79\,keV energy band. The vertical dotted black lines indicate the barycentric arrival times of the radio bursts shown in Fig.~\ref{fig:radio_bursts} (Extended Data Table~\ref{tab:radio_burst_properties}). The horizontal dashed black line corresponds to the \text{X-ray} luminosity of the Crab Nebula. The shaded pink, grey, and green regions indicate the range of \text{X-ray} luminosities of ULX sources, Galactic LMXBs, and Galactic HMXBs spanning the persistent \text{X-ray} luminosity limits from FRB~20200120E. Panel~\textbf{b}: Histograms showing the persistent \text{X-ray} luminosity distributions of ULX sources~(pink), Galactic LMXBs~(grey), Galactic HMXBs~(green), and magnetars~(brown). The dashed blue line shows the most constraining 3$\sigma$ persistent \text{X-ray} luminosity limit from a \textit{Chandra} observation of FRB~20200120E on MJD~59200 (17~December~2020). The dotted red lines indicate the overlapping range of persistent \text{X-ray} luminosities from panel~\textbf{b} that are shown in panel~\textbf{a}.}
	\label{fig:xray_luminosity_upper_limits}
	
\end{figure*}


\clearpage

\clearpage

\textbf{References} \\

\bigskip

\bibliographystyle{naturemag}
\bibliography{references}


\newpage

\section*{Supplementary Information}
\label{sec:supplementary_information}


\bigskip

\subsection{Gamma-ray Bursts~(GRBs) from Extragalactic Magnetar Giant Flares}
\label{sec:grbs}

Over 8,000~GRBs have been detected so far using a variety of high-energy instruments (e.g., see the online GRBweb catalogue\,\citesupplementary{GRBWeb}). Several of these GRBs (GRB~970110\,\cite{Crider2006}, GRB~051103\,\cite{Ofek+2006, Frederiks+2007}, GRB~070201\,\cite{Mazets+2008}, GRB~070222\,\cite{Burns+2021}, and GRB~200415A\,\cite{Svinkin+2021}) have been localized to host galaxies in the local Universe, and they are believed to be associated with giant flares from extragalactic magnetars based on their temporal characteristics, energetics, and spectral properties. The absorbed \text{X-ray} pseudo-fluences of GRB~970110, GRB~051103, GRB~070201, and GRB~200415A (Fig.~\ref{fig:xray_transient_histogram}) are all several orders of magnitude larger than our 3$\sigma$ \text{X-ray} fluence limits from FRB~20200120E, measured using \textit{NICER} and \textit{XMM-Newton} at the times of bursts~B4 and~B9 (Table~\ref{tab:nicer_xmm_b4_b9_xray_limits}). Based on this, we conclude that a soft \text{X-ray} burst, with similar properties to these extragalactic magnetar flares, was not produced near the times of these radio bursts from FRB~20200120E. We are unable to place strong constraints on the possibility of a GRB~070222-like flare being emitted from FRB~20200120E, since the absorbed \text{X-ray} pseudo-fluence of GRB~070222 is slightly (a factor of $\sim$2) below our 3$\sigma$ \text{X-ray} fluence limits.

Additionally, we note that the position of GRB~051103 was triangulated to the M81/M82 galaxy group using the Interplanetary Network~(IPN)\,\citesupplementary{Hurley+2010b}. The position of GRB~051103 is significantly offset from the VLBI position of FRB~20200120E by $\sim$0.6$^{\circ}$. Based on the reported localization regions, we conclude that GRB~051103 and FRB~20200120E are not associated (at a confidence level greater than 418$\sigma$).


\bigskip

\subsection{\text{X-ray} Pseudo-Fluence Distribution of FRB-Emitting Magnetars}
\label{sec:frb_emitting_magnetar_fluences}

Ordinary magnetars, with similar energetics and activity levels as SGR~1935+2154, are believed to be unable to account for the repetition rates observed from the cosmological repeating FRB population\,\citesupplementary{Margalit+2020b}. If repeating FRB sources are powered by magnetars, then they must be younger than typical \text{1--10-kyr-old} magnetars found in the Milky Way, possess larger magnetic fields, and have larger energy reservoirs\,\cite{Margalit+2019}\textsuperscript{,}\citesupplementary{Margalit+2020b}. An FRB-emitting magnetar may therefore be capable of powering more energetic high-energy bursts than ordinary Galactic magnetars\,\citesupplementary{Margalit+2020b}. In this case, the distribution of magnetar burst \text{X-ray} pseudo-fluences shown in Fig.~\ref{fig:xray_transient_histogram} would be shifted towards larger values for such FRB-emitting magnetars by an amount proportional to the square of its magnetic field. Our prompt \text{X-ray} fluence limits would then reside in the bulk of the resulting \text{X-ray} pseudo-fluence distribution, indicating that it would be possible to detect many short bursts from these highly energetic, FRB-emitting magnetars in the soft \text{X-ray} band with \textit{NICER}, \textit{XMM-Newton}, and possibly other \text{X-ray} detectors. Although the repetition rate of FRB~20200120E is comparable to that of other active repeating FRBs\,\cite{Nimmo+2023}, the energies of radio bursts detected from FRB~20200120E (Table~\ref{tab:radio_burst_properties}) are several orders of magnitude lower than those from most active repeating FRBs\,\cite{Majid+2021, Nimmo+2022}, such as FRB~20121102A and FRB~20180916B. This suggests that, if FRB~20200120E is a magnetar, it could be powered by a source with a lower energy reservoir than other repeating FRBs.


\bigskip

\subsection{\text{X-ray} Emission from Giant Radio Pulse-Emitting Pulsars}
\label{sec:xray_emission_gp_pulsars}

Enhanced \text{X-ray} emission has been detected during giant radio pulses from some Galactic pulsars, such as PSR~B1937+21 and the Crab pulsar. \text{X-ray} observations of PSR~B1937+21 have revealed that its \text{X-ray} pulse profile peak is closely aligned with its radio giant pulses, rather than the peak of its normal radio pulse profile\,\citesupplementary{Cusumano+2003}. This provides strong support for a common origin between giant pulses from radio MSPs and their pulsed \text{X-ray} emission. Our \text{X-ray} observations were not sensitive to detecting pulsed \text{X-ray} emission from a \text{PSR~B1937+21-like} object at the distance of FRB~20200120E. The 3$\sigma$ persistent \text{X-ray} flux upper limit from our deepest \textit{Chandra} observation of FRB~20200120E (Extended Data Table~\ref{tab:xray_obs}) was $\sim$10$^{\text{3}}$ times larger than the expected pulsed \text{X-ray} flux of PSR~B1937+21 ($F_{\text{X, 3.63\,Mpc}}^{\text{B1937+21}}$\,$=$\,4\,$\times$10$^{\text{--18}}$\,erg\,s$^{\text{--1}}$\,cm$^{\text{--2}}$\,\citesupplementary{Takahashi+2001}), if it were placed at the distance of FRB~20200120E. An \text{X-ray} enhancement of 3.8\,$\pm$\,0.7\% has also been observed in the Crab pulsar's time-averaged pulsed \text{X-ray} emission, coinciding with giant radio pulses\,\citesupplementary{Enoto+2021a}, but we cannot rule out a similar \text{X-ray} enhancement from FRB~20200120E due to limited sensitivity during our \text{X-ray} observations.


\bigskip

\subsection{Low-mass \text{X-ray} Binaries~(LMXBs)}
\label{sec:lmxbs}

\text{X-ray} observations have revealed large numbers of LMXBs in the globular clusters of nearby galaxies\,\citesupplementary{Fabbiano2006}. Many LMXBs likely also reside in FRB~20200120E's globular cluster. The distribution of persistent \text{X-ray} luminosities from Galactic LMXBs shown in Fig.~\ref{fig:xray_luminosity_upper_limits}b indicates that our \text{X-ray} observations would have been sensitive to detecting persistent or transient \text{X-ray} emission from up to $\sim$30\% of the brightest, Milky~Way-like LMXBs\,\cite{Avakyan+2023} from FRB~20200120E's location. Based on our deepest \text{X-ray} observation with \textit{Chandra}, we rule out an association between FRB~20200120E and LMXBs with persistent \text{X-ray} luminosities larger than $L_{\text{X}}$\,$>$\,9.8\,$\times$\,10$^{\text{36}}$\,erg\,s$^{\text{--1}}$.

There are two types of \text{X-ray} bursts produced by LMXBs: type~I and type~II \text{X-ray} bursts. Type~I \text{X-ray} bursts are thermonuclear explosions that are triggered by unstable ignition of accreted hydrogen and/or helium on neutron stars~(NSs) in LMXBs\,\citesupplementary{Galloway+2008}. The typical durations of type~I \text{X-ray} bursts range from seconds to minutes, and very bright type~I \text{X-ray} bursts can reach Eddington luminosities at the NS surface. The absorbed \text{X-ray} pseudo-fluence distribution of type~I \text{X-ray} bursts (Fig.~\ref{fig:xray_transient_histogram}) shows that a small fraction (0.1--0.2\%) of type~I \text{X-ray} bursts have \text{X-ray} pseudo-fluences that are above our 3$\sigma$ \text{X-ray} fluence limits, derived from \textit{NICER} and \textit{XMM-Newton} observations of FRB~20200120E at the times of bursts~B4 and~B9 (Table~\ref{tab:nicer_xmm_b4_b9_xray_limits}). Thus, type~I \text{X-ray} bursts with similar or larger \text{X-ray} \text{pseudo-fluences} than our \text{X-ray} fluence limits would have been detectable from FRB~20200120E near the times of these radio bursts.

Type~II \text{X-ray} bursts are caused by accretion instabilities during mass transfer onto NSs in binary systems, and they are thought to originate from the sudden release of gravitational potential energy during the accretion process. Type~II \text{X-ray} bursts have been detected from only a few LMXB sources\,\cite{Younes+2015, Bagnoli+2015}. We find that the most energetic type~II \text{X-ray} bursts from MXB~1730--335 have absorbed \text{X-ray} pseudo-fluences that are slightly above our best 3$\sigma$ \text{X-ray} fluence limit, derived at the time of burst~B4 from \textit{NICER} observations of FRB~20200120E (Fig.~\ref{fig:xray_transient_histogram} and Table~\ref{tab:nicer_xmm_b4_b9_xray_limits}). Thus, \text{X-ray} bursts similar to these bright type~II \text{X-ray} bursts would have been detectable from FRB~20200120E during our \textit{NICER} observations. However, the absorbed \text{X-ray} pseudo-fluences of type~II \text{X-ray} bursts from other LMXB sources, such as GRO~J1744--28, were below the \textit{NICER} and \textit{XMM-Newton} \text{X-ray} fluence detection thresholds. Therefore, an LMXB capable of producing type~I or type~II \text{X-ray} bursts cannot be entirely excluded as a possible source type for FRB~20200120E.


\bigskip

\subsection{High-mass \text{X-ray} Binaries~(HMXBs)}
\label{sec:hmxbs}

Although FRB models involving HMXBs\,\citesupplementary{Lyutikov+2020a, Ioka+2020} have been invoked to explain the 16\,day periodic radio activity\,\citesupplementary{CHIME+2020b} observed from the repeating FRB~20180916B and its positional offset from the nearest region of active star formation in its host galaxy\,\citesupplementary{Tendulkar+2021}, it is unlikely that FRB~20200120E is powered by an active HMXB. The massive stellar~OB companions of HMXBs have lifespans less than $\sim$1--10\,Myr, which is significantly shorter than the age ($\sim$10\,Gyr) of FRB~20200120E's globular cluster\,\cite{Kirsten+2022}. Approximately 5\% of Galactic HMXBs have measured \text{X-ray} luminosities above 10$^{\text{37}}$\,erg\,s$^{\text{--1}}$ in the soft \text{X-ray} band\,\cite{Neumann+2023}. Only a few of our deepest \text{X-ray} observations would have been sensitive to detecting persistent or transient \text{X-ray} emission from a bright, \text{Milky Way-like} HMXB at the high-end of the Galactic luminosity distribution (Fig.~\ref{fig:xray_luminosity_upper_limits}b and Extended Data Table~\ref{tab:xray_obs}), if placed at the distance of FRB~20200120E.


\bigskip

\subsection{Relativistic Shock Models}
\label{sec:shock_models}

There are two popular classes of models involving NSs that are commonly invoked to explain the coherent radio emission observed from FRB sources: magnetospheric (pulsar-like) models\,\citesupplementary{Kumar+2017, Lu+2020a} and relativistic shock (GRB-like) models\,\cite{Metzger+2019, Margalit+2020a}\textsuperscript{,}\citesupplementary{Lyubarsky2014, Beloborodov2017}. In the relativistic shock model described in refs.~\citen{Metzger+2019} and~\citen{Margalit+2020a}, the emission from FRBs is thought to be produced by synchrotron maser emission from ultrarelativistic, magnetized shocks that are generated from the collision of flare ejecta (produced by a central engine) and an upstream medium at large distances. Here, we examine the theoretical predictions for multiwavelength emission from the relativistic shock model\,\cite{Metzger+2019, Margalit+2020a} (Methods) and compare these predictions to our results from simultaneous \text{X-ray} and radio observations of FRB~20200120E.

Our best prompt 3$\sigma$ \text{X-ray} fluence limit with \textit{NICER} at the time of burst~B4 constrains the relativistic flare energy from FRB~20200120E to be $E_{\text{flare}}$\,$<$\,3\,$\times$\,10$^{\text{40}}$\,erg in the \text{0.5--10}\,keV energy band. This limit is lower than the range of flare energies~(10$^{\text{41}}$--10$^{\text{48}}$\,erg) predicted in ref.~\citen{Margalit+2020a} for other repeating and non-repeating FRBs, based on the properties of their radio bursts. Larger flare energies are predicted for other FRB sources, using the relativistic shock model\,\cite{Metzger+2019, Margalit+2020a}, because their radio bursts are typically longer in duration and have larger energies than those observed from FRB~20200120E (equation~(\ref{eqn:flare_energy_shock_model})).

From equation~(\ref{eqn:flare_energy_shock_model}), we find that a relativistic flare energy of $E_{\text{flare}}$\,$\approx$\,5.3\,$\times$\,10$^{\text{37}}$\,erg is predicted near the time of burst~B4. At the distance of FRB~20200120E, the corresponding \text{X-ray} fluence for this flare would be approximately 3.4\,$\times$\,10$^{\text{--14}}$\,erg\,cm$^{\text{--2}}$, which is below our 3$\sigma$ \text{X-ray} fluence limits (Fig.~\ref{fig:xray_transient_histogram} and Table~\ref{tab:nicer_xmm_b4_b9_xray_limits}). Using equation~(\ref{eqn:fluence_ratio_shock_model}) and our best 3$\sigma$ \text{X-ray} fluence limit for burst~B4 in Table~\ref{tab:nicer_xmm_b4_b9_xray_limits}, we place the following constraint on the ratio of the relativistic flare energy to radio burst energy for FRB~20200120E in the \text{0.5--10}\,keV energy band: ($\eta_{\text{shock}}$\,$\approx$\,1.9\,$\times$\,10$^{\text{4}}$)\,$<$\,(1.1\,$\times$\,10$^{\text{7}}$\,$=$\,$\eta^{\text{B4}}_{\text{lim}}$), where $\eta^{\text{B4}}_{\text{lim}}$ is our most constraining upper limit on the \text{X-ray}-to-radio fluence ratio of FRB~20200120E from simultaneous \text{X-ray} and radio measurements at the time of burst~B4. The predicted \text{X-ray} flare to radio burst fluence ratio ($\eta_{\text{shock}}$\,$\approx$\,1.9\,$\times$\,10$^{\text{4}}$) obtained for FRB~20200120E from the relativistic shock model is \text{2--3} times smaller than the \text{X-ray}-to-radio fluence ratio ($\eta_{\text{SGR~1935+2154}}$\,$\approx$\,4.6\,$\times$\,10$^{\text{4}}$) derived from the FRB-like burst detected from SGR~1935+2154 on 28~April~2020, based on the reported \text{X-ray} and radio fluences from \textit{Insight-HXMT}\,\cite{Li+2021}, STARE2\,\cite{Bochenek+2020c}, and CHIME/FRB\,\cite{CHIME+2020a}. To probe the fluence ratio predicted by equation~(\ref{eqn:fluence_ratio_shock_model}) for a relativistic shock from FRB~20200120E, using \textit{NICER} in the \text{0.5--10}\,keV energy band, a more energetic radio burst ($E_{\text{R}}$\,$\gtrsim$\,580\,$\times$\,$E_{\text{R}}^{\text{B4}}$) would be required for radio bursts with widths comparable to~B4.

However, the ultra-short-timescale variability ($\delta t$\,$\lesssim$\,100\,ns) previously reported in some radio bursts detected between 1.2 and 2.3\,GHz from FRB~20200120E\,\cite{Majid+2021, Nimmo+2022} disfavours models that require the FRB emission to be produced by relativistic shocks located at large distances ($\gtrsim$\,10$^{\text{10}}$\,cm) from a compact object\,\citesupplementary{Beniamini+2020b, Lu+2022} and instead suggests that the emission may be produced within the magnetosphere of a NS.


\bigskip

\subsection{Comparisons between FRB~20200120E and Other Repeating FRBs}
\label{sec:frb_comparison}

There are remarkable differences between the observed properties of FRB~20200120E and other repeating FRBs. The isotropic-equivalent luminosities of radio bursts detected from FRB~20200120E are, on average, several orders of magnitude fainter than the luminosities of radio bursts from other repeating FRBs\,\cite{Majid+2021, Nimmo+2022}. Some repeating FRB sources, such as FRB~20121102A\,\citesupplementary{Chatterjee+2017, Bassa+2017, Kokubo+2017} and FRB~20180916B\,\citesupplementary{Marcote+2020, Tendulkar+2021}, have been localized to star-forming galaxies and spatially associated with nearby knots of star formation within their host galaxies. This is consistent with expectations from FRB models involving young pulsars or magnetars formed through core-collapse supernovae\,\cite{Kulkarni+2014, Lyutikov+2016}\textsuperscript{,}\citesupplementary{Metzger+2017}. On the other hand, it is unlikely that a massive star has recently undergone a core-collapse supernova in the $\sim$10-Gyr-old globular cluster associated with FRB~20200120E, which provides strong evidence for an alternative, delayed formation channel for FRB~20200120E. Together, this suggests that FRB~20200120E may be an atypical source among the repeating FRB population.

The proximity of FRB~20180916B (luminosity distance of $d_{\text{L}}$\,$=$\,149\,Mpc)\,\citesupplementary{Marcote+2020}, along with its 16\,day radio activity period\,\citesupplementary{CHIME+2020b} and precise localization\,\citesupplementary{Marcote+2020}, has enabled targeted multiwavelength searches for emission outside of the radio band\,\cite{Pilia+2020, Scholz+2020a, Tavani+2021}\textsuperscript{,}\citesupplementary{Tavani+2020, Trudu+2023}. The prompt, high-energy upper limits from observations of FRB~20180916B have yielded constraints that challenge models invoking giant magnetar flares\,\cite{Pilia+2020, Scholz+2020a, Tavani+2021}. However, previous \text{X-ray} observations of FRB~20180916B have lacked the sensitivity to place constraints on the emission of magnetar-like intermediate flares or short \text{X-ray} bursts. On the other hand, deep optical limits at the time of an energetic radio burst from FRB~20180916B yielded an upper limit of $\sim$10$^{\text{2}}$ on the optical-to-radio fluence ratio, ruling out strong optical emission from a blast wave in the wind of a hyperactive magnetar~\citesupplementary{Beloborodov+2020}. FRB models involving giant magnetar flares are also ruled out for FRB~20200120E, based on our simultaneous \text{X-ray} and radio observations, and GRB-like shocks are disfavored based on the temporal variability observed in previously reported radio bursts from FRB~20200120E\,\cite{Majid+2021, Nimmo+2022}\textsuperscript{,}\citesupplementary{Beniamini+2020b, Lu+2022}.

Simultaneous \text{X-ray} and radio observations of more distant repeating FRB sources, such as \break FRB~20201124A\,\citesupplementary{Piro+2021} (luminosity distance of $d_{\text{L}}$\,$=$\,453\,Mpc) and FRB~20121102A\,\cite{Scholz+2017a} (luminosity distance of $d_{\text{L}}$\,$=$\,972\,Mpc), have provided upper limits that are consistent with a flaring magnetar origin. These upper limits also disfavor a link between extremely energetic GRB-like events, with \text{X-ray} energies larger than 10$^{\text{45}}$--10$^{\text{47}}$\,erg, and FRB~20201124A\,\citesupplementary{Piro+2021} or FRB~20121102A\,\cite{Scholz+2017a}. It has not yet been possible to rule out less energetic high-energy transients from these repeating FRB sources due to the limited sensitivity of currently operating telescopes. Since the origins of FRB~20200120E and other repeating FRBs may be different, the constraints derived from observations of other repeating FRBs\,\citesupplementary{Cook+2024} provide complementary insight into the nature of the sources comprising the repeating FRB population.


\bigskip

\subsection{Future Searches for \text{X-ray} Emission from FRBs}
\label{sec:future_xray_counterparts}

We find that the prospects for detecting \text{X-ray} bursts from extragalactic FRB sources, located within \text{1--10}\,Mpc, are particularly promising in the soft \text{X-ray} band using both current and future \text{X-ray} telescopes. In Extended Data Fig.~\ref{fig:m81r_fluence_distance_constraints}, we show how the predicted \text{X-ray} burst fluences in the soft~(\text{0.5--10}\,keV) and hard~(\text{10--250}\,keV) \text{X-ray} bands vary with distance for different emission models and sources that are known to or could produce observable high-energy emission and FRB-like radio emission from extragalactic distances. We also show nominal 3$\sigma$ \text{X-ray} burst fluence detection thresholds for a selection of currently operating \text{X-ray} telescopes (\textit{AstroSat}, \textit{Chandra}, \textit{Fermi}, \textit{INTEGRAL}, \textit{NICER}, \textit{NuSTAR}, \textit{Swift}, and \textit{XMM-Newton}). These detection thresholds are conservative estimates, based on the in-orbit sensitivities of each instrument under nominal operating conditions, and are consistent with typical prompt high-energy results from FRB observations performed using these \text{X-ray} telescopes. The sensitivity at the time of observations may be affected by the instrument's background level, off-axis angle between the detector and the source, and possibly other factors.

Using currently operating \text{X-ray} telescopes, such as \textit{NICER}, {X-ray} bursts similar to the \text{X-ray} burst that accompanied the \text{FRB-like} radio burst from SGR~1935+2154 on 28~April~2020\,\cite{Li+2021} are detectable from extragalactic FRB sources located within $\sim$1\,Mpc (Extended Data Fig.~\ref{fig:m81r_fluence_distance_constraints}). More energetic \text{X-ray} bursts may be detectable from larger distances. Modest improvements in the sensitivity of currently operating high-energy \text{X-ray} detectors would facilitate deeper studies of the parameter space of magnetar-like bursts produced from nearby extragalactic FRB sources (Fig.~\ref{fig:xray_transient_histogram}).

Future \text{X-ray} telescopes with much larger effective areas, such as \textit{NewAthena}\,\citesupplementary{Barret+2020} and \textit{Strobe-X}\,\citesupplementary{Wilson-Hodge+2017}, would be able to detect \text{X-ray} bursts with properties similar to the \text{X-ray} burst associated with FRB-like emission from SGR~1935+2154\,\cite{Li+2021} from sources located as far as $\sim$5--10\,Mpc in the soft \text{X-ray} band, if the bursts are not heavily absorbed. These same types of bursts may only be detectable from FRB sources located within $\sim$300\,kpc in the hard \text{X-ray} band due to the sensitivity limitations of current and planned \text{X-ray} telescopes. Our predictions for \textit{HEX-P}, \textit{NewAthena}, and \textit{Strobe-X} are based on the currently available sensitivity information for these \text{X-ray} telescopes\,\citesupplementary{Wilson-Hodge+2017, Madsen+2019, Barret+2020}.

In summary, we showed that current and future \text{X-ray} telescopes will have the greatest sensitivity to detecting \text{X-ray} bursts in the soft \text{X-ray} band from extragalactic FRB sources within \text{1--10}\,Mpc. Many models predict multiwavelength emission from FRBs. However, no electromagnetic radiation has yet been detected from any extragalactic FRB source at any wavelength outside of the radio band (between radio frequencies of roughly 100\,MHz\,\citesupplementary{Pleunis+2021b, Pastor-Marazuela+2021} and 8\,GHz\,\citesupplementary{Gajjar+2018}). Additional multiwavelength observations of nearby FRBs with sensitive \text{X-ray} instruments in the future will provide further insights into the emission mechanisms of FRBs, opportunities for testing predictions from FRB models, and valuable information that will aid in illuminating the origins of the sources comprising the FRB population.


\clearpage

\bibliographystylesupplementary{naturemag}
\bibliographysupplementary{references}



\renewcommand{\figurename}{\textbf{Extended Data Fig.}}
\setcounter{figure}{0}


\clearpage

\begin{figure*}
	
	\hspace*{-1.5cm}
	\includegraphics[trim=0cm 0cm 0cm 0cm, clip=false, scale=0.85, angle=0]{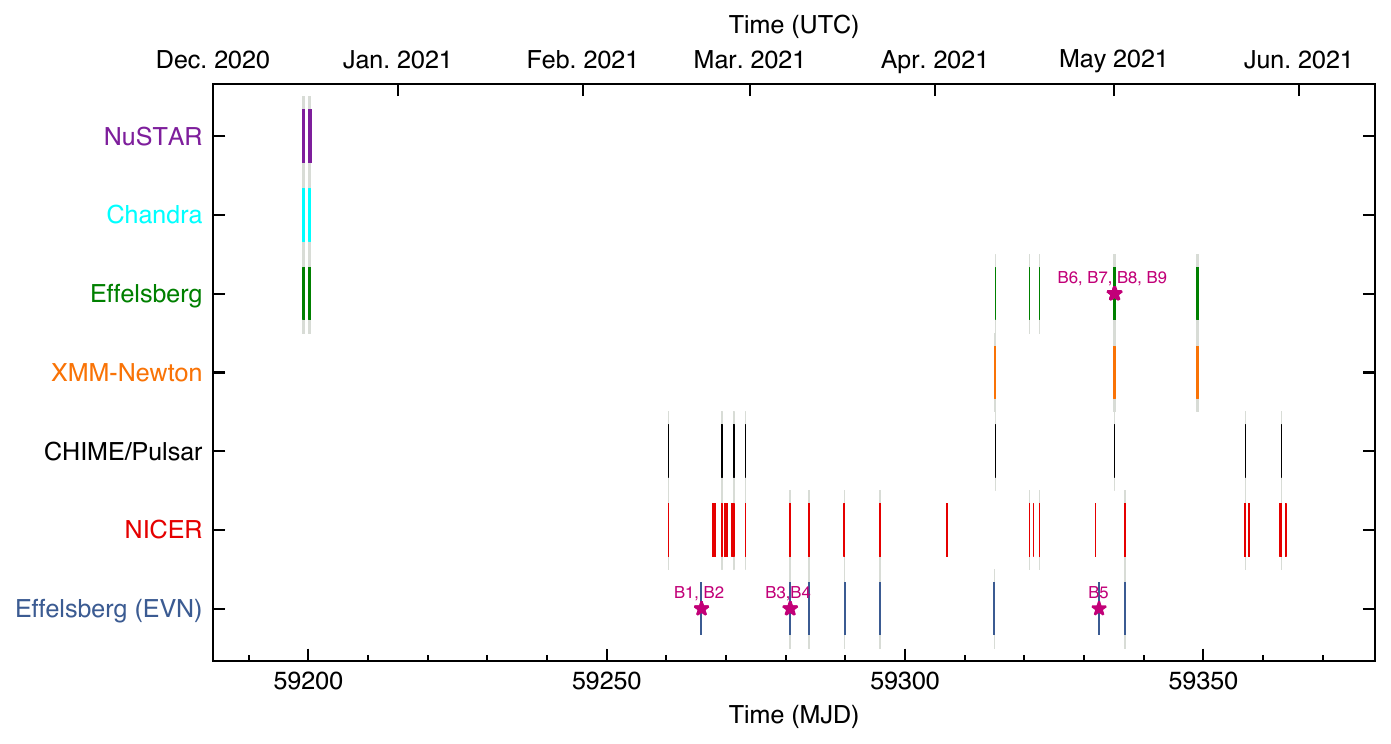}
	
	\caption{\textbf{Timeline of \text{X-ray} and radio observations of FRB~20200120E performed between 2020~December and 2021~June.} The coloured rectangles indicate the start time, end time, and duration of each observation. \text{X-ray} observations carried out with \textit{NuSTAR}, \textit{Chandra}, \textit{XMM-Newton}, and \textit{NICER} are labelled using purple, cyan, orange, and red rectangles, respectively. The green rectangles correspond to independent observations carried out with the Effelsberg radio telescope, and the blue rectangles indicate radio observations performed with Effelsberg during interferometric observations with radio telescopes from the European Very Long Baseline Interferometry~(VLBI) Network~(EVN). Radio observations carried out with the CHIME/Pulsar system are shown using black rectangles. The shaded grey regions highlight times when multiple X-ray and radio instruments were used to perform simultaneous observations. The barycentric arrival times of the radio bursts, shown in Fig.~\ref{fig:radio_bursts}, are labelled using magenta stars.}
	\label{fig:xray_radio_obs_timeline}
	
\end{figure*}


\clearpage

\begin{figure*}
	
	\centering
	
	\vspace*{-0.8cm}
	\includegraphics[trim=0cm 0cm 0cm 0cm, clip=false, scale=0.60, angle=0]{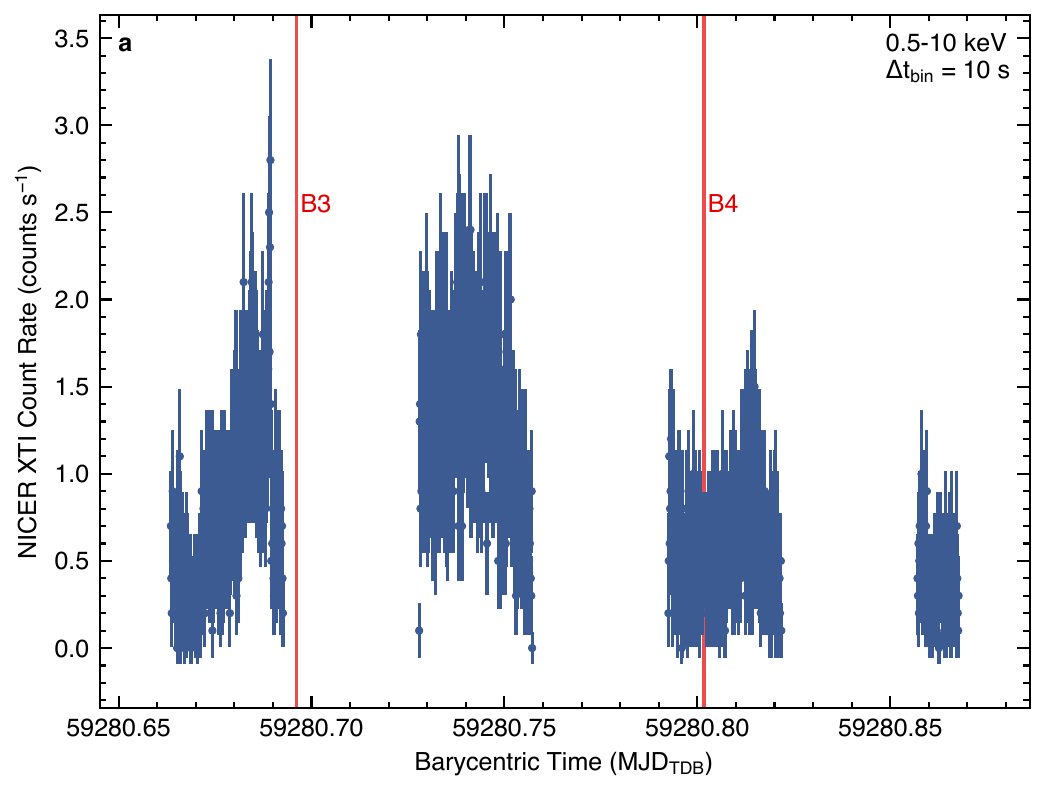}
	
	\includegraphics[trim=0cm 0cm 0cm 0cm, clip=false, scale=0.49, angle=0]{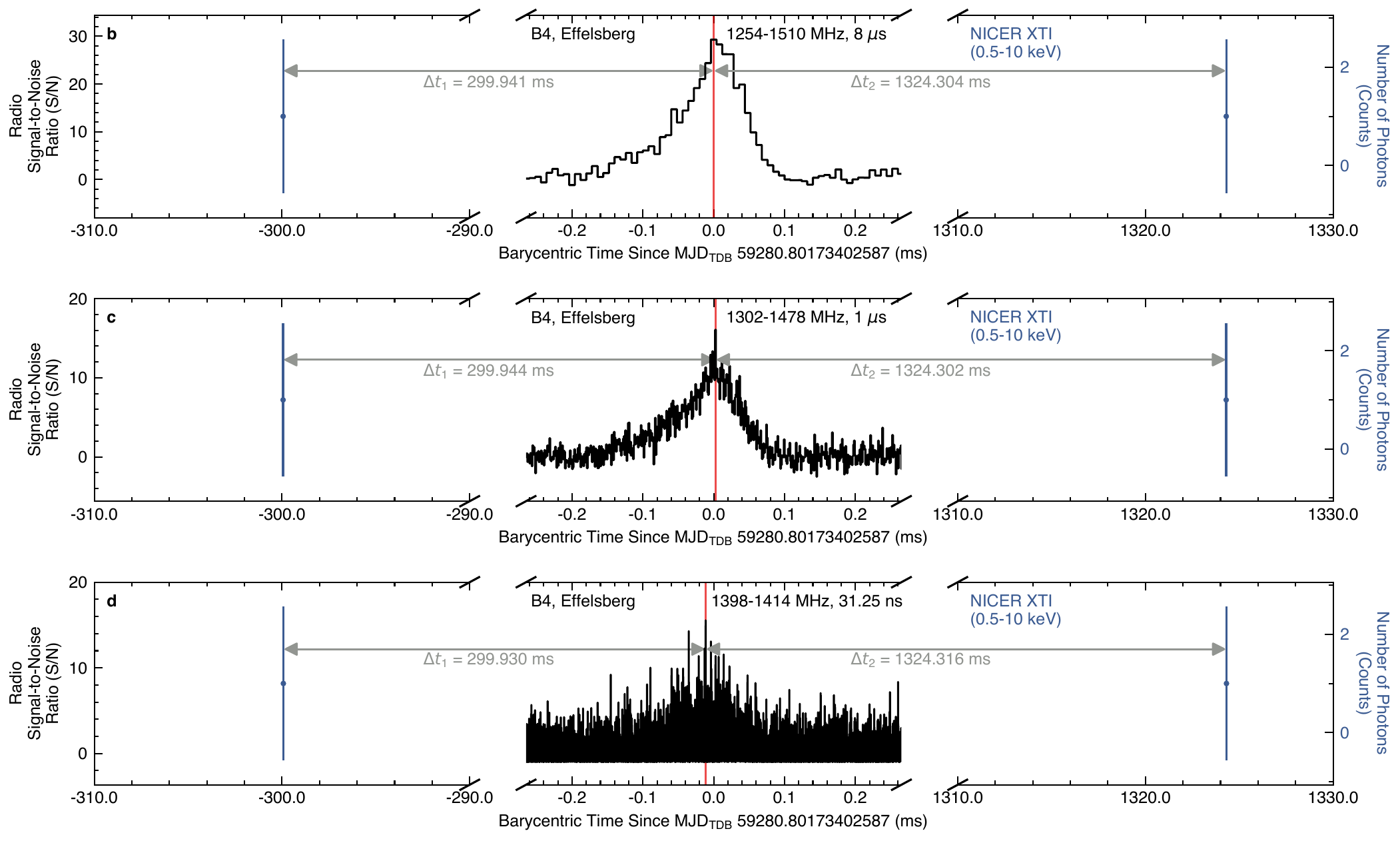}
	
	\caption{\textbf{\text{X-ray} and radio light curves from simultaneous \textit{NICER} and Effelsberg observations of FRB~20200120E, covering bursts~B3 and~B4.} Panel~\textbf{a}: \textit{NICER} \text{0.5--10}\,keV light curve, shown with time bin widths of 10\,s. The error bars correspond to 1$\sigma$ Poisson uncertainties on the count rates. The barycentric arrival times of bursts~B3 and~B4 are indicated by the vertical red lines. Panel~\textbf{b}: Frequency-summed burst profile of~B4, detected using the Effelsberg radio telescope in the 1254--1510\,MHz frequency range and shown at a time resolution of 8\,$\mu$s. Panel~\textbf{c}: Frequency-summed burst profile of~B4 in the 1302--1478\,MHz frequency range, shown at a time resolution of 1\,$\mu$s. Panel~\textbf{d}: Frequency-summed burst profile of~B4 in the 1398--1414\,MHz frequency range, shown at a time resolution of 31.25\,ns. The \text{X-ray} photons detected closest in time to burst~B4 with \textit{NICER} in the \text{0.5--10}\,keV energy band are also shown in panels~\textbf{b--d}, along with 1$\sigma$ Poisson uncertainties based on the photon counts. The time differences between the nearest \text{X-ray} photons and the peak barycentric arrival time of burst~B4 are labelled using grey arrows. The vertical red lines in panels~\textbf{b--d} indicate the peak barycentric arrival time of burst~B4 in each frequency range.}
	
	\label{fig:nicer_b4_xray_radio_light_curves}
	
\end{figure*}


\clearpage

\begin{figure*}
	
	\centering
	
	\includegraphics[trim=0cm 0cm 0cm 0cm, clip=false, scale=0.78, angle=0]{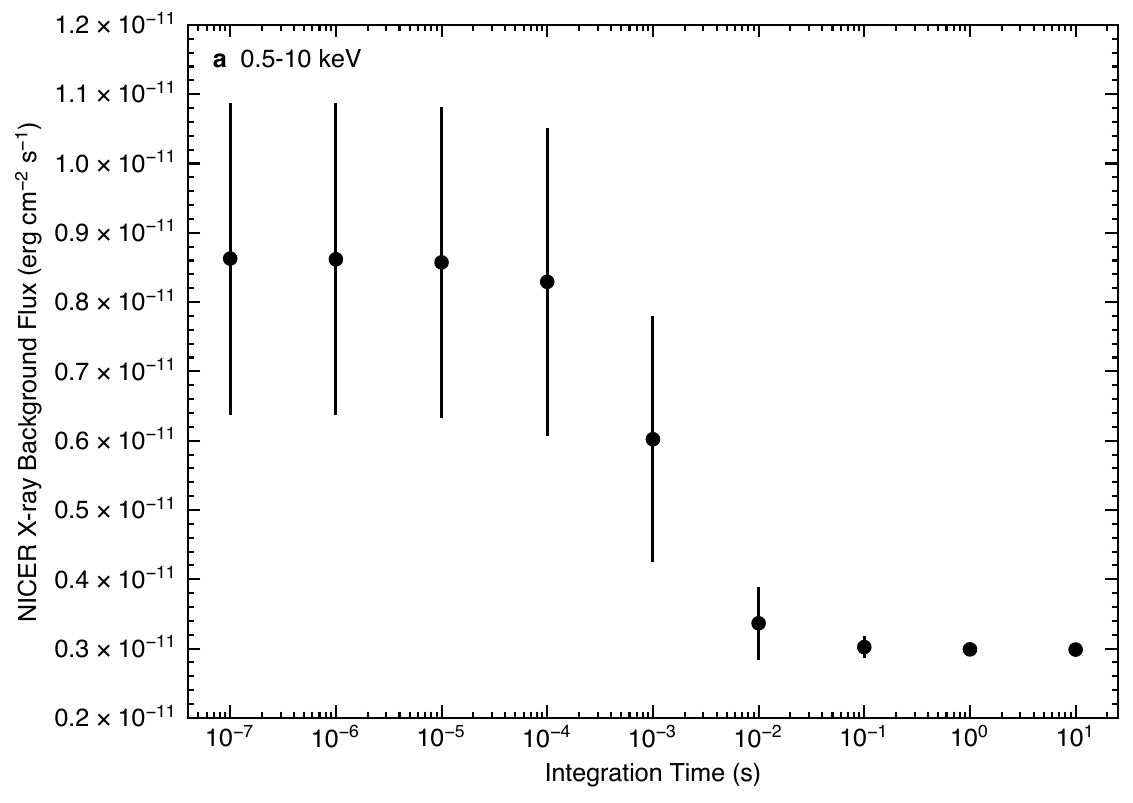}
	
	\includegraphics[trim=0cm 0cm 0cm 0cm, clip=false, scale=0.78, angle=0]{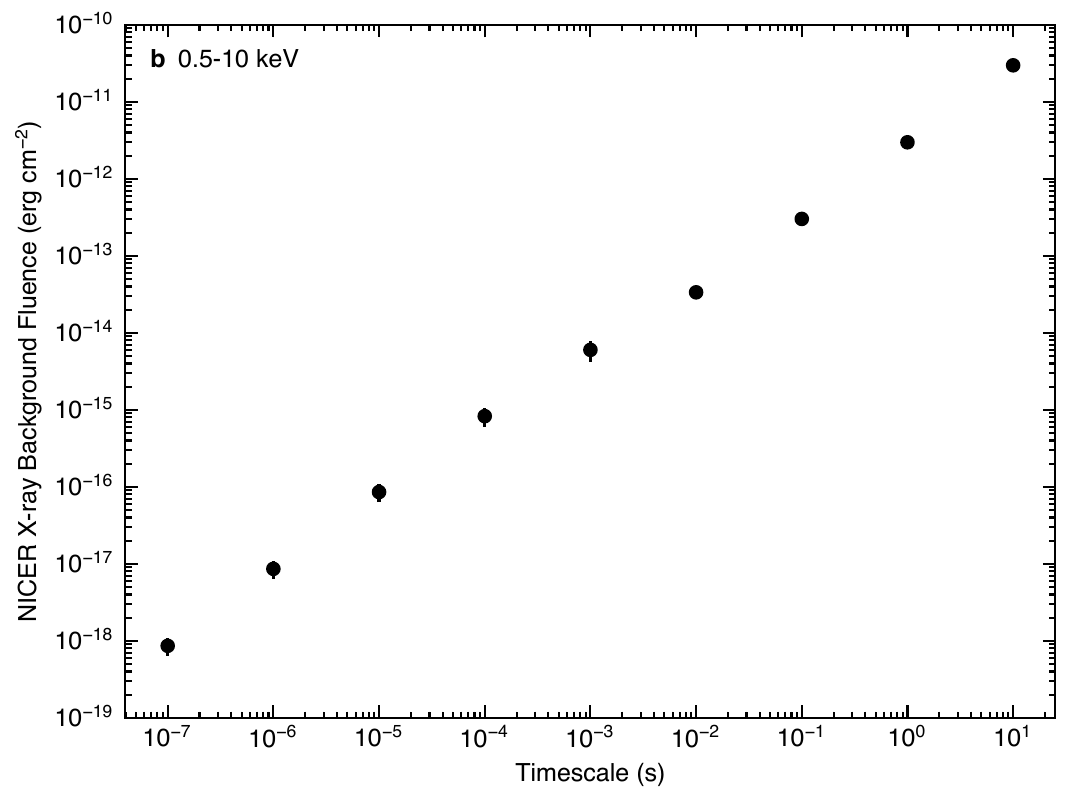}
	
	\caption{\textbf{Average background \text{X-ray} flux and fluence from \textit{NICER} observations of blank sky regions.} Panel~\textbf{a} shows the average absorbed background \text{X-ray} flux from \textit{NICER} in the \text{0.5--10}\,keV energy band on timescales ranging from 100\,ns to 10\,s, and panel~\textbf{b} shows the corresponding \text{X-ray} fluence on these timescales. These measurements were obtained by assuming an \text{X-ray} spectrum with a photon index of $\Gamma$\,$=$\,1.4, similar to that of the diffuse \text{X-ray} background\,\cite{Mushotzky+2000}, and a hydrogen column density of $N_{\text{H}}$\,$=$\,6.73\,$\times$\,10$^{\text{20}}$\,cm$^{\text{--2}}$ towards FRB~20200120E\,\cite{HI4PI+2016}. The error bars shown in panels~\textbf{a} and~\textbf{b} correspond to 1$\sigma$ uncertainties (Methods).}
	\label{fig:nicer_bkgd_xray_flux}
	
\end{figure*}


\clearpage

\begin{figure*}
	
	\centering
	
	\vspace*{-0.8cm}
	\includegraphics[trim=0cm 0cm 0cm 0cm, clip=false, scale=0.63, angle=0]{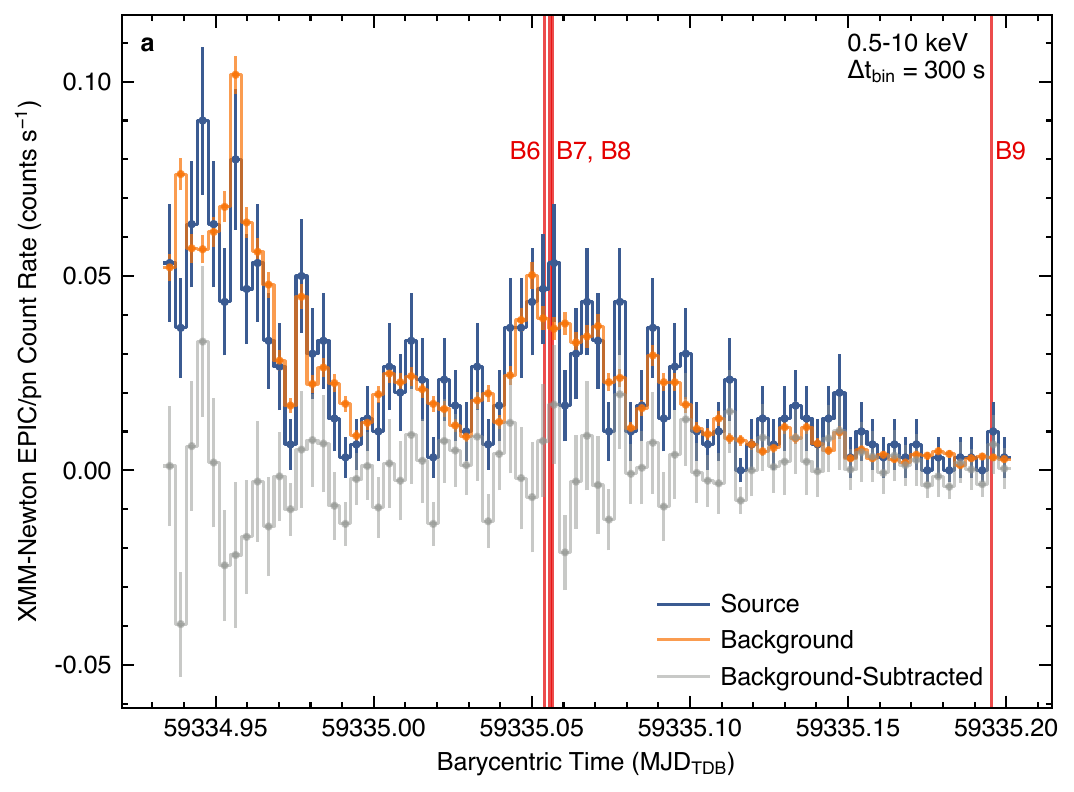}
	
	\includegraphics[trim=0cm 0cm 0cm 0cm, clip=false, scale=0.5, angle=0]{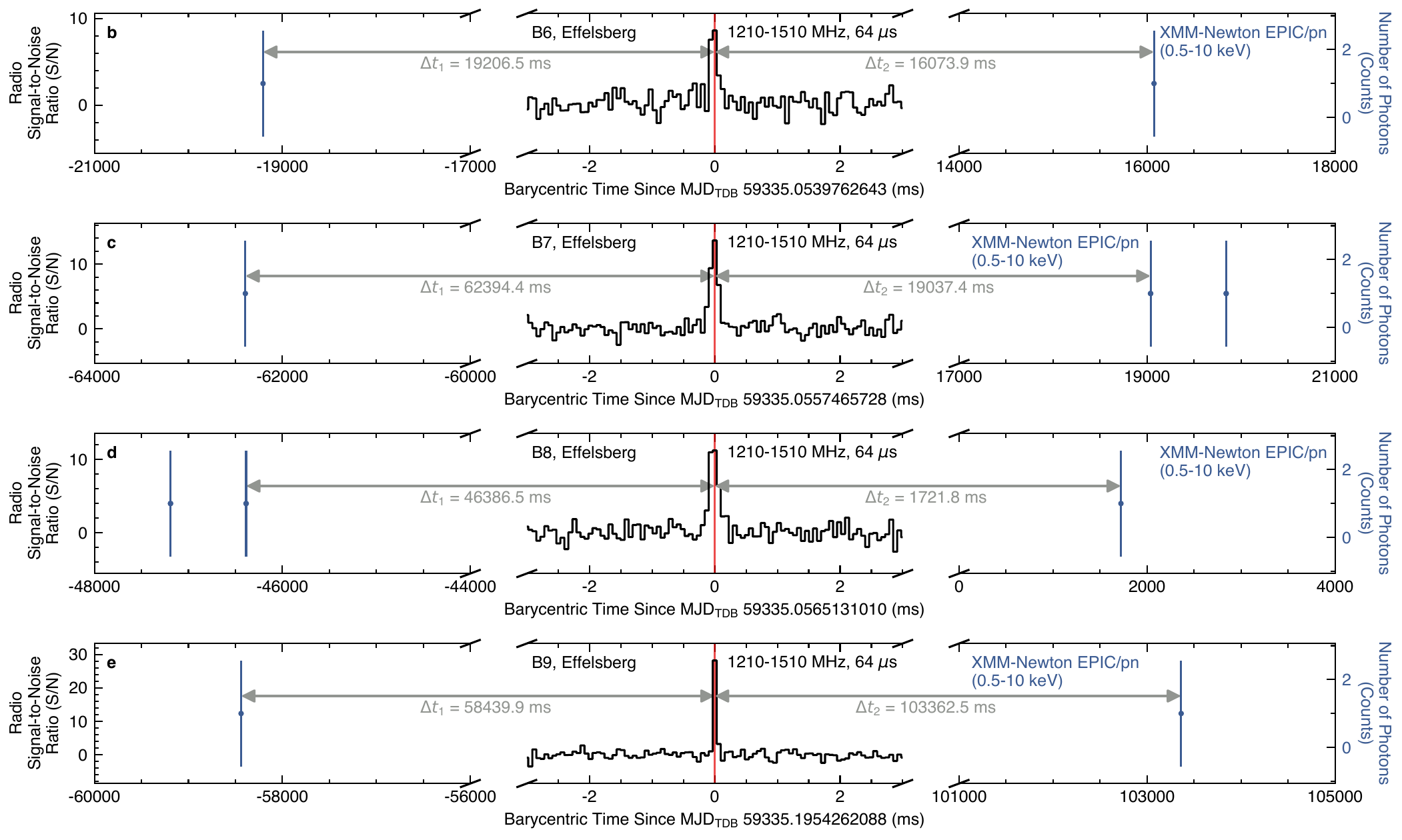}
	
	\caption{\textbf{\text{X-ray} and radio light curves from simultaneous \textit{XMM-Newton} EPIC/pn and Effelsberg observations of FRB~20200120E, covering bursts~B6, B7, B8, and~B9.} Panel~\textbf{a}: Source, background, and background-subtracted \text{0.5--10}\,keV light curves of FRB~20200120E from \textit{XMM-Newton}'s EPIC/pn camera. The light curves are shown with time bin widths of 300\,s, and the error bars correspond to 1$\sigma$ Poisson uncertainties on the count rates. The barycentric arrival times of bursts~B6, B7, B8, and~B9 are indicated by the vertical red lines. Panels~\textbf{b--e}: Frequency-summed burst profiles of~B6, B7, B8, and~B9, detected using the Effelsberg radio telescope in the 1210--1510\,MHz frequency range and shown at a time resolution of 64\,$\mu$s. The \text{X-ray} photons detected closest in time to these radio bursts with \textit{XMM-Newton}'s EPIC/pn camera in the \text{0.5--10}\,keV energy band are also shown in panels~\textbf{b--e}, along with 1$\sigma$ Poisson uncertainties based on the photon counts. The time differences between the nearest \text{X-ray} photons and the peak barycentric arrival times of the radio bursts are labelled using grey arrows. The vertical red lines indicate the peak barycentric arrival times of the radio bursts.}
	
	\label{fig:xmm_b6_b7_b8_b9_xray_radio_light_curves}
	
\end{figure*}


\clearpage

\begin{figure*}
	
	\centering
	
	\vspace*{-0.8cm}
	\includegraphics[trim=0cm 0cm 0cm 0cm, clip=false, scale=0.57, angle=0]{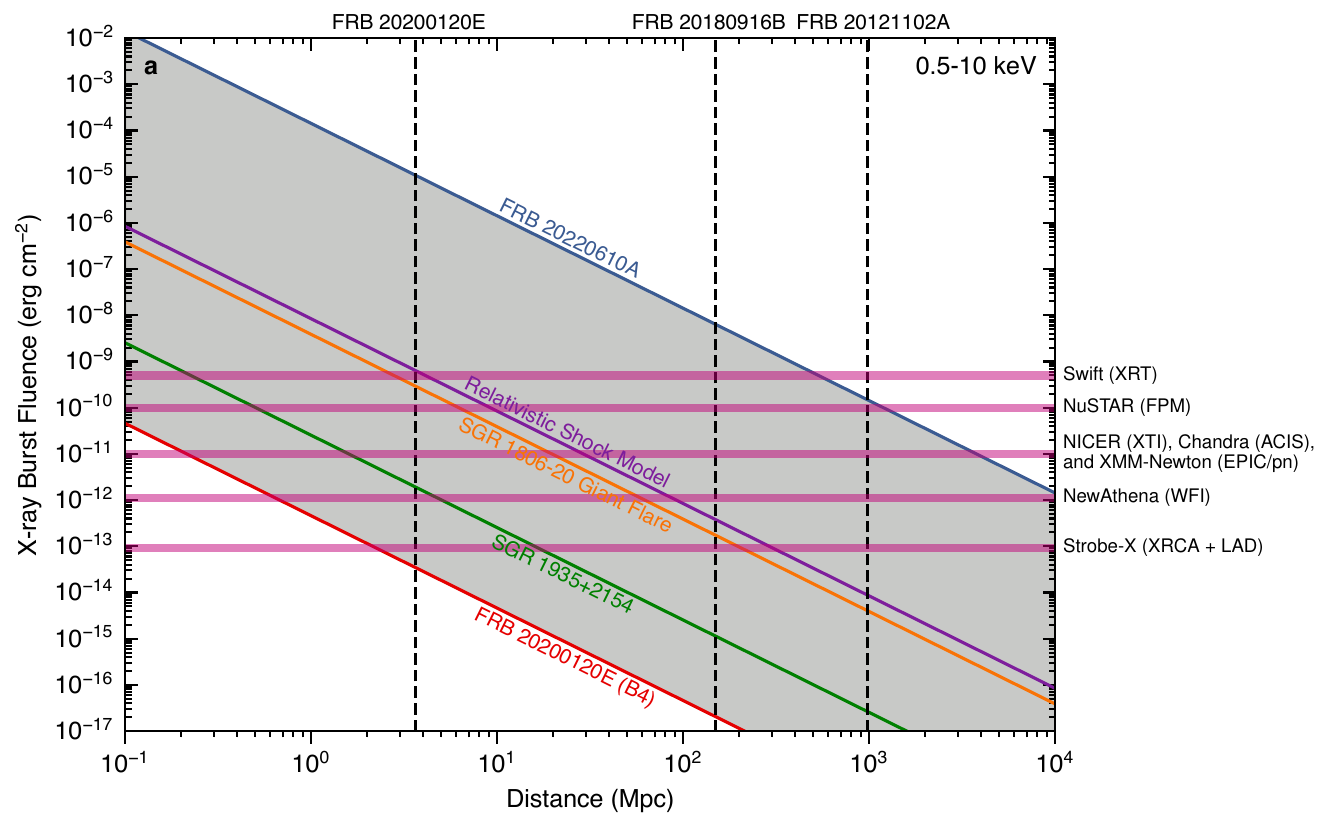}
	
	\includegraphics[trim=0cm 0cm 0cm 0cm, clip=false, scale=0.57, angle=0]{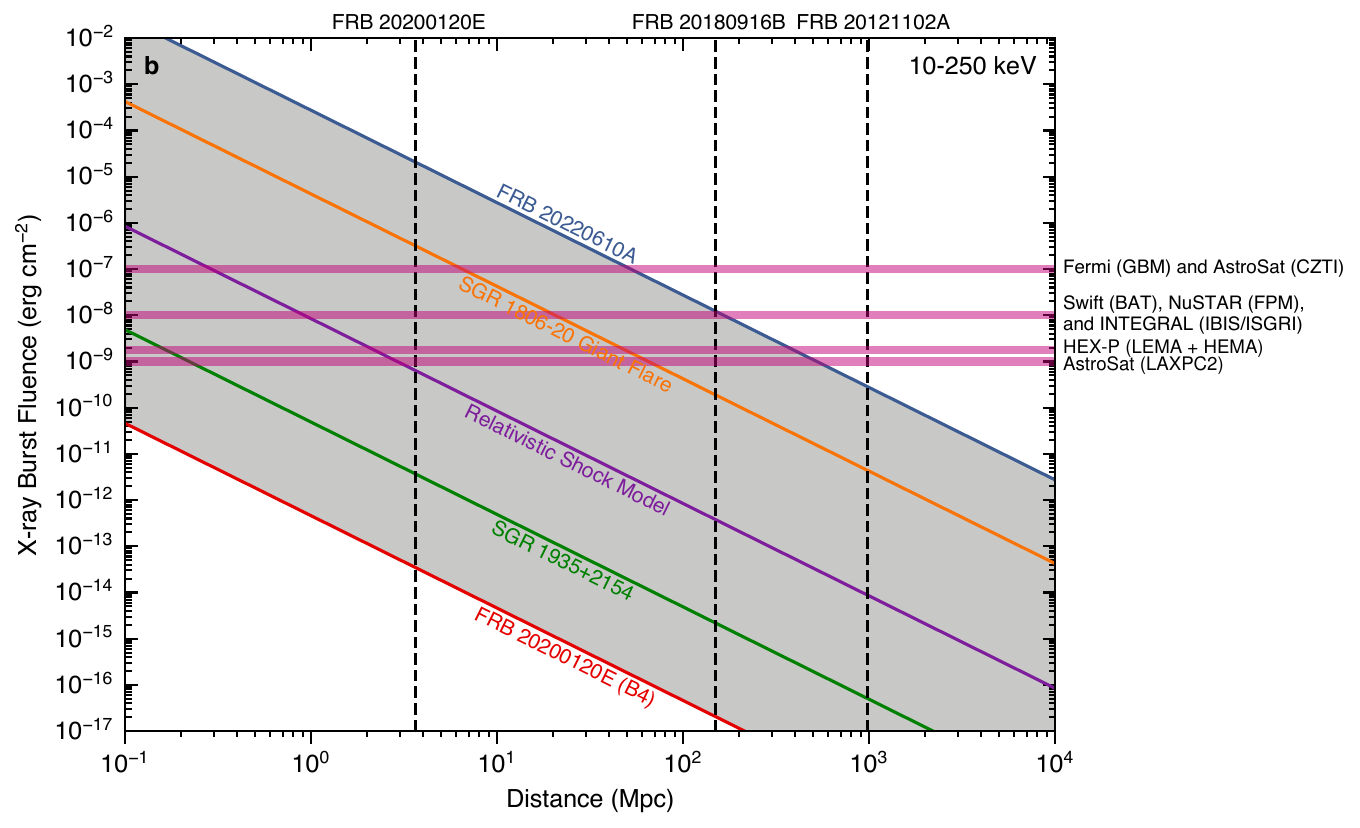}
	
	\caption{\textbf{Detectability of \text{X-ray} counterparts from FRB sources located at distances between 100\,kpc and 10\,Gpc with current and future \text{X-ray} instruments.} Panel~\textbf{a} shows the predicted \text{X-ray} burst fluences in the soft \text{X-ray} band (0.5--10\,keV), and panel~\textbf{b} shows the predicted \text{X-ray} burst fluences in the hard \text{X-ray} band (10--250\,keV). The red lines show predictions for bursts similar to~B4 from FRB~20200120E, and the blue lines indicate the expected \text{X-ray} fluences for \text{FRB~20220610A-like} radio bursts\,\cite{Ryder+2023}. The red and blue lines in each energy band are derived by scaling the measured radio fluences by an \text{X-ray}-to-radio fluence ratio based on the \text{X-ray} and radio fluences observed from the \text{FRB-like} burst detected on 28~April~2020 from SGR~1935+2154\,\cite{Li+2021, Bochenek+2020c, CHIME+2020a}. The green lines are also derived from \text{X-ray} and radio burst fluence measurements at the time of FRB-like activity on 28~April~2020 from SGR~1935+2154\,\cite{Li+2021, Bochenek+2020c, CHIME+2020a}. The orange lines show the predicted \text{X-ray} fluences for giant flares similar to the 27~December~2004 flare observed from SGR~1806--20\,\cite{Palmer+2005}. The purple lines correspond to predictions from the relativistic shock model\,\cite{Metzger+2019, Margalit+2020a} for a fiducial \text{X-ray} flare energy of $E_{\text{flare}}$\,$=$\,10$^{\text{42}}$\,erg. The distances of three repeating FRB sources (FRB~20200120E, FRB~20180916B, and FRB~20121102A) are indicated by the vertical dashed black lines. The horizontal magenta lines correspond to nominal 3$\sigma$ \text{X-ray} burst fluence detection thresholds for a selection of current (\textit{AstroSat}, \textit{Chandra}, \textit{Fermi}, \textit{INTEGRAL}, \textit{NICER}, \textit{NuSTAR}, \textit{Swift}, and \textit{XMM-Newton}) and future (\textit{HEX-P}, \textit{NewAthena}, and \textit{Strobe-X}) \text{X-ray} instruments.}
	
	\label{fig:m81r_fluence_distance_constraints}
	
\end{figure*}



\renewcommand{\tablename}{\textbf{Extended Data Table}}
\setcounter{table}{0}


\clearpage

\begin{table*}
	\caption{Radio observations of FRB~20200120E.}
	
	\hspace{-1.45cm}
	\resizebox{1.15\textwidth}{!}{
		\begin{tabular}{ccccccccc}
			\toprule
			Telescope & Start Time$^{\mathrm{a}}$ & End Time$^{\mathrm{b}}$ & Mid-time$^{\mathrm{c}}$ & Exposure$^{\mathrm{d}}$ & Time Resolution$^{\mathrm{e}}$ & Frequency Resolution$^{\mathrm{f}}$ & Observing Backend$^{\mathrm{g}}$ & Number of Radio \\
			& & & & & & & & Bursts Detected \\
			& (UTC) & (UTC) & (MJD) & (min) & ($\mu$s) & (MHz) & \\
			\hline
			Effelsberg & 2020-12-16~02:07:43 & 2020-12-16~10:08:54 & 59199.25577 & 480.0 & 54.0 & 0.59 & PFFTS \& EDD & 0 \\ 
			Effelsberg & 2020-12-17~02:13:59 & 2020-12-17~09:14:57 & 59200.23921 & 420.0 & 54.0 & 0.59 & PFFTS \& EDD & 0 \\ 
			
			CHIME & 2021-02-15~07:56:44 & 2021-02-15~08:35:07 & 59260.34439 & 38.4 & 327.68 & 0.390625 & Pulsar & 0 \\ 
			
			\textbf{Effelsberg} & 2021-02-20~17:07:57 & 2021-02-20~21:47:06 & 59265.81078 & 162.0 & 0.03125 & 16.0 & PSRIX \& VLBI & 2 \\ 
			
			CHIME & 2021-02-24~07:21:19 & 2021-02-24~07:59:47 & 59269.31983 & 38.5 & 327.68 & 0.390625 & Pulsar & 0 \\ 
			
			CHIME & 2021-02-26~07:13:28 & 2021-02-26~07:51:54 & 59271.31436 & 38.4 & 327.68 & 0.390625 & Pulsar & 0 \\ 
			CHIME & 2021-02-28~07:05:40 & 2021-02-28~07:44:03 & 59273.30893 & 38.4 & 327.68 & 0.390625 & Pulsar & 0 \\ 
			
			\textbf{Effelsberg} & 2021-03-07~15:52:58 & 2021-03-07~20:32:09 & 59280.75872 & 162.2 & 0.03125 & 16.0 & PSRIX \& VLBI & 2 \\ 
			Effelsberg & 2021-03-10~19:08:05 & 2021-03-10~23:48:28 & 59283.89464 & 161.2 & 0.03125 & 16.0 & PSRIX \& VLBI & 0 \\ 
			Effelsberg & 2021-03-16~18:20:01 & 2021-03-16~22:43:46 & 59289.85548 & 148.9 & 0.03125 & 16.0 & PSRIX \& VLBI & 0 \\ 
			Effelsberg & 2021-03-22~16:20:37 & 2021-03-22~20:44:06 & 59295.77247 & 149.3 & 0.03125 & 16.0 & PSRIX \& VLBI & 0 \\ 
			
			Effelsberg & 2021-04-10~21:18:40 & 2021-04-10~23:18:12 & 59314.92947 & 73.3 & 0.03125 & 16.0 & PSRIX \& VLBI & 0 \\ 
			
			CHIME & 2021-04-11~04:20:14 & 2021-04-11~04:58:37 & 59315.19404 & 38.4 & 40.96 & 0.390625 & Pulsar & 0 \\ 
			
			Effelsberg & 2021-04-11~04:35:02 & 2021-04-11~05:28:26 & 59315.20954 & 53.4 & 65.5 & 0.24 & PSRIX & 0 \\ 
			Effelsberg & 2021-04-16~19:52:19 & 2021-04-16~22:05:31 & 59320.87425 & 133.2 & 65.5 & 0.24 & PSRIX & 0 \\ 
			
			Effelsberg & 2021-04-18~12:20:10 & 2021-04-18~14:20:10 & 59322.55567 & 120.0 & 65.5 & 0.24 & PSRIX & 0 \\ 
			
			\textbf{Effelsberg} & 2021-04-28~11:07:57 & 2021-04-28~15:42:43 & 59332.55926 & 172.6 & 0.03125 & 16.0 & PSRIX \& VLBI & 1 \\ 
			
			\textbf{Effelsberg} & 2021-04-30~21:55:38 & 2021-05-01~05:15:39 & 59335.06642 & 436.6 & 64.0 & 0.78125 & EDD & 4 \\ 
			
			CHIME & 2021-05-01~03:01:50 & 2021-05-01~03:40:16 & 59335.13962 & 38.4 & 40.96 & 0.390625 & Pulsar & 0 \\ 
			
			Effelsberg & 2021-05-02~17:08:07 & 2021-05-02~23:35:57 & 59336.84863 & 240.5 & 0.03125 & 16.0 & PSRIX \& VLBI & 0 \\ 
			
			Effelsberg & 2021-05-14~21:21:06 & 2021-05-15~05:22:30 & 59349.05681 & 480.0 & 64.0 & 0.78125 & EDD & 0 \\ 
			
			CHIME & 2021-05-23~01:35:06 & 2021-05-23~02:13:29 & 59357.07937 & 38.4 & 40.96 & 0.390625 & Pulsar & 0 \\ 
			
			CHIME & 2021-05-29~01:11:31 & 2021-05-29~01:49:54 & 59363.06299 & 38.4 & 40.96 & 0.390625 & Pulsar & 0 \\ 
			
			\bottomrule
			\multicolumn{9}{l}{Radio bursts were detected from FRB~20200120E during the observations highlighted in bold.} \\
			\multicolumn{9}{l}{$^{\mathrm{a}}$ Start time of the observation.} \\
			\multicolumn{9}{l}{$^{\mathrm{b}}$ End time of the observation.} \\
			\multicolumn{9}{l}{$^{\mathrm{c}}$ Mid-time of the observation.} \\
			\multicolumn{9}{l}{$^{\mathrm{d}}$ Total exposure time.} \\
			\multicolumn{9}{l}{$^{\mathrm{e}}$ Time resolution of the channelized data used to search for radio bursts. For Effelsberg observations performed during the PRECISE observing campaign, we list the native} \\
			\multicolumn{9}{l}{time resolution~(31.25\,ns) of the voltage data recorded with the VLBI backend.} \\
			\multicolumn{9}{l}{$^{\mathrm{f}}$ Frequency resolution of the channelized data used to search for radio bursts. For Effelsberg observations performed during the PRECISE observing campaign, we list the} \\
			\multicolumn{9}{l}{bandwidth of the subbands (16\,MHz) recorded with the VLBI backend.} \\
			\multicolumn{9}{l}{$^{\mathrm{g}}$ Observing backend used during the observation. Observations with two listed backends indicate that data were simultaneously recorded using both backends.}			
	\end{tabular}}
	
	\label{tab:radio_obs}
\end{table*}


\clearpage

\begin{table*}
	\caption{\text{X-ray} observations of FRB~20200120E.}
	
	\hspace{-1.5cm}
	\resizebox{1.15\textwidth}{!}{
		\begin{tabular}{ccccccccc}
			\toprule
			Telescope & Observation ID & Start Time$^{\mathrm{a}}$ & End Time$^{\mathrm{b}}$ & Mid-time$^{\mathrm{c}}$ & Exposure$^{\mathrm{d}}$ & Count Rate$^{\mathrm{e}}$ & Off-axis Angle$^{\mathrm{f}}$ & 3$\sigma$ Persistent \text{X-ray} \\
			& & & & & & & & Flux Upper Limit$^{\mathrm{g}}$ \\
			& & (UTC) & (UTC) & (MJD) & (s) & (counts\,s$^{\text{--1}}$) & (arcmin) & ($F_{\text{X}}$, 10$^{\text{--13}}$\,erg\,cm$^{\text{--2}}$\,s$^{\text{--1}}$) \\
			\hline
			\textit{Chandra} & 9540 & 2008-08-24~09:51:41 & 2008-08-24~17:35:52 & 54702.57207 & 25724 & 1.40\,$\times$\,10$^{\text{--3}}$ & 13.78 & $<$\,0.1 \\
			
			\textit{NuSTAR} & 80701506002 & 2020-12-15~23:26:09 & 2020-12-16~10:01:09 & 59199.19698 & 23774 & 0.13 & 0.17 & $<$\,1 \\
			
			\textit{Chandra} & 23544 & 2020-12-16~01:02:35 & 2020-12-16~09:53:28 & 59199.22780 & 30695 & 1.63\,$\times$\,10$^{\text{--4}}$ & 0.30 & $<$\,0.06 \\
			
			\textit{NuSTAR} & 80701506004 & 2020-12-17~02:11:09 & 2020-12-17~12:51:09 & 59200.31330 & 21963 & 0.13 & 0.17 & $<$\,2 \\
			
			\textit{Chandra} & 24894 & 2020-12-17~02:17:05 & 2020-12-17~09:17:29 & 59200.24117 & 24071 & 8.31\,$\times$\,10$^{\text{--5}}$ & 0.28 & $<$\,0.06 \\
			
			\textit{NICER} & 3658040101 & 2021-02-15~08:17:39 & 2021-02-15~10:00:20 & 59260.38124 & 2011 & 0.61 & 0.58 & $<$\,3 \\
			\textit{NICER} & 3658040102 & 2021-02-22~07:54:16 & 2021-02-22~23:09:00 & 59267.64697 & 5899 & 0.60 & 0.56 & $<$\,0.5 \\
			\textit{NICER} & 3658040103 & 2021-02-23~00:22:19 & 2021-02-23~06:54:00 & 59268.15149 & 4850 & 0.78 & 0.58 & $<$\,4 \\
			\textit{NICER} & 3658040104 & 2021-02-24~07:21:38 & 2021-02-24~23:11:20 & 59269.63645 & 6840 & 3.24 & 0.57 & $<$\,11 \\
			\textit{NICER} & 3658040105 & 2021-02-25~00:18:16 & 2021-02-25~23:59:40 & 59270.50623 & 6299 & 1.27 & 0.55 & $<$\,6 \\
			\textit{NICER} & 3658040106 & 2021-02-26~01:02:19 & 2021-02-26~07:45:40 & 59271.18332 & 6343 & 0.54 & 0.56 & $<$\,0.7 \\
			\textit{NICER} & 3658040107 & 2021-02-28~07:21:19 & 2021-02-28~07:55:40 & 59273.31840 & 1933 & 0.58 & 0.66 & $<$\,2 \\
			\textbf{\textit{NICER}} & \textbf{3658040108} & 2021-03-07~15:48:32 & 2021-03-07~20:45:10 & 59280.76170 & 8503 & 0.81 & 0.55 & $<$\,1 \\
			\textit{NICER} & 4689010101 & 2021-03-10~19:44:49 & 2021-03-10~23:27:00 & 59283.89993 & 5925 & 0.44 & 0.54 & $<$\,0.4 \\
			\textit{NICER} & 4689010102 & 2021-03-16~18:16:05 & 2021-03-16~22:03:20 & 59289.84007 & 5363 & 0.51 & 0.53 & $<$\,0.1 \\
			\textit{NICER} & 4689010103 & 2021-03-22~16:44:00 & 2021-03-22~20:26:30 & 59295.77448 & 7087 & 0.65 & 0.50 & $<$\,0.6 \\
			\textit{NICER} & 4689010104 & 2021-04-02~23:47:35 & 2021-04-03~03:20:10 & 59307.06519 & 6068 & 1.63 & 0.49 & $<$\,0.4 \\
			
			\textit{XMM-Newton} & 0872392401 & 2021-04-10~22:59:32 & 2021-04-11~05:04:49 & 59315.08484 & 17300$^{\mathrm{h}}$ & 5.66\,$\times$\,10$^{\text{--3}}$ & 0.53 & $<$\,0.07 \\
			
			\textit{NICER} & 4689010105 & 2021-04-16~20:46:09 & 2021-04-16~21:15:47 & 59320.87567 & 1360 & 18.81$^{\mathrm{i}}$ & 0.35 & $<$\,3 \\
			\textit{NICER} & 4689010106 & 2021-04-17~12:28:59 & 2021-04-17~12:51:40 & 59321.52800 & 1242 & 1.45 & 0.01 & $<$\,1 \\
			\textit{NICER} & 4689010107 & 2021-04-18~13:01:19 & 2021-04-18~13:38:00 & 59322.55532 & 1772 & 1.04 & 0.35 & $<$\,3 \\
			\textit{NICER} & 4689010108 & 2021-04-27~22:53:04 & 2021-04-27~23:19:00 & 59331.96252 & 1198 & 0.45 & 0.02 & $<$\,0.3 \\
			
			\textbf{\textit{XMM-Newton}} & \textbf{0872392501} & 2021-04-30~21:51:57 & 2021-05-01~05:03:57 & 59335.06108 & 7285$^{\mathrm{h}}$ & 8.78\,$\times$\,10$^{\text{--3}}$ & 0.53 & $<$\,0.2 \\
			
			\textit{NICER} & 4689010109 & 2021-05-02~17:23:04 & 2021-05-02~22:22:30 & 59336.82832 & 2108 & 0.37 & 0.06 & $<$\,0.08 \\
			
			\textit{XMM-Newton} & 0872392601 & 2021-05-14~21:00:24 & 2021-05-15~05:02:20 & 59349.04262 & 15429$^{\mathrm{h}}$ & 7.19\,$\times$\,10$^{\text{--3}}$ & 0.53 & $<$\,0.1 \\
			
			\textit{NICER} & 3658040109 & 2021-05-22~20:35:00 & 2021-05-22~22:44:27 & 59356.90259 & 4172 & 0.68 & 0.09 & $<$\,1 \\
			\textit{NICER} & 3658040110 & 2021-05-22~23:38:55 & 2021-05-23~17:20:22 & 59357.35392 & 12616 & 0.74 & 0.06 & $<$\,0.6 \\
			\textit{NICER} & 3658040111 & 2021-05-28~19:03:15 & 2021-05-28~22:54:09 & 59362.87410 & 7557 & 0.48 & 0.02 & $<$\,0.7 \\
			\textit{NICER} & 3658040112 & 2021-05-28~23:41:41 & 2021-05-29~20:02:40 & 59363.41123 & 10211 & 0.66 & 0.03 & $<$\,0.7 \\
			\bottomrule
			\multicolumn{9}{l}{Radio bursts were detected from FRB~20200120E during the observations highlighted in bold. FRB~20200120E was observed with \textit{NICER} for a total exposure time} \\
			\multicolumn{9}{l}{of 109.4\,ks between 15~February~2021 and 30~May~2021. The total exposure times of the \textit{Chandra}, \textit{NuSTAR}, and \textit{XMM-Newton} observations of FRB~20200120E were} \\
			\multicolumn{9}{l}{80.5\,ks, 45.7\,ks, and 40.0\,ks, respectively.} \\
			\multicolumn{9}{l}{$^{\mathrm{a}}$ Start time of the observation.} \\
			\multicolumn{9}{l}{$^{\mathrm{b}}$ End time of the observation.} \\
			\multicolumn{9}{l}{$^{\mathrm{c}}$ Mid-time of the observation.} \\
			\multicolumn{9}{l}{$^{\mathrm{d}}$ Total exposure time during Good Time Intervals~(GTIs).} \\
			\multicolumn{9}{l}{$^{\mathrm{e}}$ Average (non-background-subtracted) count rate during GTIs. The count rates are derived in the \text{0.5--10}\,keV energy band for the \textit{Chandra}, \textit{NICER}, and} \\
			\multicolumn{9}{l}{\textit{XMM-Newton} observations and in the \text{3--79}\,keV energy band for the \textit{NuSTAR} observations. A circular region around the source, centred on the position of} \\
			\multicolumn{9}{l}{FRB~20200120E from VLBI\,\cite{Kirsten+2022}, was used to measure the count rates for the \textit{Chandra}, \textit{NuSTAR}, and \textit{XMM-Newton} observations.} \\
			\multicolumn{9}{l}{$^{\mathrm{f}}$ Off-axis angle, calculated relative to the nominal pointing direction and the VLBI position of FRB~20200120E\,\cite{Kirsten+2022}.} \\
			\multicolumn{9}{l}{$^{\mathrm{g}}$ 3$\sigma$ persistent \text{X-ray} flux upper limits are derived in the \text{0.5--10}\,keV energy band for the \textit{Chandra}, \textit{NICER}, and \textit{XMM-Newton} observations and in the \text{3--79}\,keV} \\
			\multicolumn{9}{l}{energy band for the \textit{NuSTAR} observations. A fiducial absorbed power-law spectral model, with $N_{\text{H}}$\,$=$\,6.73\,$\times$\,10$^{\text{20}}$\,cm$^{\text{--2}}$\,\cite{HI4PI+2016} and $\Gamma$\,$=$\,2, was used to calculate the} \\
			\multicolumn{9}{l}{persistent \text{X-ray} flux upper limits.} \\
			\multicolumn{9}{l}{$^{\mathrm{h}}$ Total exposure time after removal of instrumental flares.} \\
			\multicolumn{9}{l}{$^{\mathrm{i}}$ The higher count rate compared to other \textit{NICER} observations was due to higher background, as indicated by larger housekeeping parameter} \\
			\multicolumn{9}{l}{(\texttt{FPM\_OVERONLY\_COUNT}) values.}
	\end{tabular}}
	
	\label{tab:xray_obs}
\end{table*}


\clearpage

\begin{table*}
	
	\caption{Measured properties of radio bursts detected from FRB~20200120E with the Effelsberg radio telescope at 1.4\,GHz.}
	
	\hspace{-1.45cm}
	\resizebox{1.15\textwidth}{!}{
		\begin{tabular}{cccccccccc}
			\toprule
			Burst ID$^{\mathrm{a}}$ & Time of Arrival~(ToA) & Peak S/N$^{\mathrm{c}}$ & Dispersion Measure$^{\mathrm{d}}$ & Width$^{\mathrm{e}}$ & Frequency Extent$^{\mathrm{f}}$ & Peak Flux Density$^{\mathrm{g}}$ & Fluence$^{\mathrm{g,h}}$ & Luminosity$^{\mathrm{i}}$ & Energy$^{\mathrm{j}}$ \\
			& at Burst Peak$^{\mathrm{b}}$ & & & & & & & & \\
			& ($t_{\text{peak}}$, MJD$_{\text{TDB}}$) & (S/N)$_{\text{peak}}$ & (DM, pc\,cm$^{\text{--3}}$) & ($w_{\text{t}}$, $\mu$s) & ($w_{\text{f}}$, MHz) & ($F_{\text{R}}$, Jy) & ($\mathcal{F}_{\text{R}}$, Jy~ms) & ($L_{\text{R}}$, 10$^{\text{37}}$\,erg\,s$^{\text{--1}}$) & ($E_{\text{R}}$, 10$^{\text{32}}$\,erg)\\
			\hline
			B1 & 59265.88304442095 & 6.6 & 87.75 & 156\,$\pm$\,1 & 140\,$\pm$\,1 & 1.59\,$\pm$\,0.32 & 0.13\,$\pm$\,0.03 & 0.3\,$\pm$\,0.1 & 5.1\,$\pm$\,1.5 \\
			B2 & 59265.88600917244 & 36.1 & 87.75 & 62\,$\pm$\,1, 93\,$\pm$\,0.5$^{\mathrm{k}}$ & 103\,$\pm$\,1, 89\,$\pm$\,1$^{\mathrm{k}}$ & 8.71\,$\pm$\,1.74 & 0.63\,$\pm$\,0.13 & 1.6\,$\pm$\,0.5 & 25.1\,$\pm$\,7.0 \\
			B3 & 59280.69618750235 & 64.8 & 87.75 & 46.7\,$\pm$\,0.1 & 94\,$\pm$\,1 & 15.6\,$\pm$\,3.12 & 0.53\,$\pm$\,0.11 & 4.5\,$\pm$\,1.2 & 20.9\,$\pm$\,5.8 \\
			B4 & 59280.80173402587 & 29.3 & 87.75 & 117\,$\pm$\,1 & 134\,$\pm$\,1 & 7.07\,$\pm$\,1.41 & 0.71\,$\pm$\,0.14 & 2.4\,$\pm$\,0.7 & 28.0\,$\pm$\,7.6 \\
			B5 & 59332.50446582513 & 6.9 & 87.75 & 56.6\,$\pm$\,0.1 & 86\,$\pm$\,1 & 1.66\,$\pm$\,0.33 & 0.09\,$\pm$\,0.02 & 0.6\,$\pm$\,0.2 & 3.6\,$\pm$\,1.0 \\
			B6 & 59335.0539762643 & 8.6 & 87.76 & 92\,$\pm$\,11 & 53.9\,$\pm$\,0.8 & 0.66\,$\pm$\,0.10 & 0.10\,$\pm$\,0.02 & 0.5\,$\pm$\,0.1 & 4.9\,$\pm$\,1.2 \\
			B7 & 59335.0557465728 & 13.7 & 87.76 & 107\,$\pm$\,8 & 72.7\,$\pm$\,0.8 & 1.05\,$\pm$\,0.16 & 0.17\,$\pm$\,0.03 & 0.7\,$\pm$\,0.2 & 7.9\,$\pm$\,1.9 \\
			B8 & 59335.0565131010 & 11.2 & 87.76 & 124\,$\pm$\,10 & 87.5\,$\pm$\,0.8 & 0.86\,$\pm$\,0.13 & 0.17\,$\pm$\,0.02 & 0.6\,$\pm$\,0.2 & 7.8\,$\pm$\,1.9 \\
			B9 & 59335.1954262088 & 28.2 & 87.76\,$\pm$\,0.06 & 46\,$\pm$\,4 & 135.2\,$\pm$\,0.8 & 2.16\,$\pm$\,0.32 & 0.19\,$\pm$\,0.03 & 2.0\,$\pm$\,0.5 & 9.1\,$\pm$\,2.2 \\
			\bottomrule
			\multicolumn{10}{l}{$^{\mathrm{a}}$ The properties of bursts B1, B2, B3, B4, and~B5 were measured using data sampled at a time resolution of 8\,$\mu$s\,\cite{Nimmo+2022}. The properties of bursts B6, B7, B8, and~B9 were determined} \\
			\multicolumn{10}{l}{from data recorded with a time resolution of 64\,$\mu$s.} \\
			\multicolumn{10}{l}{$^{\mathrm{b}}$ Barycentric time at the peak of the burst, determined after removing the time delay due to dispersion and correcting to infinite frequency. The barycentric times were derived using} \\
			\multicolumn{10}{l}{the Jet Propulsion Laboratory~(JPL) DE405 ephemeris and the position of FRB~20200120E ($\alpha_{\text{J2000}}$\,$=$\,09$^{\text{h}}$57$^{\text{m}}$54.69935$^{\text{s}}$,  $\delta_{\text{J2000}}$\,$=$\,68$^{\circ}$49$\arcmin$00.8529$\arcsec$)\,\cite{Kirsten+2022}, measured in the International} \\
			\multicolumn{10}{l}{Celestial Reference Frame~(ICRF) from VLBI. A dispersion constant of $\mathcal{D}$\,$=$\,1/(2.41\,$\times$\,10$^{\text{--4}}$)\,MHz$^{\text{2}}$\,pc$^{\text{--1}}$\,cm$^{\text{3}}$\,s was used to correct the burst ToAs for dispersion. The burst ToAs were} \\
			\multicolumn{10}{l}{calculated using the Barycentric Dynamical Time~(TDB) standard. We present refined barycentric ToAs for bursts~B1, B2, B3, B4, and~B5, compared to those listed in refs.~\citen{Kirsten+2022}} \\
			\multicolumn{10}{l}{and~\citen{Nimmo+2022}, since an incorrect position for FRB~20200120E was used during the barycentring procedure in these earlier studies.} \\
			\multicolumn{10}{l}{$^{\mathrm{c}}$ Signal-to-noise ratio~(S/N) at the burst peak.} \\
			\multicolumn{10}{l}{$^{\mathrm{d}}$ The properties of bursts B1, B2, B3, B4, and~B5 were measured using a dispersion measure~(DM) of 87.75\,pc\,cm$^{\text{--3}}$\,\cite{Nimmo+2022}. The DM of burst~B9 corresponds to the value that maximized} \\
			\multicolumn{10}{l}{the peak S/N. The properties of bursts B6, B7, and~B8 were derived using the S/N-optimized~DM obtained from~B9.} \\
			\multicolumn{10}{l}{$^{\mathrm{e}}$ Temporal width of each burst. The width of bursts B1, B2, B3, B4, and~B5 is defined as the full-width at half-maximum~(FWHM) divided by $\sqrt{\text{2}}$ of a two-dimensional~(2D)} \\
			\multicolumn{10}{l}{Gaussian fit to the 2D~autocorrelation function~(ACF)\,\cite{Nimmo+2022}. The width of bursts B6, B7, B8, and B9 is defined as the FWHM divided by $\sqrt{\text{2}}$ of a Gaussian fit to the dedispersed,} \\
			\multicolumn{10}{l}{frequency-summed burst profile.} \\
			\multicolumn{10}{l}{$^{\mathrm{f}}$ Frequency extent of each burst. The frequency extent of bursts B1, B2, B3, B4, and~B5 is defined as the FWHM divided by $\sqrt{\text{2}}$ of a 2D Gaussian fit to the 2D ACF\,\cite{Nimmo+2022}. The} \\
			\multicolumn{10}{l}{frequency extent of bursts B6, B7, B8, and~B9 was determined by eliminating frequency channels until the `on-pulse' data were consistent with noise.} \\
			\multicolumn{10}{l}{$^{\mathrm{g}}$ The errors associated with these measurements are based on a conservative estimate of the uncertainty on the system equivalent flux density~(SEFD) of the Effelsberg radio telescope} \\
			\multicolumn{10}{l}{at the time of these observations. We estimate a 20\% uncertainty on Effelsberg's SEFD at the time of bursts B1, B2, B3, B4, and~B5, and a 15\% uncertainty on Effelsberg's SEFD at} \\
			\multicolumn{10}{l}{the time of bursts B6, B7, B8, and~B9.} \\
			\multicolumn{10}{l}{$^{\mathrm{h}}$ Fluences of the radio bursts were computed using the $\pm$2$\sigma$ temporal width region.} \\
			\multicolumn{10}{l}{$^{\mathrm{i}}$ Isotropic-equivalent radio luminosity values were calculated using the expression $L_{\text{R}}$\,$=$\,$4\pi d^{\text{2}}\mathcal{F}_{\text{R}}/(w_{\text{t}}(1+z))$, where $d$\,$=$\,3.63\,$\pm$\,0.34\,Mpc\,\cite{Freedman+1994} corresponds to the distance of} \\
			\multicolumn{10}{l}{FRB~20200120E, $\mathcal{F}_{\text{R}}$ is the fluence of the radio burst, $w_{\text{t}}$ is the temporal width of the radio burst, and $z$ is the redshift of FRB~20200120E.} \\
			\multicolumn{10}{l}{$^{\mathrm{j}}$ Isotropic-equivalent energy values were calculated using the expression $E_{\text{R}}$\,$=$\,$4\pi d^{\text{2}}\mathcal{F}_{\text{R}}/(1+z)$, where $d$ is the distance of FRB~20200120E, $\mathcal{F}_{\text{R}}$ is the fluence of the radio burst,} \\
			\multicolumn{10}{l}{and $z$ is the redshift of FRB~20200120E.} \\
			\multicolumn{10}{l}{$^{\mathrm{k}}$ Values measured from each burst component.}
	\end{tabular}}
	
	\label{tab:radio_burst_properties}
	
\end{table*}


\clearpage

\begin{table*}
	\caption{\textit{NICER} background measurements on 100\,ns to 10\,s timescales in the \text{0.5--10}\,keV energy band.}
	
	\begin{center}
		
		\begin{tabular}{cccc}
			\toprule
			Timescale & $\langle N_{\text{bkg}}^{\text{blank}}\rangle$$^{\mathrm{a}}$ & $N_{\text{bkg}}^{\text{sw}}$$^{\mathrm{b}}$ & $N_{\text{bkg}}^{\text{100\,s}}$\,$^{\mathrm{c}}$ \\
			& (counts) & (counts) & (counts) \\
			\hline
			10\,s & 8.53\,$\pm$\,0.04 & 8.1\,$\pm$\,0.2 & 4.9\,$\pm$\,0.5 \\
			1\,s & 0.85 \,$\pm$\,0.01 & 0.81\,$\pm$\,0.2 & 0.49\,$\pm$\,0.05 \\
			100\,ms & 0.086 \,$\pm$\,0.004 & 0.081\,$\pm$\,0.02 & 0.049\,$\pm$\,0.005 \\
			10\,ms & 0.010 \,$\pm$\,0.001 & 0.0081\,$\pm$\,0.002 & 0.0049\,$\pm$\,0.0005 \\
			1\,ms & 0.0017 \,$\pm$\,0.0005 & (8.1\,$\pm$\,0.2)\,$\times$\,10$^{\text{--4}}$ & (4.9\,$\pm$\,0.5)\,$\times$\,10$^{\text{--4}}$ \\
			100\,$\mu$s & (2.4\,$\pm$\,0.6)\,$\times$\,10$^{\text{--4}}$ & (8.1\,$\pm$\,0.2)\,$\times$\,10$^{\text{--5}}$ & (4.9\,$\pm$\,0.5)\,$\times$\,10$^{\text{--5}}$ \\
			10\,$\mu$s & (2.4\,$\pm$\,0.6)\,$\times$\,10$^{\text{--5}}$ & (8.1\,$\pm$\,0.2)\,$\times$\,10$^{\text{--6}}$ & (4.9\,$\pm$\,0.5)\,$\times$\,10$^{\text{--6}}$ \\
			1\,$\mu$s & (2.5\,$\pm$\,0.6)\,$\times$\,10$^{\text{--6}}$ & (8.1\,$\pm$\,0.2)\,$\times$\,10$^{\text{--7}}$ & (4.9\,$\pm$\,0.5)\,$\times$\,10$^{\text{--7}}$ \\
			100\,ns & (2.5\,$\pm$\,0.6)\,$\times$\,10$^{\text{--7}}$ & (8.1\,$\pm$\,0.2)\,$\times$\,10$^{\text{--8}}$ & (4.9\,$\pm$\,0.5)\,$\times$\,10$^{\text{--8}}$ \\
			\bottomrule
			\multicolumn{4}{l}{$^{\mathrm{a}}$ Average number of background photons in the \text{0.5--10}\,keV energy} \\
			\multicolumn{4}{l}{band, derived from \textit{NICER} observations of blank sky fields.} \\
			\multicolumn{4}{l}{$^{\mathrm{b}}$ Number of background photons in the \text{0.5--10}\,keV energy band,} \\
			\multicolumn{4}{l}{derived from the \textit{NICER} `Space Weather' model\,\cite{Remillard+2022}.} \\
			\multicolumn{4}{l}{$^{\mathrm{c}}$ Inferred number of background photons based on the count rate} \\
			\multicolumn{4}{l}{within $\pm$100\,s of the barycentric, infinite frequency peak time of} \\
			\multicolumn{4}{l}{burst~B4 ($t_{\text{peak}}^{\text{B4}}$), assuming all detected photons are attributed to the} \\
			\multicolumn{4}{l}{background.}
		\end{tabular}
		
	\end{center}
	
	\label{tab:nicer_xray_bkg}
\end{table*}


\clearpage

\begin{table*}
	\caption{Observed fluences of the \text{X-ray} burst associated with the FRB-like radio burst detected from SGR~1935+2154 on 28~April~2020 and predicted \text{X-ray} fluences in the \text{0.5--10}\,keV and \text{3--79}\,keV energy bands for a similar \text{X-ray} burst emitted from the location of FRB~20200120E.}
	
	\hspace{-1.5cm}
	\resizebox{1.15\textwidth}{!}{
		\begin{tabular}{ccccc}
			\toprule
			Instrument & Observed \text{X-ray} Fluence & Predicted \text{X-ray} Fluence & Predicted \text{X-ray} Fluence & References \\
			& from SGR~1935+2154$^{\mathrm{a}}$ & from FRB~20200120E$^{\mathrm{b}}$ & from FRB~20200120E$^{\mathrm{c}}$ & \\
			& (Absorbed / Unabsorbed) & (Absorbed / Unabsorbed) & (Absorbed / Unabsorbed) & \\
			& ($\mathcal{F}_{\text{X}}$, 10$^{\text{--7}}$\,erg\,cm$^{\text{--2}}$) & ($\mathcal{F}_{\text{X}}$, 10$^{\text{--12}}$\,erg\,cm$^{\text{--2}}$, \text{0.5--10}\,keV) & ($\mathcal{F}_{\text{X}}$, 10$^{\text{--12}}$\,erg\,cm$^{\text{--2}}$, \text{3--79}\,keV) & \\
			\hline
			\textit{AGILE} & (5$^{\text{+10}}_{\text{--4}}$)$^{\mathrm{d}}$ / (5$^{\text{+10}}_{\text{--4}}$)$^{\mathrm{d}}$ & (0.6$^{\text{+0.6}}_{\text{--0.3}}$) / (0.6$^{\text{+0.6}}_{\text{--0.3}}$) & (6$^{\text{+11}}_{\text{--5}}$) / (6$^{\text{+11}}_{\text{--5}}$) & \citen{Tavani+2021} \\
			\textit{Insight-HXMT} & (6.4$^{\text{+0.4}}_{\text{--0.3}}$)$^{\mathrm{e}}$ / (7.1$^{\text{+0.4}}_{\text{--0.4}}$)$^{\mathrm{e}}$ & (1.9$^{\text{+0.1}}_{\text{--0.1}}$) / (2.0$^{\text{+0.1}}_{\text{--0.1}}$) & (3.9$^{\text{+0.2}}_{\text{--0.2}}$) / (3.9$^{\text{+0.2}}_{\text{--0.2}}$) & \citen{Li+2021} \\
			\textit{INTEGRAL} & (6.1$^{\text{+0.3}}_{\text{--0.3}}$)$^{\mathrm{f}}$ / (6.1$^{\text{+0.3}}_{\text{--0.3}}$)$^{\mathrm{f}}$ & (0.55$^{\text{+0.03}}_{\text{--0.03}}$) / (0.57$^{\text{+0.03}}_{\text{--0.03}}$) & (4.1$^{\text{+0.2}}_{\text{--0.2}}$) / (4.1$^{\text{+0.2}}_{\text{--0.2}}$) & \citen{Mereghetti+2020a} \\
			\textit{Konus-Wind} & (9.7$^{\text{+0.4}}_{\text{--0.4}}$)$^{\mathrm{g}}$ / (9.7$^{\text{+0.4}}_{\text{--0.4}}$)$^{\mathrm{g}}$ & (0.60$^{\text{+0.02}}_{\text{--0.02}}$) / (0.61$^{\text{+0.03}}_{\text{--0.03}}$) & (5.0$^{\text{+0.2}}_{\text{--0.2}}$) / (5.0$^{\text{+0.2}}_{\text{--0.2}}$) & \citen{Ridnaia+2021} \\
			\bottomrule
			\multicolumn{5}{l}{Predicted \text{X-ray} fluence values at the distance of FRB~20200120E (3.63\,$\pm$\,0.34\,Mpc\,\cite{Freedman+1994}) were calculated using the best-fit} \\
			\multicolumn{5}{l}{(exponentially cut-off power-law) spectral models in refs.~\citen{Tavani+2021}, \citen{Li+2021}, \citen{Mereghetti+2020a}, and~\citen{Ridnaia+2021} for SGR~1935+2154. We assumed a} \\
			\multicolumn{5}{l}{fiducial distance of 10\,kpc to SGR~1935+2154 and hydrogen column densities of $N_{\text{H}}$\,$=$\,2.79\,$\times$\,10$^{\text{22}}$\,cm$^{\text{--2}}$\,\cite{Li+2021} and} \\
			\multicolumn{5}{l}{$N_{\text{H}}$\,$=$\,6.73\,$\times$\,10$^{\text{20}}$\,cm$^{\text{--2}}$\,\cite{HI4PI+2016} towards SGR~1935+2154 and FRB~20200120E, respectively.} \\
			\multicolumn{5}{l}{$^{\mathrm{a}}$ Absorbed and unabsorbed \text{X-ray} fluences are derived from best-fit spectral models for SGR~1935+2154.} \\
			\multicolumn{5}{l}{$^{\mathrm{b}}$ Predicted absorbed and unabsorbed \text{X-ray} fluences in the \text{0.5--10}\,keV energy band.} \\
			\multicolumn{5}{l}{$^{\mathrm{c}}$ Predicted absorbed and unabsorbed \text{X-ray} fluences in the 3--79\,keV energy band.} \\
			\multicolumn{5}{l}{$^{\mathrm{d}}$ Measured \text{X-ray} fluence in the \text{18--60}\,keV energy band.} \\
			\multicolumn{5}{l}{$^{\mathrm{e}}$ Measured \text{X-ray} fluence in the \text{1--250}\,keV energy band.} \\
			\multicolumn{5}{l}{$^{\mathrm{f}}$ Measured \text{X-ray} fluence in the \text{20--200}\,keV energy band.} \\
			\multicolumn{5}{l}{$^{\mathrm{g}}$ Measured \text{X-ray} fluence in the \text{20--500}\,keV energy band.}
	\end{tabular}}
	
	\label{tab:sgr1935_xray_fluences}
\end{table*}


\end{document}